\documentclass[11pt,letterpaper]{article}
\usepackage[T1]{fontenc}
\usepackage[utf8]{inputenc}
\usepackage[letterpaper,margin=1in]{geometry}
\usepackage{amsmath,amsthm,amsfonts,amssymb,mathtools}
\usepackage[dvipsnames]{xcolor}
\usepackage{lmodern}
\usepackage{bm}
\usepackage{microtype,booktabs,enumitem}
\usepackage{multirow}
\usepackage{float}
\usepackage{mdframed}
\usepackage{tikz}
\usetikzlibrary{arrows.meta,calc}
\usepackage[backend=biber,style=alphabetic,sorting=nyt,maxbibnames=99,maxalphanames=5,minalphanames=3]{biblatex}
\usepackage[colorlinks=true,linkcolor=MidnightBlue,citecolor=MidnightBlue,urlcolor=MidnightBlue]{hyperref}
\hypersetup{pdftitle={Softmax Attention with Any Temperature in Fixed Dimension: Matching the Best Known Maximum-Inner-Product Bounds},pdfauthor={Zixuan Huang and Guozhang Chen},pdfsubject={Fine-grained complexity of softmax attention in fixed dimension},pdfkeywords={softmax attention, maximum inner product, ray shooting, halfspace range searching, fine-grained complexity}}
\usepackage[capitalize,noabbrev]{cleveref}
\AtBeginBibliography{\raggedright}

\allowdisplaybreaks
\setlist{itemsep=2pt,topsep=5pt}
\newmdenv[
  linecolor=black!45,linewidth=0.5pt,backgroundcolor=white,
  skipabove=9pt,skipbelow=9pt,
  innerleftmargin=8pt,innerrightmargin=8pt,
  innertopmargin=6pt,innerbottommargin=6pt,
  splittopskip=6pt,splitbottomskip=6pt,
  needspace=6\baselineskip
]{procedurebox}

\theoremstyle{plain}
\newtheorem{theorem}{Theorem}[section]
\newtheorem{lemma}[theorem]{Lemma}
\newtheorem{proposition}[theorem]{Proposition}
\newtheorem{corollary}[theorem]{Corollary}
\newtheorem{conjecture}[theorem]{Conjecture}
\theoremstyle{definition}
\newtheorem{definition}[theorem]{Definition}
\theoremstyle{remark}
\newtheorem{remark}[theorem]{Remark}
\numberwithin{equation}{section}

\DeclareMathOperator{\Att}{Att}
\DeclareMathOperator{\ATT}{Attention}
\DeclareMathOperator{\BMIP}{BatchMaxIP}
\DeclareMathOperator{\MIP}{MaxIP}
\DeclareMathOperator{\softmax}{softmax}
\DeclareMathOperator{\conv}{conv}
\DeclareMathOperator{\poly}{poly}
\DeclareMathOperator{\polylog}{polylog}

\DeclareMathOperator{\vertset}{vert}
\DeclareMathOperator{\diag}{diag}
\newcommand{\eps}{\varepsilon}
\newcommand{\vct}[1]{\bm{#1}}
\newcommand{\mat}[1]{\bm{#1}}
\newcommand{\e}{\mathrm{e}}
\let\Gamma\varGamma
\let\Phi\varPhi
\newcommand{\R}{\mathbb R}
\newcommand{\Q}{\mathbb Q}
\newcommand{\Z}{\mathbb Z}
\newcommand{\E}{\mathbb E}
\newcommand{\Prob}[1]{\Pr\!\left[#1\right]}
\newcommand{\wt}{\widetilde}
\newcommand{\Oh}{\mathrm{O}}
\newcommand{\softO}{\widetilde{\Oh}}
\newcommand{\wh}{\widehat}
\newcommand{\cH}{\mathcal H}
\newcommand{\cR}{\mathcal R}
\newcommand{\KP}{\mathrm{KP}}
\newcommand{\BP}{\mathrm{BP}}
\newcommand{\Cenv}{C_{\mathrm{cell}}}
\newcommand{\Ztop}{Z_{\mathrm{top}}}
\newcommand{\Zcap}{Z_{\mathrm{cap}}}
\newcommand{\etablk}{\eta_{\mathrm{blk}}}
\newcommand{\nint}{n_{\mathrm{int}}}
\newcommand{\ngrp}{n_{\mathrm{grp}}}
\newcommand{\sown}{s_{\mathrm{own}}}
\newcommand{\cconf}{c_{\mathrm{conf}}}
\newcommand{\Csamp}{C_{\mathrm{samp}}}
\newcommand{\dpert}{\delta_{\mathrm{pert}}}
\newcommand{\acap}{a_{\mathrm{cap}}}
\newcommand{\cpt}{c_{\mathrm{pert}}}
\newcommand{\cP}{\mathcal P}
\DeclareMathOperator{\aff}{aff}
\DeclareMathOperator{\intr}{int}

\title{Softmax Attention with Any Temperature in Fixed Dimension:\\ Matching the Best Known Maximum-Inner-Product Bounds}
\author{%
Zixuan Huang$^{1,2}$\qquad
Guozhang Chen$^{1,3}$\thanks{Corresponding author.}\\[4pt]
\normalsize $^1$\,State Key Laboratory of Multimedia Information Processing, School of Computer Science\\
\normalsize $^2$\,School of Physics\\
\normalsize $^3$\,National Engineering Research Center of Visual Technology, School of Computer Science\\
\normalsize Peking University, Beijing 100871, China\\[4pt]
\normalsize\texttt{2400011442@stu.pku.edu.cn}\qquad
\normalsize\texttt{guozhang.chen@pku.edu.cn}
}
\date{}

\begin{document}
\pagenumbering{roman}
\maketitle
\thispagestyle{empty}
\begin{abstract}
\normalsize
Attention is central to modern AI models for language, vision, and scientific modeling. However, direct evaluation compares every query with every key, requiring quadratic time in the context length, which limits its scalability to long contexts. Consequently, characterizing the complexity bounds of attention is of fundamental importance.

For every fixed dimension~$d$, we give an algorithm for $m$ queries, $n$ key--value pairs, and additive error $\eps$, with expected time
$$\softO\bigl((mn)^{k/(k+1)}+m+n\bigr),\qquad k=\lfloor d/2\rfloor.$$
For fixed dimension and a fixed number of value channels, soft-O hides factors of the form $\poly_d\bigl(b,\log(m+n),\log B,\log(1/\eps)\bigr),$
where~$b$~bounds input encoding lengths and~$B$~bounds coordinate and value magnitudes. In particular, the dependence on $B$ and $1/\eps$ is polylogarithmic, which allows arbitrary positive temperature. When $m=n$ and~$d\ge2$, the bound is~$\softO(n^{2-2/(\lfloor d/2\rfloor+1)})$: near-linear in dimensions two and three, with exponent $4/3$ in dimensions four and five and $3/2$ in dimensions six and seven. It improves the previous $\softO(n^{2-1/d})$ upper bound and matches the best known batch maximum-inner-product bound in each fixed dimension. Additionally, we reduce integer batch maximum inner product computation to one attention call with the same dimension and input sizes $m,n$.
\end{abstract}
\clearpage
\setcounter{tocdepth}{3}
\tableofcontents
\clearpage
\pagenumbering{arabic}

\section{Introduction}
\label{sec:intro}

Transformers and related attention-based architectures are central to modern AI, with wide applications in areas such as language processing~\cite{vaswani2017}, visual recognition~\cite{dosovitskiyEtAl2021}, and biomolecular structure prediction~\cite{abramsonEtAl2024}. By exponentiating query--key inner products into a sharp distribution concentrated on the most relevant keys, attention also exhibits close connections to classical problems in Theoretical Computer Science (TCS), such as Maximum Inner Product (Max-IP) problem, making it a fundamental subject of study in TCS.

As frontier problems grow in complexity, extending the context length of Transformers has become essential. However, the cost of direct attention evaluation grows quadratically, with $n^2$ pairwise interactions for a context of length $n$ even in fixed dimension. This quadratic bottleneck restricts the deployment of Transformers in a wide range of long-context applications,  such as repository-level code generation or full-document analysis.

Practical algorithms reduce the cost of attention through memory-efficient exact evaluation, as in FlashAttention~\cite{daoEtAl2022}, and random-feature approximation with provable error guarantees, as in Performer~\cite{choromanskiEtAl2021}. Nevertheless, these methods did not establish a non-trivial asymptotic upper bound for attention that faithfully preserves the exact softmax semantics. Establishing theoretical upper bounds for attention helps us delineate the fundamental boundaries of where attention can be applied.

Formally, softmax attention maps queries, keys, and values to normalized exponential averages. For query vectors~$\vct{q}_1,\ldots,\vct{q}_m\in\R^d$, key vectors~$\vct{k}_1,\ldots,\vct{k}_n\in\R^d$, and scalar values~$v_1,\ldots,v_n$, the output associated with query~$\vct{q}_i$~is
$$\frac{\sum_{j=1}^n \e^{\vct{q}_i\cdot \vct{k}_j}v_j}{\sum_{j=1}^n \e^{\vct{q}_i\cdot \vct{k}_j}}.$$
More generally, each key~$\vct{k}_j$~carries a vector~$\vct{v}_j=(v_{j1},\ldots,v_{jr})$~of~$r$~values, one for each value channel, and the output of a query has one such average for each channel. Direct evaluation processes all~$mn$~query--key pairs, although the input and output sizes are only linear in $m+n$ when the dimension and number of value channels are fixed.

An inverse temperature~$\beta>0$~rescales each score to~$\beta\,\vct{q}_i\cdot\vct{k}_j$~and is absorbed into the query coordinates. The polylogarithmic dependence on coordinate magnitudes therefore extends the computational bounds to every positive rational temperature (Remark~\ref{rem:temperature}).

Gupta, Huang, Saha, Xu, and Ye~\cite{gupta2026} gave an algorithm with running time
$$\softO(n^{2-1/d})$$
for every fixed dimension~$d$. For each query, their algorithm retains keys whose scores lie within a logarithmic window of the row maximum and approximates the exponential on that window by a polynomial. It evaluates the resulting monomial sums with a simplex range-searching data structure~\cite{matousek1992}, whose query cost yields the exponent~$2-1/d$. Their reduction from maximum inner product also established conditional hardness in dimension~$2^{\Oh(\log^*n)}$ via the lower bound of Chen~\cite{chen2018}. They further observe that attention whose query or key matrix has rank at most~$d$~reduces to dimension~$d$~through a rank factorization~\cite[Theorem~1.2]{gupta2026}, so fixed-dimensional algorithms also apply to low-rank attention with a large head dimension. Additionally, they gave a reduction from $d$-dimensional Max-IP to $(d+1)$-dimensional attention, where Max-IP is the natural geometric core of attention: given query vectors~$\vct{x}_i$~and points~$\vct{y}_j$, the batch problem asks for all maxima~$M_i=\max_j \vct{x}_i\cdot \vct{y}_j$.

However, the above upper bound is not sufficiently tight, considering that in fixed dimension~$d$, the best known algorithms answer~$m$~maximum-inner-product queries on~$n$~points in~$\softO((mn)^{k/(k+1)}+m+n)$~time, where~$k=\lfloor d/2\rfloor$~\cite{matousekSchwarzkopf1993,chan1996output}; for~$d\le3$~this is near-linear~\cite{preparataShamos1985,dobkinKirkpatrick1990}. Furthermore, because the reduction increases the dimension by one, one cannot obtain non-trivial hardness for $4$-dimensional attention.

While Max-IP only seeks the extreme inner product, computing attention requires aggregating the exponentially weighted contributions of all keys. This leaves the question whether the global aggregation in softmax attention makes it strictly harder than Max-IP, or if we can still match the best known complexity of batch Max-IP and give a  dimension-preserving reduction in every fixed dimension.

In this work, we study the bit complexity of softmax attention, with prescribed entrywise additive error~$\eps$. Our running-time bounds depend polynomially on the input encoding length and on $\log(1/\eps)$, so inverse-polynomial error contributes only polylogarithmic factors.
\subsection{Results}

In every fixed dimension, our attention algorithm matches the best known dependence on $m$ and $n$ for batch maximum inner product. To state the result, let $N=m+n$, suppose the input coordinates are rationals of magnitude at most $B\ge1$ and encoding length at most $b$, and let $\eps$ be the target additive error.

\begin{theorem}[Attention in every fixed dimension; proof in Section~\ref{subsec:mainproof}]
\label{thm:main}
Fix integers~$d\ge1$~and~$r\ge1$, and let~$k=k(d)=\lfloor d/2\rfloor$. There is a Las Vegas algorithm for~$\ATT_d$~(Definition~\ref{def:att}) with~$r$~value channels whose expected running time is
$$\softO\bigl((mn)^{k/(k+1)}+m+n\bigr).$$
Here soft-O suppresses factors of the form
$$\poly_d\bigl(b,\log(m+n),\log B,\log(1/\eps)\bigr),$$
with constants that may also depend on the fixed number~$r$~of value channels.
\end{theorem}

For~$d\ge2$,~$m=n$,~$b=\polylog n$~and~$B,\eps^{-1}=n^{\Oh(1)}$, the expected running time is~$\softO(n^{2-2/(\lfloor d/2\rfloor+1)})$. This is near-linear for~$d\in\{2,3\}$; Table~\ref{tab:results} lists the exponents for~$d\le9$~and compares them with the previous bounds. For~$d=1$~the bound is~$\softO(m+n)$; an~$\softO(n)$~algorithm for this case was already known~\cite[Lemma~A.1]{gupta2026}. Since~$2/(\lfloor d/2\rfloor+1)>1/d$~for every~$d\ge2$, Theorem~\ref{thm:main} improves the bound~$\softO(n^{2-1/d})$~of~\cite{gupta2026} in every dimension~$d\ge2$, while retaining their polylogarithmic dependence on $B$ and $1/\eps$. Moreover, in the plane ($d=2$), our algorithm is completely deterministic and runs in $\softO(n)$ time (Theorem~\ref{thm:d2}).

Writing $F_d(m,n)=(mn)^{k(d)/(k(d)+1)}+m+n$, the same algorithm supports a variable number $r$ of value channels in expected time $r\,\softO(F_d(m,n))$, where soft-O suppresses factors polynomial in $\log(r+2)$.

\begin{table}[t]
\centering
\caption{Time bounds and the offline counting baseline derived in Section~\ref{subsec:overview-planar} by combining polynomial approximation with offline halfspace range counting. Here $d$ is fixed, $m=n$, coordinate encoding lengths are polylogarithmic, magnitudes are polynomial, and $\eps=n^{-\Oh(1)}$. The exponent is the same in dimensions $2k$ and $2k+1$, where $k=\lfloor d/2\rfloor$; the algorithm is Las Vegas for $d\ge3$ and deterministic for $d=2$. The last column lists the best known deterministic real-RAM bound for exact batch maximum inner product~\cite{preparataShamos1985,dobkinKirkpatrick1990,matousekSchwarzkopf1993,chan1996output}, which reduces to attention by Theorem~\ref{thm:red}.}
\label{tab:results}
\begin{tabular}{@{}ccccc@{}}
\toprule
\multirow[c]{2}{*}[-7pt]{Dimension} & \multicolumn{4}{c}{Time upper bounds $\softO(\cdot)$}\\
\cmidrule(lr){2-5}
& $\vcenter{\hbox{\shortstack{Prior work\\\cite{gupta2026}}}}$ & $\vcenter{\hbox{\shortstack{Offline counting\\baseline}}}$ & \textbf{This work} & Batch Max-IP\\
\midrule
$2$&$n^{3/2}$&$n^{4/3}$&$n$&$n$\\
$3$&$n^{5/3}$&$n^{3/2}$&$n$&$n$\\
$4$&$n^{7/4}$&$n^{8/5}$&$n^{4/3}$&$n^{4/3}$\\
$5$&$n^{9/5}$&$n^{5/3}$&$n^{4/3}$&$n^{4/3}$\\
$6$&$n^{11/6}$&$n^{12/7}$&$n^{3/2}$&$n^{3/2}$\\
$7$&$n^{13/7}$&$n^{7/4}$&$n^{3/2}$&$n^{3/2}$\\
$8$&$n^{15/8}$&$n^{16/9}$&$n^{8/5}$&$n^{8/5}$\\
$9$&$n^{17/9}$&$n^{9/5}$&$n^{8/5}$&$n^{8/5}$\\
\midrule
$d\ge2$&$n^{2-1/d}$&$n^{2-2/(d+1)}$&$n^{2-2/(k+1)}$&$n^{2-2/(k+1)}$\\
\bottomrule
\end{tabular}
\end{table}

Batch maximum inner product also reduces to attention in the same dimension. For integer matrices~$\mat{X}\in\Z^{m\times d}$~and~$\mat{Y}\in\Z^{n\times d}$~with rows~$\vct{x}_i$~and~$\vct{y}_j$, the problem~$\BMIP_d$~asks for every~$M_i=\max_j \vct{x}_i\cdot \vct{y}_j$.

When the values equal the keys, the inner product of the attention output with the original query is a softmax average of the scores. Its distance from the maximum is at most $(\ln n)/\beta_0$, so integer rounding determines every maximum exactly at a suitable precision; Section~\ref{subsec:exact-reduction} contains the formal reduction and proof. Whereas the earlier reduction of~\cite{gupta2026} used polylogarithmically many calls in dimension $d+1$, Theorem~\ref{thm:red-informal} reduces batch maximum inner product to a single attention call within the same dimension $d$.

\begin{samepage}
\begin{theorem}[Dimension-preserving reduction, informal]
\label{thm:red-informal}
The problem~$\BMIP_d$~reduces to one attention call with~$d$~value channels, namely
$$\mat{Q}=\beta_0\mat{X},\qquad \mat{K}=\mat{Y},\qquad \mat{V}=\mat{Y},\qquad\beta_0=\Theta(\log(n+1)).$$
For a suitable absolute constant in $\beta_0$, the attention instance has the same~$m,n,d$ and uses inverse-polynomial error for polynomially bounded integer inputs; the reduction requires deterministic near-linear additional bit operations for fixed~$d$. Scaling~$\mat{V}$~by~$dB$, where~$B$~bounds the integer coordinates, yields a constant-error version.
\end{theorem}
\end{samepage}

The reduction also transfers the conditional quadratic lower bound for maximum inner product~\cite{sahaXuYe2026} under the Orthogonal Vectors Hypothesis (OVH)~\cite{williams2005} to every constructible dimension~$d(n)$~with~$d(n)\to\infty$~and~$d(n)=n^{\mathrm{o}(1)}$. This hardness consequence was already observed in~\cite{sahaXuYe2026} through the earlier reduction. Section~\ref{subsec:superconstant} gives the precise statement.

\paragraph{Consequences in fixed dimension.}
Theorems~\ref{thm:main} and~\ref{thm:red-informal} show that, in every fixed dimension, the best known exponents for attention and for batch maximum inner product coincide. Hence, for~$d\ge4$, an attention algorithm with exponent below~$2-2/(\lfloor d/2\rfloor+1)$~on instances with~$d$~value channels, polynomial magnitudes, and inverse-polynomial or constant error would improve the exponent for integer batch maximum inner product with polynomially bounded coordinates, and hence for offline halfspace emptiness, in the same dimension. Applying a single-channel attention algorithm to each of the~$d$~value columns gives this implication as well (Theorem~\ref{thm:red}(c)), since~$d$~is fixed. Both problems admit near-linear algorithms for~$d\le3$, whereas whether improvements for batch maximum inner product in higher dimensions transfer back to attention remains open (Section~\ref{sec:conclusion}).

The planar algorithm also yields a static data structure that preprocesses fixed keys, values, numerical bounds, and error tolerance in $\softO(n)$ time and space and answers each query satisfying these bounds in $\softO(1)$ time (Corollary~\ref{cor:online-d2}). For exact halfplane aggregation over a faithful semigroup with near-linear storage, a query may require $\wt\Omega(n^{1/3})$ operations~\cite{aryaMountXia2012} (Section~\ref{sec:semigroup}). The attention data structure avoids this barrier by using exponential decay to sum over a geometrically convenient superset of near-maximal keys with a small omitted contribution. Similarly, while relative-error halfplane counting admits near-linear storage through random sampling~\cite{aronovHarPeled2008}, with query time polylogarithmic in the input size and polynomial in the reciprocal of the relative error, the attention bound is polylogarithmic in both the input size and $1/\eps$.

\subsection{Technical Overview}
\label{subsec:techniques}
\label{subsec:organization}

Subtracting the maximum score from each row leaves the attention output unchanged and puts the exponential weights in $(0,1]$, with a weight of $1$ in every row. By exponential decay, the total contribution from keys outside a logarithmic window of the maximum is small, so it suffices to aggregate the potentially numerous keys within that window.

In the plane, we describe these keys by radial and boundary ranges over which polynomial moments can be summed in near-linear time. In higher dimensions, we cover the significant pairs by pairwise disjoint Cartesian blocks $I\times J$ and evaluate them through lower-dimensional kernels, in total expected time $\softO(F_d(m,n))$. For $d\ge3$, the reductions are
$$\ATT_d\to\KP_d
\xrightarrow{\text{temperature interpolation}}\BP_{d-1}
\xrightarrow{\text{anchor blocks}}\KP_{d-1}\to\cdots\to\KP_2.$$
Here $\KP_d$ denotes shifted-kernel products in dimension $d$, and $\BP_{d-1}$ denotes biased exponential products with $(d-1)$-dimensional inner products.

\subsubsection{Normalization and the Planar Base Case}
\label{subsec:overview-planar}
For one value channel, write the shifted numerator and denominator as
$$F_i=\sum_j \e^{\vct{q}_i\cdot\vct{k}_j-M_i}v_j,\qquad
D_i=\sum_j \e^{\vct{q}_i\cdot\vct{k}_j-M_i},\qquad
M_i=\max_j\vct{q}_i\cdot\vct{k}_j.$$
Since $D_i\ge1$ and $|F_i|\le BD_i$, approximating both sums to a small constant multiple of $\eps/(B+1)$ ensures error at most $\eps$ in their ratio (Lemma~\ref{lem:shift}).

\begin{figure}[t]
\centering
\begin{tikzpicture}[x=1cm,y=1cm,font=\normalsize]
\fill[MidnightBlue!10] (7.0,-0.3) rectangle (12.7,1.6);
\draw[->,line width=0.7pt] (0,0) -- (13.5,0);
\draw[dashed,MidnightBlue] (7.0,-0.3) -- (7.0,1.6);
\draw[line width=0.7pt] (12.7,-0.12) -- (12.7,1.6);
\foreach \x in {0.6,1.5,2.7,3.4,4.8,5.7,6.5}
  \fill[black!45] (\x,0) circle (2pt);
\foreach \x in {7.3,7.8,8.2,8.9,9.5,10.0,10.5,10.9,11.3,11.7,12.1,12.4,12.7}
  \fill[MidnightBlue] (\x,0) circle (2pt);
\node[below] at (7.0,-0.18) {$M_i-L$};
\node[below] at (12.7,-0.18) {$M_i$};
\node[align=center] at (3.3,1.0)
  {Discardable tail\\absolute contribution at most $nB\e^{-L}$};
\node[align=center,MidnightBlue] at (9.85,1.0)
  {Near-maximal scores\\possibly all $n$ keys};
\node[below,align=center] at (3.3,-0.18)
  {increasing score $\vct{q}_i\cdot\vct{k}_j$};
\draw[<->,MidnightBlue] (7.1,1.9) -- node[above] {window width $L$} (12.6,1.9);
\end{tikzpicture}
\caption{Scores in one attention row. The shaded interval contains the near-maximal keys; each omitted key contributes at most $B\e^{-L}$ to the shifted numerator.}
\label{fig:score-window}
\end{figure}
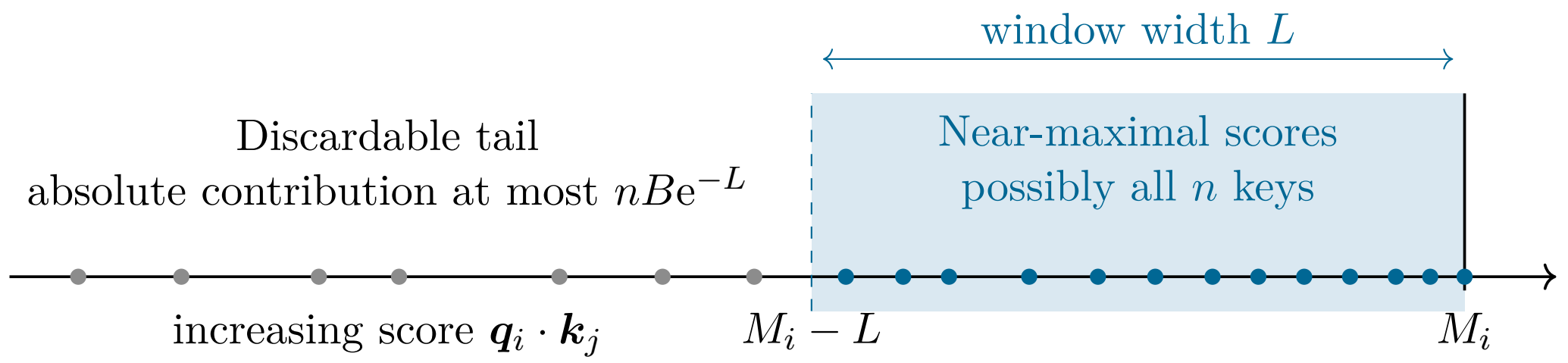

For a window width~$L>0$, every key whose score is below~$M_i-L$~has shifted weight at most~$\e^{-L}$. Discarding any subset of such keys changes~$F_i$~by at most~$nB\e^{-L}$~and~$D_i$~by at most~$n\e^{-L}$. Thus a logarithmic window suffices (Figure~\ref{fig:score-window}), as in~\cite{gupta2026}. We may retain any superset of this window, which allows us to use radial ranges in the plane and common-anchor blocks in higher dimensions.

The score threshold defines a halfspace range, over which the approach of~\cite{gupta2026} evaluates monomial sums. If~$b=\polylog(m+n)$~and~$B,\eps^{-1}=(m+n)^{\Oh(1)}$, combining this approach with offline range counting and the standard space/query tradeoff~\cite{agarwalErickson1999} gives time~$\softO((mn)^{d/(d+1)}+m+n)$. For $m=n$, we obtain the offline counting baseline $\softO(n^{2-2/(d+1)})$ in Table~\ref{tab:results}, with exponent~$8/5$~in dimension four compared with~$4/3$~for batch maximum inner product.

To go beyond the baseline and achieve the maximum-inner-product exponent determined by~$\lfloor d/2\rfloor$, we use the geometry of halfspace emptiness and convex polytopes to organize near-maximal pairs. Polynomial moments and common-anchor blocks yield the corresponding weighted sums within this bound, including for near-maximal sets containing almost all keys.

In the plane, we center the key polygon $\mathcal C=\conv\{\vct{k}_1,\ldots,\vct{k}_n\}$ at its area centroid $\vct c$. This translation adds the same scalar $-\vct q_i\cdot\vct c$ to every score in a row and preserves the attention output. We continue to write $\vct k_j,\mathcal C$ for the centered keys and polygon.

Centering ensures that every score lies between $-2h_i$ and the support value~$h_i=\max_j\vct{q}_i\cdot\vct{k}_j$. Thus, when $h_i\le L$, all shifted scores lie in $[-3L,0]$ and we retain every key. For $h_i>L$, write each nonzero key as~$\vct{k}_j=\rho_j\vct{k}_j^{\partial}$, where~$\vct{k}_j^{\partial}$~lies on the boundary of~$\mathcal C$~and~$0<\rho_j\le1$. Separate thresholds on~$\rho_j$~and~$\vct{q}_i\cdot\vct{k}_j^{\partial}$~retain every key within $L$ of the maximum and only keys within $2L$ of it. The boundary threshold selects a single arc, so the selected range is specified by a radial threshold and at most two intervals in the boundary order.

Expanding the approximating polynomial in query and key coordinates reduces the product to range sums of weighted key monomials. An offline sweep with binary indexed trees~\cite{fenwick1994} computes these moments, which we combine with the query coefficients to obtain the kernel product. Reversing the updates and queries computes the transpose product, and storing successive versions of the sweep in persistent segment trees~\cite{driscollEtAl1989} yields the online data structure.

\subsubsection{Temperature Interpolation and Common-Anchor Blocks}
\label{subsec:overview-induction}
For general~$d$~we induct on the dimension. For rows~$\vct{q}_1,\ldots,\vct{q}_m\in\Q^d$~and columns~$\vct{k}_1,\ldots,\vct{k}_n\in\Q^d$, the shifted kernel is the~$m\times n$~matrix~$\mat{T}$~with entries
$$T_{ij}=\exp(\vct{q}_i\cdot \vct{k}_j-M_i),\qquad M_i=\max_{j\in[n]}\vct{q}_i\cdot \vct{k}_j.$$

Attention is the ratio of a product~$\mat T\vct w$ for a value channel and the product for the all-ones vector. The shifted-kernel problem~$\KP_d$~(Definition~\ref{def:KP}) asks for~$\mat{T}\vct{w}$~with a given vector~$\vct{w}\in\Q^n$, or~$\mat{T}^{\top}\vct{z}$~with a given vector~$\vct{z}\in\Q^m$, to additive error~$\eta$; the entries of~$\vct{w}$~and~$\vct{z}$~have absolute value at most~$W$. We prove by induction that a Las Vegas algorithm computes either product in expected time~$\softO(F_d(m,n))$. The suppressed numerical parameters include the weight bound $W$ and error tolerance $\eta$.

The induction includes both the forward product $\mat T\vct w$ and the transpose product $\mat T^\top\vct z$, since the anchor's orientation determines which lower-dimensional product evaluates a block. The planar algorithm computes both products deterministically in $\softO(m+n)$ time and serves as the base case (Lemma~\ref{lem:KPbase}).

For the first reduction, group nonzero queries by the position and sign of a largest coordinate. After applying the corresponding signed permutation to queries and keys, write $\vct q_i=\beta_i(\vct u_i,1)$ and $\vct k_j=(\vct p_j,c_j)$, where $\beta_i=\|\vct q_i\|_\infty$ and $\vct u_i\in[-1,1]^{d-1}$. Then
$$T_{ij}=\e^{-\beta_i\mu_{ij}},\qquad
\mu_{ij}=h_i-\vct u_i\cdot\vct p_j-c_j\ge0,\qquad h_i=M_i/\beta_i.$$
Here the row-dependent factor $\beta_i$ still multiplies the inner product. To decouple them, we partition the rows into dyadic intervals of $\beta_i$ and use Lagrange interpolation on each interval to approximate $\mathrm{e}^{-\beta\mu}$ by a linear combination of exponentials $\mathrm{e}^{-\tau\mu}$ at fixed nodes $\tau$. By Lemma~\ref{lem:entrywise}, the error decreases exponentially with the number of nodes, uniformly over $\mu\ge0$, and the sum of the absolute interpolation coefficients grows at most exponentially in that number. Consequently, $\softO(1)$ nodes and precision bits suffice. At a fixed node $\tau$, we rescale $(\vct p_j,c_j)$ by $\tau$ to write the exponent as
$$x_{ij}=\vct u_i\cdot\vct p_j+b_i+c_j\le0,$$
with $d-1$ bilinear coordinates and a bias on each side, defining the intermediate product problem $\BP_{d-1}$ for the matrix $(\e^{x_{ij}})$. In Proposition~\ref{prop:KPstep}, we combine its forward outputs using the interpolation coefficients, or exchange rows and columns and absorb those coefficients into the input weights for a transpose product.

Directly factoring the entire biased matrix into a row-shifted kernel and diagonal matrices would give diagonal factors $\e^{b_i+M'_i}$ and $\e^{c_j}$, where $M'_i=\max_j\vct u_i\cdot\vct p_j$. However, these factors can be exponentially large in the input magnitudes. Instead, we obtain polynomially bounded factors by partitioning the significant pairs into blocks with a common anchor, a row or column whose scores across the block lie in a window of width $L=\Theta(\log(NW/\eta))$ below zero, with constant slack, where $N=m+n$. If column $j_0$ anchors $I\times J$, subtracting its score cancels the row bias, so that
$$x_{ij}-x_{ij_0}=\vct u_i\cdot(\vct p_j-\vct p_{j_0})+c_j-c_{j_0}.$$
After shifting these inner products by their maxima, the block factors as
$$\mat S_{IJ}=(\e^{x_{ij}})_{i\in I,\,j\in J}=\diag(\vct d_{\mathrm{row}})\,\mat T^{\top}\diag(\vct d_{\mathrm{col}}),$$
where $\mat T$ is a shifted kernel in dimension $d-1$, with rows $\vct p_j-\vct p_{j_0}$ and columns $\vct u_i$. For a row anchor, the analogous factorization uses $\mat T$ in place of $\mat T^{\top}$. In both cases, the anchor condition bounds the diagonal factors by $\e^{\Oh(L)}$, giving expected evaluation time $\softO(F_{d-1}(|I|,|J|))$.

We construct disjoint anchor blocks with total expected evaluation time $\softO(F_d(m,n))$ by sampling one side and grouping the other side according to its first near-maximal sample.

\subsubsection{Random Prefix Chains and Heavy/Light Recursion}
\label{subsec:overview-prefixes}

To construct the blocks, we use one side as sources and the other as targets. When columns serve as sources, the points $(\vct p_s,c_s)\in\Q^d$ define affine functions $\vct u\mapsto\vct u\cdot\vct p_s+c_s$ evaluated at the row targets. A small rational perturbation shared by all samples puts these functions in general position and changes each score by at most a constant slack in the window. We then use the resulting geometry to define the blocks while retaining the original scores to evaluate them.

For a sample of $h$ sources, we triangulate their maximizing regions into $\Oh_d(h^k)$ canonical cells, where $k=\lfloor d/2\rfloor$, so that each cell $\Delta$ has a sampled owner $s_\Delta$ whose function equals the envelope on all of $\Delta$. After $\softO(h^k)$ preprocessing, ray shooting locates each target in $\softO(1)$ time. The cell's conflict set $C_\Delta$ consists of the sources that exceed its owner somewhere on the cell, a condition we test at its vertices using a constant number of strict halfspace-reporting queries.

Start with logarithmically many sources and expose doubling prefixes $P_0,P_1,\ldots$ of a random permutation, with batches $B_\ell=P_{\ell+1}\setminus P_\ell$. For a source set of size $n$, the sampling estimates imply
$$|C_\Delta|=\softO(n/|P_\ell|),\qquad
|C_\Delta\cap B_\ell|=\softO(1)$$
simultaneously for all prefix cells with constant probability, by a union bound over the possible cells fixed by the shared perturbation. We test both inequalities and resample upon failure.

A source is a hit for a target if its perturbed score is at least $-L$. We assign each target to the first of $P_0,B_0,B_1,\ldots$ containing a hit and use a maximizing source in that set as its anchor. For a first hit in $B_\ell$, the preceding prefix's envelope lies below the hit threshold at the target, so every hit exceeds the owner of the target's prefix cell and belongs to its conflict set $C_\Delta$.

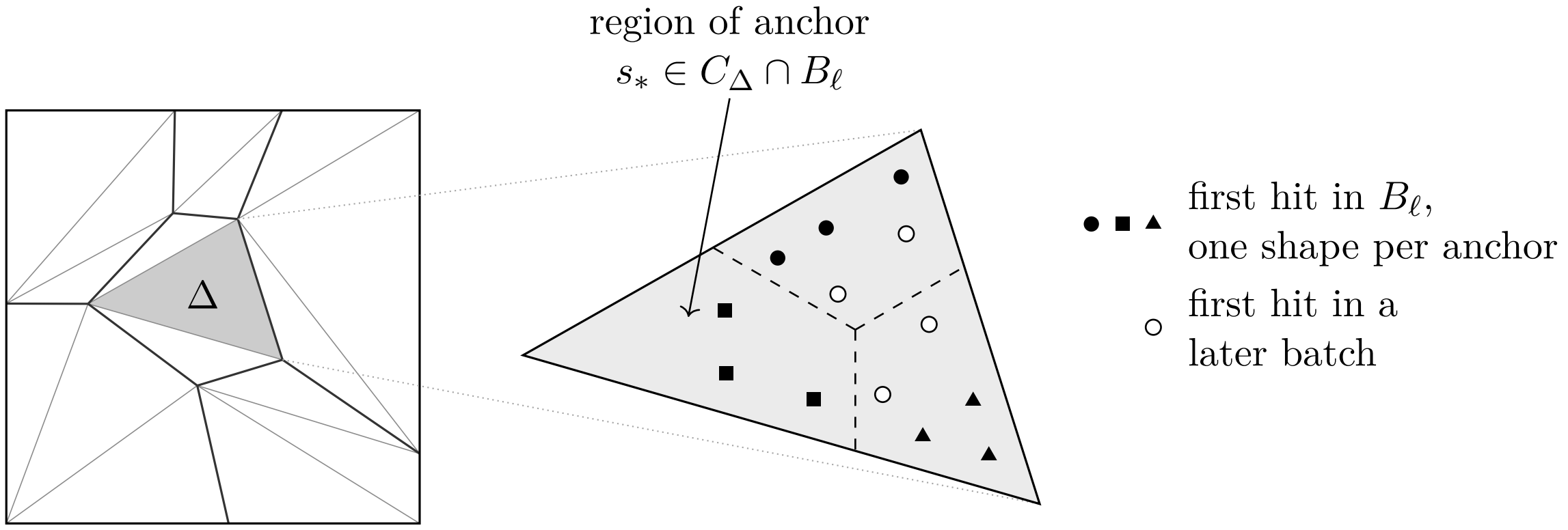
\begin{figure}[t]
\centering
\begin{tikzpicture}[font=\normalsize]
\fill[black!20] (-1.209,0.127) -- (0.673,-0.415) -- (0.240,0.948) -- cycle;
\draw[line width=0.3pt,black!45] (-2.000,-2.000) -- (-1.209,0.127);
\draw[line width=0.3pt,black!45] (-2.000,-2.000) -- (-0.152,-0.665);
\draw[line width=0.3pt,black!45] (-2.000,0.129) -- (-0.387,1.005);
\draw[line width=0.3pt,black!45] (-2.000,0.129) -- (-0.369,2.000);
\draw[line width=0.3pt,black!45] (-1.209,0.127) -- (0.240,0.948);
\draw[line width=0.3pt,black!45] (-1.209,0.127) -- (0.673,-0.415);
\draw[line width=0.3pt,black!45] (-0.387,1.005) -- (0.667,2.000);
\draw[line width=0.3pt,black!45] (-0.152,-0.665) -- (2.000,-2.000);
\draw[line width=0.3pt,black!45] (-0.152,-0.665) -- (2.000,-1.324);
\draw[line width=0.3pt,black!45] (0.240,0.948) -- (2.000,-1.324);
\draw[line width=0.3pt,black!45] (0.240,0.948) -- (2.000,2.000);
\draw[line width=0.6pt,black!80] (-2.000,0.129) -- (-1.209,0.127);
\draw[line width=0.6pt,black!80] (-1.209,0.127) -- (-0.387,1.005);
\draw[line width=0.6pt,black!80] (-1.209,0.127) -- (-0.152,-0.665);
\draw[line width=0.6pt,black!80] (-0.387,1.005) -- (-0.369,2.000);
\draw[line width=0.6pt,black!80] (-0.387,1.005) -- (0.240,0.948);
\draw[line width=0.6pt,black!80] (-0.152,-0.665) -- (0.151,-2.000);
\draw[line width=0.6pt,black!80] (-0.152,-0.665) -- (0.673,-0.415);
\draw[line width=0.6pt,black!80] (0.240,0.948) -- (0.667,2.000);
\draw[line width=0.6pt,black!80] (0.240,0.948) -- (0.673,-0.415);
\draw[line width=0.6pt,black!80] (0.673,-0.415) -- (2.000,-1.324);
\draw[line width=0.6pt] (-2.000,-2.000) rectangle (2.000,2.000);
\node at (-0.099,0.220) {$\Delta$};
\draw[black!35,densely dotted] (0.240,0.948) -- (6.850,1.811);
\draw[black!35,densely dotted] (0.673,-0.415) -- (8.000,-1.811);
\filldraw[fill=black!8,line width=0.6pt] (3.000,-0.371) -- (8.000,-1.811) -- (6.850,1.811) -- cycle;
\draw[dashed,line width=0.5pt] (6.216,-0.124) -- (4.839,0.671);
\draw[dashed,line width=0.5pt] (6.216,-0.124) -- (7.271,0.486);
\draw[dashed,line width=0.5pt] (6.216,-0.124) -- (6.216,-1.297);
\fill (4.956,0.061) +(-1.9pt,-1.9pt) rectangle +(1.9pt,1.9pt);
\fill (5.934,0.862) circle (2.1pt);
\draw[fill=white,line width=0.5pt] (6.481,-0.751) circle (2.1pt);
\draw[fill=white,line width=0.5pt] (6.709,0.805) circle (2.1pt);
\fill (7.356,-0.814) +(90:2.6pt) -- +(210:2.6pt) -- +(330:2.6pt) -- cycle;
\fill (4.965,-0.547) +(-1.9pt,-1.9pt) rectangle +(1.9pt,1.9pt);
\draw[fill=white,line width=0.5pt] (6.930,-0.072) circle (2.1pt);
\fill (5.812,-0.796) +(-1.9pt,-1.9pt) rectangle +(1.9pt,1.9pt);
\draw[fill=white,line width=0.5pt] (6.048,0.221) circle (2.1pt);
\fill (5.465,0.571) circle (2.1pt);
\fill (6.869,-1.158) +(90:2.6pt) -- +(210:2.6pt) -- +(330:2.6pt) -- cycle;
\fill (7.508,-1.343) +(90:2.6pt) -- +(210:2.6pt) -- +(330:2.6pt) -- cycle;
\fill (6.660,1.356) circle (2.1pt);
\node[align=center,inner sep=2pt] (anchorRegion) at (5.0,2.6)
  {region of anchor\\$s_*\in C_\Delta\cap B_\ell$};
\draw[->,line width=0.5pt] (anchorRegion.south) -- (4.6,0.0);
\node[below,align=center,text width=4.6cm] at (0,-2.000) {(a) Maximizing regions\\and their triangles};
\node[below,align=center,text width=4.6cm] at (5.500,-2.000) {(b) The cell $\Delta$\\with owner $s_\Delta$};
\begin{scope}[shift={(8.500,0.900)},font=\normalsize]
\fill (0,0) circle (2.1pt); \fill (0.3,0) +(-1.9pt,-1.9pt) rectangle +(1.9pt,1.9pt); \fill (0.6,0) +(90:2.6pt) -- +(210:2.6pt) -- +(330:2.6pt) -- cycle;
\node[right,align=left] at (0.8,0) {first hit in $B_\ell$,\\ one shape per anchor};
\draw[fill=white,line width=0.5pt] (0.6,-1.0) circle (2.1pt);
\node[right,align=left] at (0.8,-1.0) {first hit in a\\ later batch};
\end{scope}
\end{tikzpicture}
\caption{A schematic level of the random doubling prefix for~$d=3$, in the column orientation. (a) A square window of the maximizing regions of the perturbed functions in~$P_\ell$, with each polygon triangulated for the illustration. Thick lines bound the maximizing regions, and thin lines triangulate them. (b) The shaded cell~$\Delta$, with owner~$s_\Delta$, enlarged. Dashed lines separate the regions where different sources of~$C_\Delta\cap B_\ell$~have the largest perturbed score. The markers represent targets in~$\Delta$~whose first hit occurs after~$P_\ell$. Filled markers of the same shape share an anchor~$s_*\in C_\Delta\cap B_\ell$~and form one anchor block with source set~$C_\Delta$; hollow markers have their first hit in a later batch.}
\label{fig:prefix}
\end{figure}

Targets with the same cell and anchor form a block with source set $C_\Delta$ (Figure~\ref{fig:prefix}), and the bound on $C_\Delta\cap B_\ell$ limits each cell to $\softO(1)$ such blocks. Targets first hit in $P_0$ form blocks with the full source set, completing a collection with disjoint target sets that covers every significant pair when the chain reaches the full source set.

For $n$ sources and $m$ targets, the cell counts and conflict sizes over the complete doubling chain sum to $\softO(n^k+m)$, which is also the time for constructing the blocks. By concavity, the logarithmic number of anchors per cell and the disjoint target sets imply the same bound for anchor evaluation, so the complete algorithm is near-linear when $k=1$ or $m\ge n^k$.

When $d\geq 4$ and the two sides are balanced, however, a full prefix is too costly. For $d=4$ and two sides of size $n$, we have $k=2$ and instead truncate the prefix at $\wt\Theta(\kappa)$ sources. Suppressing logarithmic factors, the final prefix has $\Oh(\kappa^2)$ cells with $\Oh(n/\kappa)$ conflicts each, so reporting costs $\Oh(n\kappa)$ operations. In a light cell, the $\Oh((n/\kappa)^{1/2})$ unassigned targets become the sources of a full prefix, with cost $\Oh(n/\kappa)$ per cell and $\Oh(n\kappa)$ time across all light cells. We recurse on the heavy cells, whose total potential is of order at most $n(n/\kappa)^{1/2}$. The two bounds balance at $\kappa\asymp n^{1/3}$, for a total expected running time of $\softO(n^{4/3})$.

In general, at a recursive node $v$ with side sizes $m_v\le n_v$, a full prefix with the smaller side as sources runs in near-linear time when $m_v^k\le n_v$. Otherwise, at a balanced node, we use the larger side as sources and truncate at $\wt\Theta(\kappa)$ sources. Cells with $\Oh((n_v/\kappa)^{1/k})$ unassigned targets are processed by a full prefix in the reverse orientation; heavy cells are handled recursively. The reporting and light-cell cost and the total heavy-child potential have respective bounds
$$n_v\kappa^{k-1}\qquad\text{and}\qquad m_v(n_v/\kappa)^{(k-1)/k}.$$
Balancing them gives local expected time $\softO(\Phi(m_v,n_v))$, where $\Phi(m_v,n_v)=(m_vn_v)^{k/(k+1)}$. Section~\ref{subsec:truncated} specifies the integer prefix size and light-cell threshold, for which the heavy children's potentials sum to at most one quarter of their parent's. The total expected running time is therefore $\softO(F_d(m,n))$. \footnote{We used DeepSeek Harness through several rounds of interactive prompting to brainstorm proof ideas here.}

Since the evaluated blocks are disjoint, each output coordinate receives at most $N$ contributions. A block tolerance of order $\eta/N$ and a score window of order $\log(NW/\eta)$ keep the combined approximation error and omitted tail within $\eta$. Appendix~\ref{app:induction-arithmetic} shows that these parameters require polynomially many precision bits at every recursive level.

\subsection{Related Work}
\label{subsec:related}

\paragraph{Complexity of attention.}
Alman and Song~\cite{almanSong2023} established a sharp magnitude threshold for logarithmic dimension. Entries of magnitude~$\mathrm{o}(\sqrt{\log n})$~permit~$n^{1+\mathrm{o}(1)}$~time at inverse-polynomial error, and magnitude of order~$\sqrt{\log n}$~supports a conditional quadratic lower bound at the same error tolerance.\footnote{Their convention is~$\exp(\mat{Q}\mat{K}^{\top}/d)$, where~$\mat{Q}$~and~$\mat{K}$~are the matrices with rows~$\vct{q}_i$~and~$\vct{k}_j$. We use~$\exp(\mat{Q}\mat{K}^{\top})$; under it, their bounds on the query and key magnitudes are divided by~$\sqrt d$.}

Duman Keles, Wijewardena and Hegde~\cite{kelesWijewardenaHegde2023} proved quadratic lower bounds under the Orthogonal Vectors Hypothesis (OVH)~\cite{williams2005} for exact and approximate attention in every dimension~$d$~with~$d/\log n\to\infty$, and Alman and Song~\cite{almanSong2024} extended the magnitude threshold of~\cite{almanSong2023} to gradient computation. Further lower bounds concern subquadratic alternatives to attention~\cite{almanYu2025} and multi-head, multi-layer transformers~\cite{sahaXuYeYu2026}. For exact integer maximum inner product, the dimension in which OVH implies quadratic hardness was lowered to a polynomial in~$\log\log n$~by Williams~\cite{williams2018}, to~$2^{\Oh(\log^*n)}$~by Chen~\cite{chen2018}, and to every constructible superconstant dimension by Saha, Xu and Ye~\cite{sahaXuYe2026}.

\paragraph{Kernel sums and approximate attention.}
The fast multipole method and the fast Gauss transform~\cite{greengardRokhlin1987,greengardStrain1991} compute kernel sums in fixed dimension in near-linear time, with error relative to the total weight. Alman, Chu, Schild and Song~\cite{almanChuSchildSong2020} determine fine-grained limits of such methods as the dimension grows. The identity~$\e^{\vct{q}_i\cdot \vct{k}_j}=\e^{\|\vct{q}_i\|^2/2}\e^{\|\vct{k}_j\|^2/2}\e^{-\|\vct{q}_i-\vct{k}_j\|^2/2}$~expresses attention through Gaussian kernel sums with weights~$\e^{\|\vct{k}_j\|^2/2}$, whose ratios can be as large as~$\e^{dB^2/2}$. An error bound relative to the total weight can then require relative precision $\eps_{\mathrm{rel}}=\e^{-\Omega(dB^2)}$, leading to polynomial dependence on $B$ in fixed dimension even with a $\polylog(1/\eps_{\mathrm{rel}})$ dependence on precision.

KDEformer~\cite{zandiehEtAl2023} and HyperAttention~\cite{hanEtAl2024} achieve subquadratic time under bounds on entry magnitudes or other data-dependent parameters. Using halfspace reporting, Chen, Liang, Sha, Shi, and Song~\cite{chenEtAl2025hsr} locate large attention entries with error guarantees under distributional assumptions. In contrast, our algorithms achieve worst-case guarantees for arbitrary rational inputs, with factors polynomial in $\log B$, $\log(1/\eps)$, and the coordinate encoding lengths.

\paragraph{Maximum inner product and halfspace range searching.}
Following Matousek's introduction of shallow cuttings and partitions for halfspace reporting and emptiness~\cite{matousek1992reporting}, Matousek and Schwarzkopf~\cite{matousekSchwarzkopf1993} constructed static ray-shooting structures with~$\softO(n^{\lfloor d/2\rfloor})$~preprocessing for~$n$~halfspaces, logarithmic query time, and a space/query tradeoff. Matousek's linear-optimization queries~\cite{matousek1993lpq} then led to an algorithm for the extreme points of an~$n$-point set with time exponent~$2-2/(\lfloor d/2\rfloor+1)$~plus an arbitrarily small positive constant. Chan~\cite{chan1996lpq,chan1996output} simplified these query methods, derived output-sensitive applications, and observed that static data structures replace the small polynomial loss by polylogarithmic factors. Deterministic structures for ray shooting, range reporting, and linear-programming queries were subsequently given by Ramos~\cite{ramos1999,ramos2000}. For~$m$~ray-shooting queries in a polytope defined by~$n$~halfspaces in dimension~$d>3$, Chan~\cite[Corollary~2.4]{chan1996output} stated the batch bound
$$\softO\bigl((mn)^{\lfloor d/2\rfloor/(\lfloor d/2\rfloor+1)}+m+n\bigr);$$
Corollary~2.2 there treats~$d\le3$. The polylogarithmic factors were later improved for the extreme points of a point set in even dimension~\cite{chan2012partition} and for the decision version~\cite{chanZheng2022}; these remain the best known exponents. We refer to the surveys~\cite{matousek1994survey,agarwalErickson1999,agarwal2017range}.

\paragraph{Lower bounds.}
Erickson~\cite{erickson1996halfspace,erickson2000tradeoff} proved an~$\Omega(n^{4/3})$~lower bound for offline halfspace emptiness in every dimension at least five, for polyhedral partitioning algorithms. Such algorithms cover space by a constant number of constant-complexity polyhedra and recurse on the points and halfspaces that meet each of them. This lower bound matches the best known upper bound for~$d=5$ up to polylogarithmic factors, whereas establishing comparable bounds for general algorithms or in dimension four remains open.

\paragraph{Convex hulls, triangulations, and random sampling.}
Our geometric decomposition relies on the upper bound theorem~\cite{mcmullen1970}, the randomized incremental construction of convex hulls~\cite{clarksonShor1989}, with a deterministic alternative due to Chazelle~\cite{chazelle1993hull} (see~\cite{seidel2017hull}), the pulling triangulation~\cite{clarkson1988} and canonical triangulations~\cite{chazelleFriedman1990}, and the random-sampling analysis of Clarkson and Shor~\cite{clarkson1987,clarksonShor1989}. Our random prefixes follow randomized incremental constructions, with exact tests on conflict sizes and batch intersections ensuring deterministic participation bounds for every accepted decomposition. To remove degeneracies within the bit-complexity model, we apply an explicit rational moment-curve perturbation shared by all samples. For background on perturbation methods and exact computation, we refer to the symbolic schemes of~\cite{edelsbrunnerMucke1990,emirisCanny1995} and the geometric and algebraic methods in~\cite{yap1997,basuPollackRoy2006}.

\paragraph{Independent and concurrent work.} 
Very recently, during the final preparation of this manuscript, we became aware of independent and concurrent work by Mausberg~\cite{mausberg_2026_23195937} on the same problem. They improve upon the previously best-known result of Gupta, Huang, Saha, Xu and Ye~\cite{gupta2026}, obtaining an almost-linear\footnote{Following standard terminology, we use almost-linear for $\mathrm{O}(n^{1+o(1)})$ and near-linear for $\softO(n)$, i.e., linear up to polylogarithmic factors. Note that \cite{mausberg_2026_23195937} uses the term near-linear in its title to refer to an $n^{1+o(1)}$ bound.} Las Vegas algorithm for $\ATT_2$ and $\ATT_3$. In comparison, our work gives a deterministic near-linear algorithm for $\ATT_2$, a near-linear Las Vegas algorithm for $\ATT_3$, and a Las Vegas algorithm for higher dimensions with expected running time $\softO\bigl((mn)^{k/(k+1)}+m+n\bigr)$, where $k=\lfloor d/2\rfloor$. The techniques we use are also very different from those of Mausberg.

\section{Preliminaries}
\label{sec:prelim}

For a positive integer~$n$, write~$[n]=\{1,\ldots,n\}$. We use~$\ln$ for the natural logarithm and~$\log$ for the binary logarithm. For a matrix~$\mat A$, let~$\|\mat A\|_{\max}=\max_{i,j}|A_{ij}|$, and let~$\odot$ denote entrywise multiplication. Throughout, $\softO$ suppresses factors polynomial in the input encoding lengths, the logarithms of the instance size and numerical magnitudes, and the required precision in bits; Section~\ref{subsec:bit-model} collects these parameters in $\lambda$. The notation~$\wt\Omega$ suppresses polylogarithmic factors in the instance size. A subscript~$d$ permits dependence on the fixed dimension and number of value channels.

\begin{definition}
\label{def:att}
Given~$\mat{Q}\in\Q^{m\times d}$,~$\mat{K}\in\Q^{n\times d}$, and~$\mat{V}\in\Q^{n\times r}$~with rows~$\vct{q}_i$,~$\vct{k}_j$, and~$\vct{v}_j=(v_{j1},\ldots,v_{jr})$, define
$$\Att(\mat{Q},\mat{K},\mat{V})_{it}=\frac{\sum_{j=1}^n \e^{\vct{q}_i\cdot \vct{k}_j}v_{jt}}{\sum_{j=1}^n \e^{\vct{q}_i\cdot \vct{k}_j}},\qquad i\in[m],\quad t\in[r].$$
All input coordinates are rationals of magnitude at most~$B\ge1$, with signed numerators and positive denominators of bit length at most~$b$, the coordinate encoding length. Given a target additive error~$\eps=2^{-p}$ for an integer~$p\ge0$, the problem~$\ATT_d$ asks for a rational matrix~$\wh{\mat{O}}$ satisfying
$$\|\wh{\mat{O}}-\Att(\mat{Q},\mat{K},\mat{V})\|_{\max}\le\eps.$$
Our geometric algorithms allow any fixed number~$r$~of value channels; the one-call reductions use~$r=d$.
\end{definition}

Every entry of the ideal output~$\Att(\mat{Q},\mat{K},\mat{V})\in\R^{m\times r}$~is a convex combination of the corresponding value coordinates and has absolute value at most~$B$.

Extend the values by the all-ones channel $v_{j0}=1$, and define the shifted kernel by
$$M_i=\max_{j\in[n]}\vct q_i\cdot\vct k_j,\qquad
T_{ij}=\e^{\vct q_i\cdot\vct k_j-M_i},\qquad \mat T=(T_{ij}).$$
For $0\le t\le r$, write $\vct v^{(t)}=(v_{jt})_{j\in[n]}$ and
$$D_i=\sum_{j=1}^nT_{ij},\qquad F_{it}=(\mat T\vct v^{(t)})_i,\qquad F_{i0}=D_i.$$
Then $\Att(\mat Q,\mat K,\mat V)_{it}=F_{it}/D_i$ for $t\in[r]$. Each row of $\mat T$ contains a unit entry, so $1\le D_i\le n$ and $|F_{it}|\le BD_i$; we denote their approximations by $\wh D_i$ and $\wh F_{it}$. Thus $r+1$ kernel products suffice for all value channels, with the $t=0$ product equal to the denominator shared by the $r$ ratios.

A zero query has maximum zero and an all-ones shifted-kernel row, so its attention output is the average of the values. In a forward kernel product, such a row returns the sum of the input weights; in a transpose product, the weights on all zero rows contribute their sum to every output coordinate. We process these rows together in $\softO(m+n)$ time per channel before applying the algorithms below. Coordinate normalization and support queries may therefore assume nonzero query vectors.

We compare the complexity of attention with that of the batch maximum-inner-product problem, defined as follows.

\begin{definition}
\label{def:maxip}
For integer matrices~$\mat{X}\in\Z^{m\times d}$~and~$\mat{Y}\in\Z^{n\times d}$~with rows~$\vct{x}_i$~and~$\vct{y}_j$~and coordinate magnitudes at most an integer~$B\ge1$, the batch problem~$\BMIP_d$~asks for
$$M_i=\max_{j\in[n]}\vct{x}_i\cdot \vct{y}_j,\qquad i\in[m].$$
The global problem~$\MIP_d$~asks for~$\max_i M_i$. Each~$M_i$~is an integer in~$[-dB^2,dB^2]$.
\end{definition}

\subsection{Computation and Error Bounds}
\label{subsec:bit-model}

We measure complexity in bit operations, so each batch or preprocessing time bound also applies to space. Our randomized algorithms are Las Vegas, and their running times are measured in expectation.

Throughout the paper, $\lambda\ge1$ bounds an instance's rational input encoding lengths, the logarithms of its size and numerical magnitudes, and the required precision in bits. Geometric predicates are evaluated exactly. Intermediate geometric quantities, scalar exponentials, polynomial coefficients, and weighted sums use $\poly_d(\lambda)$-bit representations or approximations, as established in Appendix~\ref{app:arithmetic}. Approximate outputs lie on a common dyadic grid whose mesh is proportional to the target tolerance, uniformly in $d$ and $r$. For a data structure, the magnitude and encoding-length bounds on future queries are fixed before preprocessing; a query or target satisfying these bounds is admissible.

Subproblems formed by restricting index sets inherit the numerical bounds of their enclosing instance. For variable $r$, soft-O also suppresses factors polynomial in $\log(r+2)$, with constants and degrees depending only on $d$.

We write the running-time bounds using
\refstepcounter{equation}\label{eq:Fd}
$$k(d)=\lfloor d/2\rfloor,\qquad F_d(m,n)=(mn)^{k(d)/(k(d)+1)}+m+n\qquad(m,n\ge1).\eqno(\theequation)$$
When the dimension is clear, we abbreviate $k(d)$ to $k$; the function $F_d$ is symmetric and nondecreasing in both arguments.

\begin{remark}[Arbitrary positive temperature]
\label{rem:temperature}
A positive rational inverse temperature $\beta$ is absorbed into the queries by replacing $\mat{Q}$ with $\beta\mat{Q}$. This adds only the encoding length of $\beta$ to the coordinate encoding lengths and replaces the magnitude bound by $B\max\{1,\beta\}$. The resulting dependence on temperature is polylogarithmic in its magnitude.

An irrational inverse temperature can be replaced by a positive dyadic approximation at the precision derived in Appendix~\ref{app:basic-arithmetic}, including for the usual scaling $\beta=d^{-1/2}$.
\end{remark}

\subsection{Random Sampling}
\label{subsec:sampling}

The algorithms in Sections~\ref{sec:geom} and~\ref{sec:biased} sample uniform subsets and permutations, using standard probability estimates for sampling without replacement.

\begin{lemma}
\label{lem:hypergeom}
Let~$\Sigma$~be a nonempty finite set, let~$\Sigma''\subseteq\Sigma$, and let~$\Sigma'$~be a uniformly random subset of~$\Sigma$~of a given size~$|\Sigma'|$. Then~$\Prob{\Sigma'\cap\Sigma''=\varnothing}\le \e^{-|\Sigma'||\Sigma''|/|\Sigma|}$, and~$X_{\Sigma''}=|\Sigma'\cap\Sigma''|$~satisfies~$\Prob{X_{\Sigma''}\ge\theta}\le(\e\,\E X_{\Sigma''}/\theta)^\theta$~for every positive integer~$\theta$.
\end{lemma}

\begin{proof}
Apply Hoeffding's comparison theorem for sampling without replacement~\cite[Theorem~4]{hoeffding1963} to the indicators of membership in~$\Sigma''$. Writing $s=|\Sigma'|$, $p=|\Sigma''|/|\Sigma|$, and $\mu=sp=\E X_{\Sigma''}$ gives
$$\E \e^{tX_{\Sigma''}}\le(1-p+p\e^t)^s\le \e^{\mu(\e^t-1)}\qquad(t\in\R).$$
The first bound follows as $t\to-\infty$. For $\theta>\mu>0$, the second follows from the exponential Markov inequality with $t=\ln(\theta/\mu)$, and is immediate when $\theta\le\mu$ or $\mu=0$.
\end{proof}

\subsection{Analytic Estimates}

\begin{lemma}[Perturbation and ratio bounds]
\label{lem:softmaxpert}
\label{lem:shift}
For $\vct x,\vct x'\in\R^n$ with $\|\vct x-\vct x'\|_\infty\le\delta$, let $h=\max_jx_j$ and $h'=\max_jx'_j$. Then
$$\|\softmax(\vct x)-\softmax(\vct x')\|_1\le2\delta,\qquad |h-h'|\le\delta,$$
$$|(x_j-h)-(x'_j-h')|\le2\delta,\qquad |\e^{x_j-h}-\e^{x'_j-h'}|\le2\delta.$$
If $D\ge1$, $|F|\le WD$, and $|\wh D-D|,|\wh F-F|\le\zeta<1$, then
$$\left|\frac{\wh F}{\wh D}-\frac FD\right|\le\frac{(W+1)\zeta}{1-\zeta}.$$
\end{lemma}

\begin{proof}
The softmax Jacobian has operator norm at most $2$ from $\ell_\infty$ to $\ell_1$; the maximum and the exponential on $(-\infty,0]$ are $1$-Lipschitz. The ratio bound follows by writing its numerator as $D(\wh F-F)-F(\wh D-D)$ and using $\wh D\ge1-\zeta$.
\end{proof}

Temperature interpolation uses a single set of nodes to approximate~$\e^{-\beta\mu}$ uniformly over~$\mu\ge0$. Placing the nodes in the middle half of each dyadic interval keeps their distance from any point in the interval within three quarters of its lower endpoint, yielding exponential decay of the interpolation error.

\begin{lemma}
\label{lem:entrywise}
Let $\nint\ge2$ be an integer, fix a lower endpoint $\beta_{\mathrm{lo}}>0$, and let
$$\tau_\ell=\beta_{\mathrm{lo}}\Bigl(\frac54+\frac{\ell}{2(\nint-1)}\Bigr)\quad(0\le\ell<\nint),\qquad\mathcal L_\ell(\beta)=\prod_{\substack{0\le\ell'<\nint\\ \ell'\ne\ell}}\frac{\beta-\tau_{\ell'}}{\tau_\ell-\tau_{\ell'}}.$$
The~$\nint$~nodes define a Lagrange interpolation polynomial of degree at most~$\nint-1$.
For all $\mu\ge0$ and $\beta\in[\beta_{\mathrm{lo}},2\beta_{\mathrm{lo}}]$,
$$\Bigl|\e^{-\beta\mu}-\sum_{\ell=0}^{\nint-1}\mathcal L_\ell(\beta)\,\e^{-\tau_\ell\mu}\Bigr|\le\Bigl(\frac34\Bigr)^{\nint}\qquad\text{and}\qquad\sum_{\ell=0}^{\nint-1}|\mathcal L_\ell(\beta)|\le9^{\nint-1}.$$
\end{lemma}

\begin{proof}
For fixed $\mu\ge0$, consider the function $\beta\mapsto \e^{-\beta\mu}$, whose $\nint$-th derivative is $(-\mu)^{\nint}\e^{-\beta\mu}$. The nodes are distinct and lie in $[5\beta_{\mathrm{lo}}/4,7\beta_{\mathrm{lo}}/4]$, so $|\beta-\tau_\ell|\le3\beta_{\mathrm{lo}}/4$ for every $\beta\in[\beta_{\mathrm{lo}},2\beta_{\mathrm{lo}}]$ and every $\ell$. By the Lagrange remainder formula~\cite[Eq.~(3.3.5)]{dlmf}, for some $\beta'\in[\beta_{\mathrm{lo}},2\beta_{\mathrm{lo}}]$,
$$\e^{-\beta\mu}-\sum_{\ell=0}^{\nint-1}\mathcal L_\ell(\beta)\,\e^{-\tau_\ell\mu}=\frac{(-\mu)^{\nint}\e^{-\beta'\mu}}{\nint!}\prod_{\ell=0}^{\nint-1}(\beta-\tau_\ell).$$
The function $x\mapsto x^{\nint}\e^{-\beta_{\mathrm{lo}} x}$ on $[0,\infty)$ is maximized at $x=\nint/\beta_{\mathrm{lo}}$, so
$$\mu^{\nint}\e^{-\beta'\mu}\le\mu^{\nint}\e^{-\beta_{\mathrm{lo}}\mu}\le\Bigl(\frac{\nint}{\e\beta_{\mathrm{lo}}}\Bigr)^{\nint}.$$
Together with $\nint!\ge(\nint/\e)^{\nint}$ and $|\beta-\tau_\ell|\le3\beta_{\mathrm{lo}}/4$, the remainder has absolute value at most
$$\Bigl(\frac{\nint}{\e\beta_{\mathrm{lo}}}\Bigr)^{\nint}\Bigl(\frac{3\beta_{\mathrm{lo}}}4\Bigr)^{\nint}\Bigl(\frac{\e}{\nint}\Bigr)^{\nint}=\Bigl(\frac34\Bigr)^{\nint}.$$

For the coefficients, consecutive nodes are $\beta_{\mathrm{lo}}/(2(\nint-1))$ apart, so
$$\prod_{\ell'\ne\ell}|\tau_\ell-\tau_{\ell'}|=\Bigl(\frac{\beta_{\mathrm{lo}}}{2(\nint-1)}\Bigr)^{\nint-1}\ell!\,(\nint-1-\ell)!.$$
Each of the $\nint-1$ factors $|\beta-\tau_{\ell'}|$ in the numerator of $\mathcal L_\ell(\beta)$ is at most $3\beta_{\mathrm{lo}}/4$, hence
$$|\mathcal L_\ell(\beta)|\le\frac{\bigl(3(\nint-1)/2\bigr)^{\nint-1}}{\ell!\,(\nint-1-\ell)!}.$$
Summing over $\ell$ and using $\sum_{\ell=0}^{\nint-1}\binom{\nint-1}{\ell}=2^{\nint-1}$ and $(\nint-1)!\ge((\nint-1)/\e)^{\nint-1}$, we obtain
\begin{equation*}
\sum_{\ell=0}^{\nint-1}|\mathcal L_\ell(\beta)|\le\frac{\bigl(3(\nint-1)/2\bigr)^{\nint-1}2^{\nint-1}}{(\nint-1)!}=\frac{\bigl(3(\nint-1)\bigr)^{\nint-1}}{(\nint-1)!}\le(3\e)^{\nint-1}\le9^{\nint-1}.\qedhere
\end{equation*}
\end{proof}

\section[A Dimension-Preserving Reduction from Maximum Inner Product]{A Dimension-Preserving Reduction\\ from Maximum Inner Product}
\label{sec:reduction}

We now show that $\BMIP_d$ reduces to a single call to $\ATT_d$ in the same dimension. The idea is to set both the keys and the values equal to $\mat Y$ at a logarithmic inverse temperature $\beta_0$: taking the inner product of the resulting attention output with $\vct x_i$ yields a softmax average of the scores $\vct x_i\cdot \vct y_j$, which lies within $(\ln n)/\beta_0$ of the maximum $M_i$ and can be rounded to recover $M_i$ exactly. 

\subsection{Computing Exact Maxima from Attention}
\label{subsec:exact-reduction}

Consider an instance of $\BMIP_d$ with matrices $\mat X\in\Z^{m\times d}$ and $\mat Y\in\Z^{n\times d}$ and coordinate magnitudes at most an integer $B\ge1$. To ensure that the softmax average and the attention approximation error each contribute less than $1/4$ to the inner product with $\vct x_i$, we set $C=8$, $\beta_0=C\lceil\log_2(n+1)\rceil$, and $\eps_0=2^{-\lceil\log_2(CdB)\rceil}$, so $\eps_0=\Theta(1/(dB))$ and $\eps_0\le1/(CdB)$. The dimension may grow in this section, and soft-O suppresses only logarithmic factors in $m+n,d,B$.

\begin{theorem}[Dimension-preserving reduction]
\label{thm:red}
Set $\mat Q=\beta_0\mat X$ and $\mat K=\mat Y$. Each of the attention reductions below solves $\BMIP_d$ with the same $m,n,d$ and deterministic additional time $\softO((m+n)d)$.
\begin{enumerate}[label=(\alph*)]
\item One call with $\mat V=\mat Y$, $r=d$ value channels, magnitude bound $\beta_0B$, and error tolerance $\eps_0$.
\item One call with $\mat V=dB\,\mat Y$, $r=d$ value channels, magnitude bound $\max\{\beta_0B,dB^2\}$, and constant error tolerance $\eps_c=1/4$.
\item For either value matrix $\mat V$, $d$ calls with one value channel each, using its columns separately and the same error tolerance.
\end{enumerate}
\end{theorem}

\begin{proof}
Let $M_i=\max_j\vct x_i\cdot\vct y_j$, let $\vct o_i$ be the exact output in part~(a), and write $a_{ij}=\e^{\beta_0\vct x_i\cdot\vct y_j}/Z_i$, where $Z_i=\sum_j\e^{\beta_0\vct x_i\cdot\vct y_j}$. The entropy identity $\ln Z_i-\beta_0\vct x_i\cdot\vct o_i=-\sum_j a_{ij}\ln a_{ij}\le\ln n$, together with $Z_i\ge\e^{\beta_0 M_i}$, implies
$$E_i=\vct x_i\cdot\vct o_i
=\frac{\sum_j\e^{\beta_0\vct x_i\cdot\vct y_j}(\vct x_i\cdot\vct y_j)}{\sum_j\e^{\beta_0\vct x_i\cdot\vct y_j}},\qquad
0\le M_i-E_i\le\frac{\ln n}{\beta_0}\le\frac1C.$$
For an approximate output $\wh{\vct o}_i$ with coordinatewise error at most $\eps_0$, we have $|\vct x_i\cdot\wh{\vct o}_i-M_i|\le dB\eps_0+1/C\le2/C<1/2$. Rounding to the nearest integer determines $M_i$ exactly.

In part~(b), scaling the values scales the exact output to $dB\,\vct o_i$, so dividing the inner product of its approximation with $\vct x_i$ by $dB$ estimates $E_i$ with error at most $\eps_c$. Since $\eps_c+1/C=3/8<1/2$, the same rounding applies. All value channels share the same softmax probabilities, so the scalar calls compute the vector output in part~(c). In all three cases, the inputs and dyadic outputs have logarithmic encoding length in $m+n,d,B$, and forming the inputs and rounding the outputs require $\Oh((m+n)d)$ coordinate operations (Appendix~\ref{app:basic-arithmetic}).
\end{proof}

Computing $\max_i M_i$ also solves $\MIP_d$ with the same additional cost, which is near-linear for fixed $d$ and polynomially bounded integer inputs. The reductions are deterministic and preserve Las Vegas guarantees.

\subsection{Conditional Hardness in Superconstant Dimension}
\label{subsec:superconstant}

We use the randomized version of the Orthogonal Vectors Hypothesis (OVH)~\cite{williams2005}, which states that for every~$\delta>0$, there is a constant $c>0$ such that randomized algorithms with bounded error cannot determine in $\mathrm{O}(n^{2-\delta})$ time whether two sets of $n$ Boolean vectors in $\{0,1\}^{\lceil c\log n\rceil}$ contain an orthogonal pair. Although the attention lower bounds of~\cite{almanSong2023,gupta2026} were originally stated under the Strong Exponential Time Hypothesis (SETH)~\cite{impagliazzoPaturi2001}, their reductions factor through orthogonal vectors and hold directly under OVH; hence they are unaffected by recent subexponential-time results refuting ETH and SETH (while all algorithmic bounds and reductions elsewhere in this paper are unconditional).

We say that a dimension function~$d(n)$~is constructible if its value can be computed in~$n^{\mathrm{o}(1)}$~time. Saha, Xu, and Ye~\cite[Theorem~4.7]{sahaXuYe2026} establish conditional~$n^{2-\mathrm{o}(1)}$~hardness for integer orthogonal vectors in every constructible superconstant dimension, with~$\Oh(\log n)$-bit entries. Their Lemma~B.1, based on~\cite{williams2018,chen2018}, reduces integer orthogonal vectors in dimension~$d_{\mathrm{OV}}$~to exact integer maximum inner product in dimension~$d_{\mathrm{OV}}^2$. Applying their theorem with~$d_{\mathrm{OV}}(n)=\lfloor\sqrt{d(n)}\rfloor$, which is constructible and tends to infinity, and padding with zeros establishes the same hardness for exact integer maximum inner product in dimension~$d(n)$. They observe that this lower bound transfers to attention through the reduction of~\cite[Lemma~C.3]{gupta2026}, which uses polylogarithmically many attention calls with~$n+1$~queries and keys in dimension~$d+1$~and already works at constant error. In superconstant dimension, the extra dimension and the extra calls of that reduction are immaterial; even so, applying Theorem~\ref{thm:red} directly yields this hardness at the same dimension and input sizes.

\begin{corollary}
\label{cor:hard}
Assume the randomized OVH, and let~$d(n)$~be a constructible function with~$d(n)\to\infty$~and~$d(n)=n^{\mathrm{o}(1)}$. Then attention in dimension~$d(n)$, with~$m=n$,~$r\in\{1,d(n)\}$~value channels, and polynomial coordinate magnitudes, requires~$n^{2-\mathrm{o}(1)}$~time at either inverse-polynomial error or constant error tolerance~$\eps_c=1/4$.
\end{corollary}

\begin{proof}
Suppose attention on these instances has running time $\Oh(n^{2-\delta})$ for some constant $0<\delta<1$. Theorem~\ref{thm:red} solves the maximum-inner-product hard instances with one vector-valued call or $d(n)$ scalar calls, on inputs of polynomial magnitude and with inverse-polynomial error tolerance. Since $d(n)=n^{\mathrm o(1)}$, the additional work is $\softO(nd(n))=n^{1+\mathrm o(1)}$ and either implementation runs in total time $n^{2-\delta+\mathrm o(1)}$. Parts~(b) and~(c) imply the same complexity at constant error, so composing either reduction with the deterministic OV-to-Max-IP reductions of~\cite{williams2005,sahaXuYe2026} contradicts randomized OVH.
\end{proof}

\section{A Deterministic Algorithm in Two Dimensions}
\label{sec:d2}

We begin with the two-dimensional case ($d=2$), which both establishes Theorem~\ref{thm:d2} and serves as the base case for our dimension induction in Section~\ref{sec:plan}. Our main geometric observation is that centering the key polygon at its area centroid allows us to enclose all near-maximal keys for each query within a simple shell region specified by a radial threshold and at most two intervals along the polygon boundary, over which polynomial moments can be summed by a one-dimensional sweep. Throughout this section, the number $r$ of value channels is fixed, and $N=m+n$ denotes the total number of query and key vectors.

\begin{theorem}[Planar attention; proof in Section~\ref{subsec:d2cost}]
\label{thm:d2}
For~$d=2$~and any fixed number~$r$~of value channels, approximate attention as defined in Definition~\ref{def:att} can be computed deterministically in~$\softO(N)$~time.
\end{theorem}

\subsection{Shifted-Kernel Products in the Plane}
\label{subsec:kernel}

For query vectors~$\vct{q}_i\in\Q^2$~($i\in[m]$) and keys~$\vct{k}_j\in\Q^2$~($j\in[n]$), define the shifted kernel of the original inputs by
$$T_{ij}=\e^{\vct{q}_i\cdot \vct{k}_j-\max_{j'\in[n]}\vct{q}_i\cdot\vct{k}_{j'}},\qquad \mat{T}=(T_{ij}),\qquad N=m+n.$$
We compute either the forward product~$\mat{T}\vct{w}^{\mathrm{in}}$~or the transpose product~$\mat{T}^{\top}\vct{z}^{\mathrm{in}}$, with input weights $\vct{w}^{\mathrm{in}}\in\Q^n$ and $\vct{z}^{\mathrm{in}}\in\Q^m$.

\begin{lemma}
\label{lem:kernel}
Let~$U,W\ge1$~be integer bounds on the absolute values of the geometric coordinates and input weights, respectively, and let~$b$~bound their rational encoding lengths and that of the error tolerance~$\eta\in\Q\cap(0,1]$. Each of~$\mat{T}\vct{w}^{\mathrm{in}}$~and~$\mat{T}^{\top}\vct{z}^{\mathrm{in}}$~can be computed deterministically to coordinatewise additive error at most~$\eta$~in~$\softO(m+n)$~time.
\end{lemma}

To prove Lemma~\ref{lem:kernel}, we construct a rational approximation $\mat P$ whose rowwise $\ell_1$ error implies product error $\eta$ (Section~\ref{subsec:d2cost}), and show in Sections~\ref{subsec:d2center}--\ref{subsec:d2moment} how to evaluate its forward and transpose products through moment sweeps.

\subsection{Centering and Radial Coordinates}
\label{subsec:d2center}
\label{subsec:d2grid}

Given a target rowwise error $\eta_{\mathrm{row}}\in\Q\cap(0,1]$, we round the input coordinates to the dyadic grid of Appendix~\ref{app:planar-arithmetic}, changing each kernel row by at most $\eta_{\mathrm{row}}/16$ in $\ell_1$ norm (Lemma~\ref{lem:gridding}). Collinear rounded keys admit a one-dimensional sweep, described in the same appendix. Otherwise, when their hull has positive area, we compute its vertices in cyclic order and translate all rounded keys by the negative of its area centroid~$\vct c$. The resulting common shift of each row and its maximum leaves the rounded shifted scores unchanged.

In Sections~\ref{subsec:d2center}--\ref{subsec:d2moment}, write $\vct q_i$ for the rounded queries and $\vct k_j$ for the rounded and centered keys, and let
$$\mathcal C=\conv\{\vct{k}_1,\ldots,\vct{k}_n\}$$
be the centered key polygon, with $h_i=\max_j\vct q_i\cdot\vct k_j$. The matrix $\mat T$ always denotes the original shifted kernel. The origin is now the area centroid and lies in the interior of $\mathcal C$; since $\|\vct c\|_\infty\le U+1$, the working coordinates satisfy $\|\vct k_j\|_\infty\le2(U+1)$ and $\|\vct q_i\|_\infty\le U+1$. Computing the hull costs $\softO(n)$ time, after which each $h_i$ is computed by an $\softO(1)$-time support query~\cite{preparataShamos1985,dobkinKirkpatrick1990}, with ties resolved by a fixed vertex order.

Set the score window and polynomial degree to
$$L=\Theta\!\left(\log\frac{N}{\eta_{\mathrm{row}}}\right),\qquad g=\Theta(L),$$
with the constants specified in Appendix~\ref{app:planar-arithmetic}, which also includes all rounding errors in this section.

\begin{lemma}
\label{lem:centroid}
For the centered two-dimensional polygon~$\mathcal C$~above and~$\vct{q}\in\R^2\setminus\{0\}$, set
$$h=\max_{\vct{x}\in\mathcal C}\vct{q}\cdot \vct{x},\qquad h^-=\min_{\vct{x}\in\mathcal C}\vct{q}\cdot \vct{x}.$$
Then~$h>0$~and~$h^-\ge-2h$.
\end{lemma}

\begin{proof}
The origin lies in the interior, so~$h>0$. To bound the minimum, triangulate the polygon by a fan from a vertex~$\vct{x}^-$~with~$\vct{q}\cdot\vct{x}^-=h^-$. Each triangle then has one vertex projecting to~$h^-$~and two projecting to at most~$h$, so its centroid projects to at most~$(h^-+2h)/3$. Their area-weighted average is the polygon's centroid projection, zero, so~$0\le(h^-+2h)/3$.
\end{proof}

By Lemma~\ref{lem:centroid}, every centered key satisfies
$$-3h_i\le\vct{q}_i\cdot\vct{k}_j-h_i\le0.$$
For each nonzero key, the ray from the origin through it meets the boundary at a unique point $\vct k_j^{\partial}$, so we can write
$$\vct{k}_j=\rho_j\vct{k}_j^{\partial},\qquad0<\rho_j\le1.$$
We call~$\rho_j$~the key's radial ratio and sort the nonzero keys by the cyclic order of their boundary points, equivalently by angle about the origin. Each comparison uses a halfplane test and a cross product, with keys sharing a boundary point stored as separate consecutive entries that retain their own indices and weights.

Binary search among the angularly ordered vertices locates the ray's sector; intersecting its edge determines~$\vct{k}_j^{\partial}$~in~$\softO(1)$~time, with a ray through a vertex returning that vertex. Keys at the origin contribute only to the constant moment in Section~\ref{subsec:d2moment}.

\subsection{Radial and Boundary Ranges}
\label{subsec:d2window}

For each row, we select a range containing every key within~$L$~of the maximum. When~$h_i\le L$, all shifted scores lie in~$[-3L,0]$~by the centroid bound, so we set $\mathcal K_i=[n]$. For~$h_i>L$, on the other hand, we define
$$\mathcal K_i=\left\{j:\vct{k}_j\ne0,\quad \rho_j\ge1-\frac{L}{h_i},\quad
\vct{q}_i\cdot \vct{k}_j^{\partial}\ge h_i-L\right\}.$$

\begin{lemma}
\label{lem:sandwich}
For~$h_i>L$,
$$\{j:\vct{q}_i\cdot\vct{k}_j\ge h_i-L\}
\subseteq\mathcal K_i
\subseteq\{j:\vct{q}_i\cdot\vct{k}_j\ge h_i-2L\}.$$
\end{lemma}

\begin{proof}
If~$\vct{q}_i\cdot\vct{k}_j\ge h_i-L>0$, then~$\vct{k}_j\ne0$~and~$\vct{q}_i\cdot \vct{k}_j^{\partial}>0$. Since~$\vct{q}_i\cdot \vct{k}_j^{\partial}\le h_i$~and~$\rho_j\le1$,
$$\rho_j=\frac{\vct{q}_i\cdot\vct{k}_j}{\vct{q}_i\cdot \vct{k}_j^{\partial}}\ge1-\frac{L}{h_i},\qquad
\vct{q}_i\cdot \vct{k}_j^{\partial}\ge\vct{q}_i\cdot\vct{k}_j\ge h_i-L.$$
Conversely, both lower bounds defining~$\mathcal K_i$~are positive, so every $j\in\mathcal K_i$ satisfies
\begin{equation*}
\vct{q}_i\cdot\vct{k}_j\ge\left(1-\frac{L}{h_i}\right)(h_i-L)
=h_i-2L+\frac{L^2}{h_i}\ge h_i-2L.\qedhere
\end{equation*}
\end{proof}

Thus $\mathcal K_i$ contains every near-maximal key, including a maximizer, and widens the score window by at most a factor of two. Set
$$s_{ij}=\vct{q}_i\cdot\vct{k}_j-h_i+3L,$$
so that the centroid bound for $h_i\le L$ and Lemma~\ref{lem:sandwich} for $h_i>L$ imply
$$0\le s_{ij}\le3L\quad(j\in\mathcal K_i),\qquad
\vct{q}_i\cdot\vct{k}_j-h_i<-L\quad(j\notin\mathcal K_i).$$

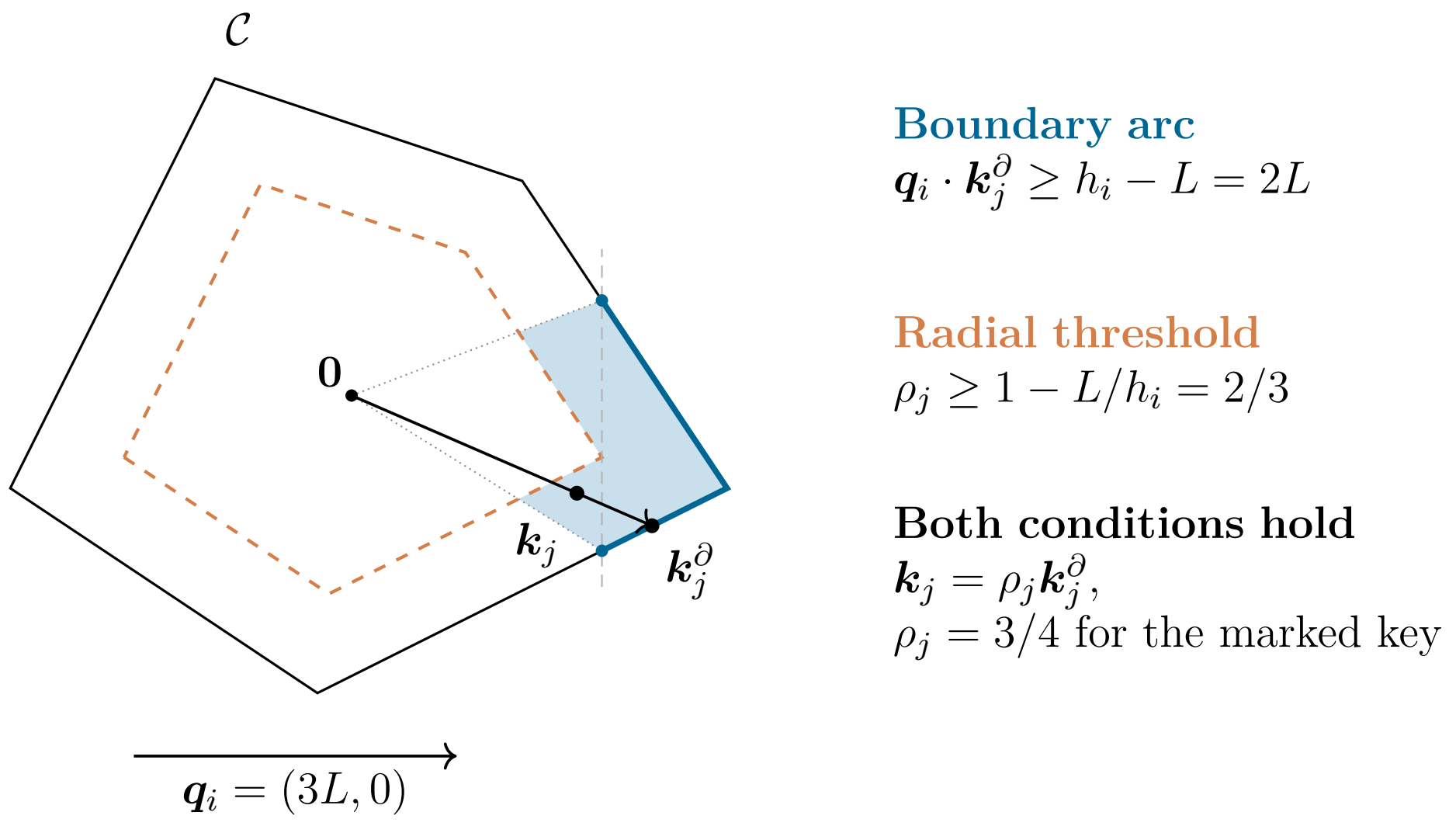
\begin{figure}[t]
\centering
\begin{tikzpicture}[x=3.2cm,y=3.2cm,font=\fontsize{11}{13.2}\selectfont]
\coordinate (arcTop) at (2/3,39/154);
\coordinate (arcBottom) at (2/3,-191/462);
\coordinate (boundaryKey) at (4/5,-267/770);
\coordinate (retainedKey) at (3/5,-801/3080);
\fill[MidnightBlue!15]
  (arcBottom) -- (1,-19/77) -- (arcTop) --
  (4/9,13/77) -- (2/3,-38/231) -- (4/9,-191/693) -- cycle;
\draw[line width=0.6pt]
  (-10/11,-19/77) -- (-1/11,-61/77) -- (1,-19/77) --
  (5/11,44/77) -- (-4/11,65/77) -- cycle;
\draw[Orange!85!black,dashed,line width=0.8pt]
  (-20/33,-38/231) -- (-2/33,-122/231) -- (2/3,-38/231) --
  (10/33,88/231) -- (-8/33,130/231) -- cycle;
\draw[black!30,densely dashed] (2/3,-0.51) -- (2/3,0.39);
\draw[black!45,densely dotted] (0,0) -- (arcTop);
\draw[black!45,densely dotted] (0,0) -- (arcBottom);
\draw[MidnightBlue,line width=1.4pt]
  (arcBottom) -- (1,-19/77) -- (arcTop);
\draw[->,line width=0.7pt] (0,0) -- (boundaryKey);
\fill[MidnightBlue] (arcTop) circle (1.5pt);
\fill[MidnightBlue] (arcBottom) circle (1.5pt);
\fill (0,0) circle (1.5pt);
\fill (retainedKey) circle (1.8pt);
\fill (boundaryKey) circle (1.8pt);
\node[above left,inner sep=2pt] at (0,0) {$\vct{0}$};
\node[below left,xshift=-3pt,yshift=-6pt,inner sep=1pt]
  at (retainedKey) {$\vct{k}_j$};
\node[below right,xshift=2pt,yshift=-3pt,inner sep=1pt]
  at (boundaryKey) {$\vct{k}_j^{\partial}$};
\node[above,inner sep=2pt] at (-0.30,0.91) {$\mathcal C$};
\draw[->,line width=0.7pt] (-0.58,-0.96) -- (0.28,-0.96);
\node[below,inner sep=3pt] at (-0.15,-0.96) {$\vct{q}_i=(3L,0)$};
\node[anchor=west,align=left,text width=6.1cm] at (1.40,0.63)
  {\textcolor{MidnightBlue}{\textbf{Boundary arc}}\\
   $\vct{q}_i\cdot\vct{k}_j^{\partial}\ge h_i-L=2L$};
\node[anchor=west,align=left,text width=6.1cm] at (1.40,0.08)
  {\textcolor{Orange!85!black}{\textbf{Radial threshold}}\\
   $\rho_j\ge1-L/h_i=2/3$};
\node[anchor=west,align=left,text width=6.1cm] at (1.40,-0.50)
  {\textbf{Both conditions hold}\\
   $\vct{k}_j=\rho_j\vct{k}_j^{\partial}$,\\
   $\rho_j=3/4$ for the marked key};
\end{tikzpicture}
\caption{A non-symmetric convex pentagon centered at its area centroid, with $h_i=3L>L$. Blue marks the eligible boundary arc, and the orange dashed polygon is $(2/3)\mathcal C$. All points in the shaded region, including its boundary, satisfy both retention conditions.}
\label{fig:planar-radial}
\end{figure}

For a row with $h_i>L$, the boundary condition selects a closed arc containing the maximizing face. Since each intermediate level line meets the boundary at the two ends of its intersection segment, projection is monotone along either chain from a maximum to a minimum. Binary search locates the arc endpoints, and cutting the cyclic order at a fixed position represents the arc by at most two disjoint intervals. All radial and boundary thresholds are inclusive, so the searches and sweeps below include equality at every endpoint. Together with the radial threshold, these intervals describe the part of the closed shell between~$(1-L/h_i)\mathcal C$~and~$\mathcal C$~in the arc's angular sector (Figure~\ref{fig:planar-radial}).

\subsection{Polynomial Approximation and Moment Sweeps}
\label{subsec:d2moment}

For $j\in\mathcal K_i$, $s_{ij}\in[0,3L]$. The degree-$g$ Taylor polynomial $\exp_g(s)=\sum_{\ell=0}^g s^\ell/\ell!$ satisfies
$$0\le\e^{-3L}\bigl(\e^s-\exp_g(s)\bigr)=\Oh(\eta_{\mathrm{row}}/N)\qquad(0\le s\le3L)$$
by the standard remainder bound (Lemma~\ref{lem:taylor}). Let $0<\omega\le\e^{-3L}$ be a rational approximation at the precision specified in Appendix~\ref{app:planar-arithmetic}.
Define
$$P_{ij}=\begin{cases}\omega\exp_g(s_{ij}),&j\in\mathcal K_i,\\0,&j\notin\mathcal K_i.\end{cases}$$
Both approximations are from below, so~$0\le P_{ij}\le\e^{\vct{q}_i\cdot\vct{k}_j-h_i}\le1$.

\begin{lemma}
\label{lem:kernelerr}
For the original shifted kernel~$\mat{T}$, the matrix $\mat P$ satisfies
$$0\le P_{ij}\le1,\qquad
\sum_{j=1}^n|T_{ij}-P_{ij}|\le\eta_{\mathrm{row}}\qquad(i\in[m]).$$
\end{lemma}

\begin{proof}
For the rounded coordinates, omitted keys have weight below $\e^{-L}$, and the Taylor and scalar approximations contribute $\Oh(\eta_{\mathrm{row}}/N)$ per selected key. Summing these errors and adding the coordinate-rounding error proves the rowwise estimate.
\end{proof}

\paragraph{Separated polynomial weights.}
Setting~$\gamma_i=3L-h_i$ and expanding first in~$\vct{q}_i\cdot\vct{k}_j$~and then in the two coordinates gives
$$\exp_g(s_{ij})=\sum_{\ell_x+\ell_y\le g}C_{i,\ell_x,\ell_y}k_{jx}^{\ell_x}k_{jy}^{\ell_y},\qquad
C_{i,\ell_x,\ell_y}=\frac{q_{ix}^{\ell_x}q_{iy}^{\ell_y}}{\ell_x!\ell_y!}\exp_{g-\ell_x-\ell_y}(\gamma_i).$$
The expansion has~$\binom{g+2}{2}$~terms indexed by nonnegative~$\ell_x,\ell_y$, each factoring into a query coefficient and a key monomial. For the rounded weights $\vct w,\vct z$, the common denominators of Appendix~\ref{app:planar-arithmetic} allow exact summation of these terms with polynomial bit lengths.

Keys at the origin belong to $\mathcal K_i$ exactly when $h_i\le L$. Their total weight contributes to the constant moment in forward products, and their transpose products involve only the $(\ell_x,\ell_y)=(0,0)$ term.

\paragraph{Forward multiplication.}
For a rounded input vector~$\vct{w}$, the required moments are
$$\mathrm{Mom}_i(\ell_x,\ell_y)=\sum_{j\in\mathcal K_i}k_{jx}^{\ell_x}k_{jy}^{\ell_y}w_j.$$
We first precompute the moments over all keys for rows with $h_i\le L$. To handle rows with $h_i>L$, a Fenwick tree~\cite{fenwick1994} on the array of keys in boundary order supports the sweep below.
\begin{enumerate}[label=(\arabic*),itemsep=1pt]
\item Sort the rows with $h_i>L$ by $1-L/h_i$ and the nonzero keys by $\rho_j$, both in nonincreasing order.
\item Before answering a row, insert the vectors of weighted monomials for every key at or above the row's radial threshold.
\item Query the row's one or two boundary intervals to obtain its moments.
\end{enumerate}
For every row, the sum of these moments weighted by $C_{i,\ell_x,\ell_y}$ and scaled by $\omega$ is $(\mat P\vct w)_i$.

\paragraph{Transpose multiplication.}
For a rounded input vector~$\vct{z}$, we reverse updates and queries to evaluate
$$(\mat{P}^{\top}\vct{z})_j=\omega\sum_{\ell_x+\ell_y\le g}k_{jx}^{\ell_x}k_{jy}^{\ell_y}
\sum_{i:\,j\in\mathcal K_i}C_{i,\ell_x,\ell_y}z_i.$$
Use a difference Fenwick tree on the same array in boundary order for interval additions and point queries.
\begin{enumerate}[label=(\arabic*),itemsep=1pt]
\item Sort the rows with $h_i>L$ by $1-L/h_i$ and the nonzero keys by $\rho_j$, both in nondecreasing order.
\item Before processing key $j$, add the vector $(C_{i,\ell_x,\ell_y}z_i)$ over the boundary intervals of every row whose threshold is at most $\rho_j$.
\item Query the boundary index of key $j$ to obtain the contribution from rows with $h_i>L$ to each inner sum.
\end{enumerate}
Adding $\sum_{i:h_i\le L}C_{i,\ell_x,\ell_y}z_i$ completes the inner sum for every key; its weighted sum with the key monomials, scaled by $\omega$, is the transpose product.
\subsection{Kernel Products and Attention}
\label{subsec:d2cost}
\label{subsec:d2error}

\begin{proof}[Proof of Lemma~\ref{lem:kernel}]
Set $\eta_{\mathrm{row}}=\eta/(8N(W+1))$. By Lemma~\ref{lem:kernelerr},
$$\| (\mat T-\mat P)\vct w^{\mathrm{in}}\|_\infty\le W\eta_{\mathrm{row}},\qquad
\| (\mat T-\mat P)^{\top}\vct z^{\mathrm{in}}\|_\infty\le mW\eta_{\mathrm{row}}.$$
The moment sweeps evaluate the products with $\mat P$ in $\softO(N)$ time. Including the input and output rounding detailed in Appendix~\ref{app:planar-arithmetic}, both errors are at most $\eta$.
\end{proof}

\begin{proof}[Proof of Theorem~\ref{thm:d2}]
By Lemma~\ref{lem:kernel}, the products in Section~\ref{sec:prelim} can be computed to tolerance $\eta=\eps/(8(B+1))$ in $\softO(N)$ time for fixed $r$. Lemma~\ref{lem:shift} then implies an attention error at most $\eps$.
\end{proof}

\subsection{Online Queries after Preprocessing the Keys}
\label{subsec:d2online}

Because the forward sweep in Section~\ref{subsec:d2moment} inserts keys in nonincreasing order of $\rho_j$ independent of the queries, we can make the sweep structure persistent~\cite{driscollEtAl1989}: storing successive versions of the boundary-order segment tree allows any arriving query to retrieve the exact moments it needs in $\softO(1)$ time. This yields the following static data structure.

\begin{corollary}
\label{cor:online-d2}
Fix a constant number~$r$~of value channels, matrices~$\mat{K}\in\Q^{n\times2}$~and~$\mat{V}\in\Q^{n\times r}$~with rows~$\vct{k}_j$~and~$\vct{v}_j$, and an error tolerance~$\eps=2^{-p}$~as in Definition~\ref{def:att}. Given bounds~$B\ge1$~and~$b$~on the magnitudes and rational encoding lengths of all key, value, and future query coordinates, there is a deterministic data structure with~$\softO(n)$~preprocessing time and storage that, for any query~$\vct{q}\in\Q^2$~satisfying these bounds, returns an~$\eps$-additive approximation to
$$\frac{\sum_{j=1}^n \e^{\vct{q}\cdot \vct{k}_j}\vct{v}_j}{\sum_{j=1}^n \e^{\vct{q}\cdot \vct{k}_j}}$$
in~$\softO(1)$~time. The suppressed factors depend on~$b$, $\log n$, $\log B$, and~$\log(1/\eps)$; the guarantee holds for every query in an arbitrarily long, possibly adaptive stream.
\end{corollary}

\begin{proof}
Set the parameters of Section~\ref{subsec:d2grid} with $U=W=\lceil B\rceil$, $N=n+1$, and $\eta_{\mathrm{row}}=\Theta(\eps/(W+1))$. Round the keys and values as in Appendix~\ref{app:planar-arithmetic}; these choices depend only on the given bounds and tolerance.

For a two-dimensional hull, we center the keys and use their radial ratios and boundary order to build a persistent binary segment tree~\cite{driscollEtAl1989} storing interval moments for every value channel and the all-ones channel. We insert the nonzero keys in nonincreasing order of radial ratio and also store the global moments over all keys.

Upon arrival, a query is rounded to the same geometric grid and its support value $h$ is computed by a support query. We read the global moments when $h\le L$, or select the version at the radial threshold when $h>L$, and read the moments from its one or two boundary intervals. Combining these moments with the row coefficients and dividing the numerator by the denominator yields the approximate attention output, with the common factor $\omega$ canceling from the ratio.

By Lemmas~\ref{lem:kernelerr} and~\ref{lem:shift}, the error is $\Oh((W+1)\eta_{\mathrm{row}})$. Appendix~\ref{app:planar-arithmetic} includes the rounding errors and establishes the preprocessing, storage, and query costs. Because the stored moments are exact, the error bound holds for every query satisfying the input bounds, including throughout an adaptive stream.
\end{proof}

\section{Shifted Kernels and Dimension Induction}
\label{sec:plan}
\label{sec:induction}

Using the planar shifted-kernel algorithm as the base case, we prove the fixed-dimensional bound for $d\geq 3$ by induction on the dimension. Each inductive step consists of two reductions (Figure~\ref{fig:induction-dependencies}):
$$\KP_d\ \longrightarrow\ \BP_{d-1}\ \longrightarrow\ \KP_{d-1}\qquad(d\ge3).$$
First, Lagrange temperature interpolation decouples the row scales from the inner products, reducing $\KP_d$ to $\BP_{d-1}$ with an additive scalar bias on each side. Second, partitioning the significant pairs into common-anchor blocks allows us to cancel these biases and reduce $\BP_{d-1}$ to $\KP_{d-1}$. Throughout Sections~\ref{sec:plan}--\ref{sec:biased}, $d$ is fixed and we use the precision convention of Section~\ref{subsec:bit-model}.

\subsection{Exact Maxima and Normalized Directions}
\label{subsec:plan-primitives}

The reduction begins by computing the row maxima exactly and normalizing each nonzero row by a coordinate of largest absolute value. Both these maxima and those used later in the anchor factorization can be computed within Chan's offline ray-shooting bound~\cite[Corollaries~2.2 and~2.4]{chan1996output}, with the bit complexity stated below.

\begin{proposition}[Exact maxima; proof in Appendix~\ref{app:geom-batch}]
\label{prop:batchsupport}
For fixed $d\ge2$, let $\vct q_i\in\Q^d$ ($i\in[m]$) be queries and $\vct k_j\in\Q^d$ ($j\in[n]$) be points. A Las Vegas algorithm computes every maximum $M_i=\max_{j\in[n]}\vct q_i\cdot\vct k_j$ and a maximizing index in expected time $\softO(F_d(m,n))$. For $d\le3$, the algorithm is deterministic and runs in $\softO(m+n)$ time.
\end{proposition}
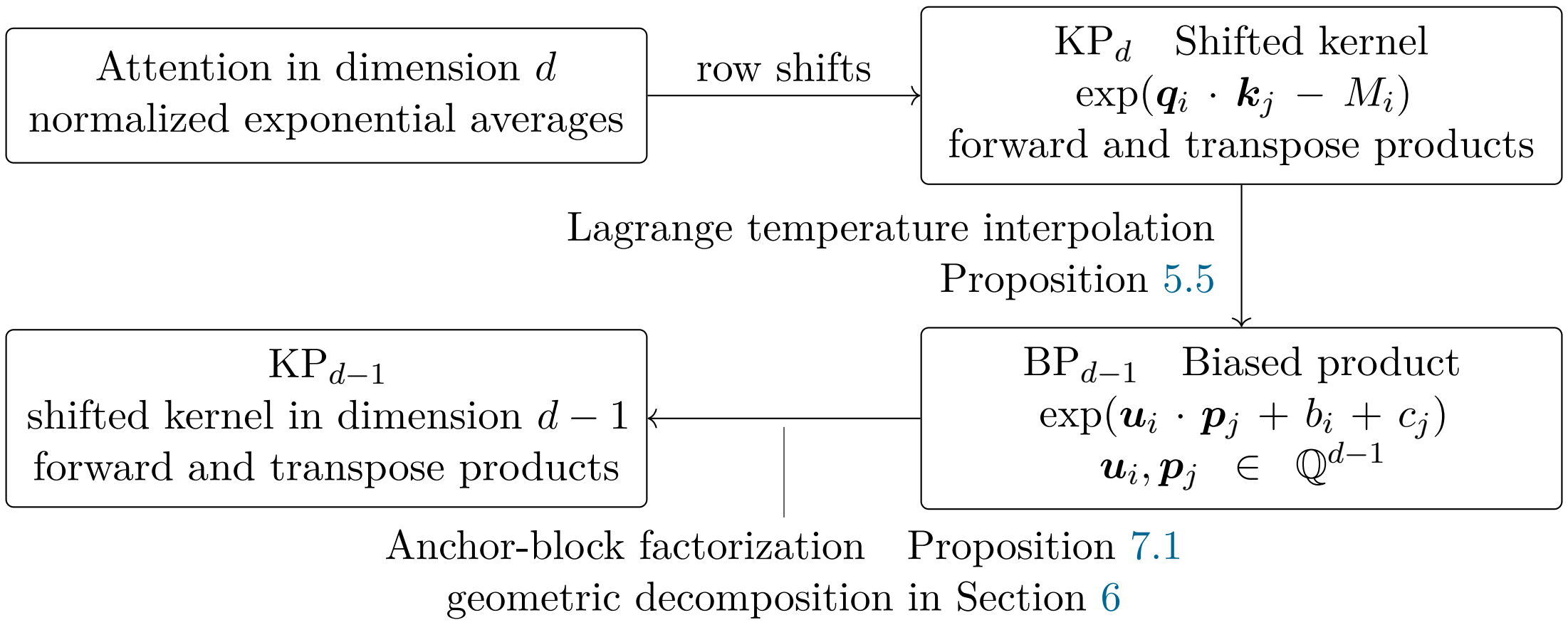
\begin{figure}[H]
  \centering
  \begin{tikzpicture}[font=\normalsize,box/.style={draw,rounded corners=2pt,align=center,text width=5.6cm,minimum height=1.25cm,inner sep=5pt}]
  \node[box] (att) at (0,0) {Attention in dimension~$d$\\normalized exponential averages};
  \node[box] (kp) at (8.5,0)
    {$\KP_d$\quad Shifted kernel\\$\exp(\vct q_i\cdot\vct k_j-M_i)$\\forward and transpose products};
  \node[box] (bp) at (8.5,-3.0)
    {$\BP_{d-1}$\quad Biased product\\$\exp(\vct u_i\cdot\vct p_j+b_i+c_j)$\\$\vct u_i,\vct p_j\in\Q^{d-1}$};
  \node[box] (lower) at (0,-3.0)
    {$\KP_{d-1}$\\shifted kernel in dimension~$d-1$\\forward and transpose products};
  \draw[->] (att) -- node[above,align=center] {row shifts} (kp);
  \draw[->] (kp) -- node[left,align=right,xshift=-3pt]
    {Lagrange temperature interpolation\\Proposition~\ref{prop:KPstep}} (bp);
  \draw[->] (bp) -- (lower);
  \node[align=center] at (4.25,-4.45)
    {Anchor-block factorization\quad Proposition~\ref{prop:bpdecomposition}\\
     geometric decomposition in Section~\ref{sec:geom}};
  \draw[-,black!60] (4.25,-3.92) -- (4.25,-3.08);
  \end{tikzpicture}
  \caption{One induction step, read clockwise from the upper left. Row shifts express attention through $\KP_d$; Lagrange temperature interpolation then gives $\BP_{d-1}$, whose anchor blocks factor through $\KP_{d-1}$. The induction ends at the planar kernel algorithm.}
  \label{fig:induction-dependencies}
  \end{figure}
\begin{definition}[Coordinate normalization]
\label{def:normalization}
For $\ell\in[d]$, let $\pi_{\ell,+}$ exchange coordinates $\ell$ and $d$, and let $\pi_{\ell,-}$ also change the sign of the last coordinate. For $\vct q\ne0$, let $\ell$ be the least index with $|q_\ell|=\|\vct q\|_\infty$ and let $\pi_{\vct q}$ be the corresponding permutation, with the sign chosen so that $\pi_{\vct q}\vct q$ has a positive last coordinate. Write
$$\pi_{\vct q}\vct q=\beta_{\vct q}(\vct u_{\vct q},1),\qquad
\beta_{\vct q}=\|\vct q\|_\infty>0,\qquad \vct u_{\vct q}\in[-1,1]^{d-1}.$$
Rows sharing $\pi_{\vct q}$ form a coordinate group; there are at most $2d$ groups. For a signed coordinate permutation $\pi$ and $\vct y\in\R^d$, define the affine function $f^\pi_{\vct y}(\vct u)=(\pi\vct y)\cdot(\vct u,1)$, and write $f_{\vct y}=f^{\mathrm{id}}_{\vct y}$. By orthogonality,
$$\vct q\cdot\vct y=\beta_{\vct q}f^{\pi_{\vct q}}_{\vct y}(\vct u_{\vct q}).$$
\end{definition}

For $\vct y=(\vct p,c)$, we have $f_{\vct y}(\vct u)=\vct u\cdot\vct p+c$.

\subsection{Shifted-Kernel Products}
\label{subsec:plan-kp}

\begin{definition}[Shifted-kernel products]
\label{def:KP}
For $d\ge2$, an instance of $\KP_d$ consists of rational rows $\vct{q}_i\in\Q^d$ ($i\in[m]$), rational columns $\vct{k}_j\in\Q^d$ ($j\in[n]$), and an error tolerance $\eta\in\Q\cap(0,1]$. Their shifted kernel is
$$M_i=\max_{j\in[n]}\vct{q}_i\cdot\vct{k}_j,\qquad T_{ij}=\e^{\vct{q}_i\cdot\vct{k}_j-M_i}.$$
The task is to compute either the forward product $\mat T\vct w$, with $\vct w\in\Q^n$, or the transpose product $\mat T^{\top}\vct z$, with $\vct z\in\Q^m$. Let integers $U,W\ge1$ bound the magnitudes of coordinates and input weights, respectively, and let $b$ bound the encoding lengths of all input rationals, including $\eta$.

A solution is a vector $\wh{\vct o}$ approximating the specified product to coordinatewise additive error $\eta$.
\end{definition}

In dimensions $d\ge3$, the algorithm shifts each row by its exact maximum, computed by Proposition~\ref{prop:batchsupport} or provided with the instance, and evaluates the products over blocks. In recursive calls from the anchor factorizations of Lemma~\ref{lem:anchor}, these maxima are already computed, so the kernel calls reuse them. Moreover, because these factorizations involve both forward and transpose products, the induction maintains both operations, which the planar algorithm computes in $\softO(m+n)$ time for the base case (Lemma~\ref{lem:KPbase}).

\subsection{Biased Products and Temperature Interpolation}
\label{subsec:plan-bp}
\label{subsec:kernel-to-biased}

The intermediate problem $\BP_{d-1}$ has $d-1$ bilinear coordinates and one scalar bias on each side. Its lifted source points lie in $\Q^d$.

\begin{definition}[Biased exponential products]
\label{def:BP}
For $d\ge3$, an instance of $\BP_{d-1}$ consists of rational rows $(\vct u_i,b_i)\in\Q^{d-1}\times\Q$ indexed by $I_0=[m]$ and columns $(\vct p_j,c_j)\in\Q^{d-1}\times\Q$ indexed by $J_0=[n]$, with nonpositive scores
$$x_{ij}=\vct u_i\cdot\vct p_j+b_i+c_j\le0,\qquad S_{ij}=\e^{x_{ij}}.$$
Given weights $\vct w\in\Q^{J_0}$ and an error tolerance $\eta\in\Q\cap(0,1]$, the task is to approximate $\mat S\vct w$ to coordinatewise additive error $\eta$. Let integers $U,W\ge1$ bound the magnitudes of coordinates and biases, and of input weights, respectively, and let $b$ bound the encoding lengths of all input rationals. Write $N=m+n$.
\end{definition}

The score formula is symmetric under exchanging $(\vct u_i,b_i)$ with $(\vct p_j,c_j)$, which transposes $\mat S$ and preserves its numerical parameters. A forward biased-product call on the exchanged instance therefore computes $\mat S^{\top}$ times a vector with the same symmetric running-time bound $F_d(m,n)$. By contrast, in $\KP_{d-1}$, rowwise normalization breaks this symmetry by fixing an orientation, so the anchor factorization of Lemma~\ref{lem:anchor} uses the forward product for a row anchor and the transpose product for a column anchor.

\begin{proposition}
\label{prop:KPstep}
Let $d\ge3$. If $\BP_{d-1}$ can be solved by a Las Vegas algorithm in expected time $\softO(F_d(m,n))$, then $\KP_d$ can be solved by a Las Vegas algorithm in the same expected time.
\end{proposition}

\begin{proof}[Proof of Proposition~\ref{prop:KPstep}]
Consider a $\KP_d$ instance with parameters $(\eta,U,W,b)$, $N=m+n$, and input weights $\vct w\in\Q^n$ for the forward product or $\vct z=(z_i)_{i\in[m]}\in\Q^m$ for the transpose product. Use the supplied row maxima or compute them by Proposition~\ref{prop:batchsupport} in expected time $\softO(F_d(m,n))$. After handling zero rows as in Section~\ref{sec:prelim}, we apply Definition~\ref{def:normalization} to write each nonzero row as $\vct q_i=\beta_i\pi_i^{-1}(\vct u_i,1)$ with $\beta_i>0$. Setting $h_i=M_i/\beta_i$, we have
$$\vct q_i\cdot\vct k_j=\beta_i f^{\pi_i}_{\vct k_j}(\vct u_i),\qquad
h_i=\max_j f^{\pi_i}_{\vct k_j}(\vct u_i),\qquad
T_{ij}=\e^{-\beta_i\mu_{ij}},\quad
\mu_{ij}=h_i-f^{\pi_i}_{\vct k_j}(\vct u_i)\ge0.$$
We partition these nonzero rows by their normalizing permutation and dyadic temperature interval into groups
$$\mathcal G_{\pi,\beta_{\mathrm{lo}}}
=\{i:\pi_i=\pi,\ \beta_{\mathrm{lo}}\le\beta_i<2\beta_{\mathrm{lo}}\},$$
where $\beta_{\mathrm{lo}}$ ranges over powers of two. Since $2^{-b}<\beta_i\le U$, the number $\ngrp$ of nonempty groups is $\softO(1)$.

Within each nonempty group $\mathcal G=\mathcal G_{\pi,\beta_{\mathrm{lo}}}$, we apply Lemma~\ref{lem:entrywise} with $\nint=\Theta(\log(NW/\eta))$ nodes $\tau_\ell$ and coefficients $\mathcal L_{i\ell}=\mathcal L_\ell(\beta_i)$. Because $\tau_\ell$ is shared by all rows in the group, we can absorb it into the column vectors and row shifts: each node defines a $\BP_{d-1}$ instance with rows $(\vct u_i,-\tau_\ell h_i)$ and columns $(\vct p_j,c_j)=\tau_\ell\pi\vct k_j$. Its matrix $\mat S^{(\mathcal G,\ell)}$ has nonpositive scores
$$(\vct u_i,1)\cdot(\tau_\ell\pi\vct k_j)-\tau_\ell h_i=-\tau_\ell\mu_{ij}\le0.$$
We solve these instances at tolerance
$$\eta_{\mathrm{BP}}=\Theta\!\left(\frac{\eta}{\ngrp\nint\,9^{\nint}}\right)$$
with a small absolute constant. For forward multiplication, we apply each matrix to $\vct w$ and combine the returned vectors $\wh{\vct o}^{(\mathcal G,\ell)}$ for $i\in\mathcal G$ as
$$\sum_\ell\mathcal L_{i\ell}\wh o_i^{(\mathcal G,\ell)}.$$
Conversely, for transpose multiplication, we exchange the two sides of each biased instance and use weights $(\mathcal L_{i\ell}z_i)_{i\in\mathcal G}$, then sum the resulting vectors over all groups and nodes. Since $|\mathcal L_{i\ell}z_i|\le9^{\nint}W$, the logarithm of the weight bound remains $\Oh(\lambda)$.

By Lemma~\ref{lem:entrywise}, the interpolation error is at most $NW(3/4)^{\nint}$ for either product, to which the biased calls add at most $9^{\nint}\eta_{\mathrm{BP}}$ per forward output coordinate and $\ngrp\nint\eta_{\mathrm{BP}}$ per transpose output coordinate. A large enough constant in $\nint$ and a small enough constant in $\eta_{\mathrm{BP}}$ give error at most $\eta/2$, leaving $\eta/2$ for the arithmetic errors analyzed in Appendix~\ref{app:induction-arithmetic}.

Since there are $\softO(1)$ groups and nodes, each defining a biased instance with side sizes at most $m,n$, possibly exchanged, the total expected cost is $\softO(F_d(m,n))$ by symmetry and monotonicity of $F_d$. Normalizing the rows, computing the interpolation coefficients, and combining the outputs require an additional $\softO(N)$ operations.
\end{proof}

To complete the induction, Section~\ref{sec:biased} partitions the significant pairs of a biased instance into common-anchor blocks. Lemma~\ref{lem:anchor} reduces each block product to a $\KP_{d-1}$ call, and the total expected evaluation time is $\softO(F_d(m,n))$.

\begin{proposition}[Proof in Section~\ref{subsec:biasederr}]
\label{prop:BP}
Let $d\ge3$ and suppose a Las Vegas algorithm solves $\KP_{d-1}$ in expected time $\softO(F_{d-1}(m,n))$. Then $\BP_{d-1}$ can be solved by a Las Vegas algorithm in expected time $\softO(F_d(m,n))$.
\end{proposition}

\subsection{The Induction on the Dimension}
\label{subsec:dimension-induction}

\begin{lemma}
\label{lem:KPbase}
$\KP_2$ can be solved deterministically in $\softO(m+n)$ time.
\end{lemma}

\begin{proof}
Lemma~\ref{lem:kernel} computes either product to error $\eta$ in $\softO(m+n)$ time.
\end{proof}

\begin{theorem}[Kernel products in fixed dimension]
\label{thm:KP}
For every fixed integer $d\ge2$, a Las Vegas algorithm solves $\KP_d$ in expected time $\softO(F_d(m,n))$.
\end{theorem}

\begin{proof}
The base case is Lemma~\ref{lem:KPbase}. Assuming the result in dimension $d-1$, Proposition~\ref{prop:BP} solves $\BP_{d-1}$ in expected time $\softO(F_d(m,n))$, and Proposition~\ref{prop:KPstep} then solves $\KP_d$ within this bound.

Across the $d-2$ reductions, the numerical parameters grow by a fixed polynomial at each step and thus remain polynomial in the original parameters for fixed $d$ (Appendix~\ref{app:induction-arithmetic}).
\end{proof}

\section{Geometric Data Structures in Fixed Dimension}
\label{sec:geom}

The biased-product algorithm uses the support, reporting, point-location, and sampling results in Lemmas~\ref{lem:support}, \ref{lem:reporttree}, \ref{lem:envelope}, and~\ref{lem:sample}. The underlying geometric constructions are usually analyzed in the real-RAM model~\cite{matousekSchwarzkopf1993,clarksonShor1989,chan1996lpq,chan1996output}; Appendices~\ref{app:arithmetic} and~\ref{app:geom-proofs} provide the rational implementations and bit-complexity bounds needed here.

In this section, $d\ge2$ is fixed and $k=\lfloor d/2\rfloor$. Input and query bounds are fixed before preprocessing, with parameter $\lambda$ as in Section~\ref{subsec:bit-model}. We use polytope for a bounded intersection $P$ of closed halfspaces $H=\{\vct{x}:\vct{a}_H\cdot\vct{x}\le b_H\}$ with $\vct{a}_H\ne\vct{0}$, and denote its interior and vertex set by $\intr P$ and $\vertset(P)$. For $\vct{x}\in\R^d$, write $\bar{\vct{x}}\in\R^{d-1}$ for its projection onto the first $d-1$ coordinates. Ties may be resolved by returning any maximizing index in a support query and any closed cell containing the point in a location query.

\begin{definition}
\label{def:genpos}
A finite indexed collection of affine hyperplanes in~$\R^d$~is in general position if the normal vectors of any at most~$d$~of its members are linearly independent and no~$d+1$~of its members have a common point. A finite indexed collection of closed halfspaces is in general position if their bounding hyperplanes are.
\end{definition}

\subsection{Classical Geometric Primitives}
\label{subsec:geom-batch}

McMullen's Upper Bound Theorem implies that there are $\Oh_d(h^k)$ vertices for a $d$-polytope defined by $h$ halfspaces~\cite{mcmullen1970,seidel2017hull}. For $d\ge3$, these vertices and their incidences can be enumerated by Clarkson and Shor's algorithm in expected $\softO(h^k)$ arithmetic operations~\cite[Theorems~4.7 and~4.8]{clarksonShor1989}. For queries, we use the high-storage ray-shooting structure of Matousek and Schwarzkopf~\cite[Lemma~7]{matousekSchwarzkopf1993}, with deterministic alternatives in dimensions two and three~\cite{preparataShamos1985,dobkinKirkpatrick1990}.

The rational version, stated in Proposition~\ref{prop:rayshoot} in Appendix~\ref{subsec:geom-cuttings}, has the same soft-O bounds. Its implementation fixes a bounding simplex and constructs each geometric object from a constant number of original constraints.

\subsection{Support Queries and Strict Reporting}
\label{subsec:geom-support}

A support query asks for a point maximizing a given linear objective, and polarity reduces it to ray shooting. The nonzero gaps between rational scores allow a general-position perturbation for which each perturbed maximizing index also maximizes the original objective.

\begin{lemma}[Proof in Appendix~\ref{app:geom-support}]
\label{lem:support}
Let~$\mathcal Y=(\vct{y}_s)_{s\in\Sigma}$~be a nonempty list of points in~$\Q^d$, with repetitions allowed. A Las Vegas algorithm builds a support-query structure in expected time~$\softO(|\Sigma|^k)$, deterministically for~$d\le3$. Given an admissible~$\vct q\in\Q^d$, the structure returns~$\max_{s\in\Sigma}\vct{q}\cdot\vct{y}_s$~and a maximizing index in time~$\softO(1)$.
\end{lemma}

Support queries also test whether a strict halfspace contains any source. Applying the standard binary-tree reduction from emptiness to reporting~\cite[Observation~2.3]{chan1996lpq} within groups of size at most $g_{\mathrm{rep}}$ yields a tradeoff between preprocessing and query time.

\begin{lemma}[Proof in Appendix~\ref{app:reporting}]
\label{lem:reporttree}
Let~$\mathcal Y=(\vct{y}_s)_{s\in\Sigma}$~be a nonempty list of points in~$\Q^d$, with repetitions allowed, and let~$g_{\mathrm{rep}}$~be an integer with~$1\le g_{\mathrm{rep}}\le|\Sigma|$. A Las Vegas algorithm builds a reporting structure in expected time~$\softO(|\Sigma|g_{\mathrm{rep}}^{k-1})$, deterministically for~$d\le3$. Given an admissible query~$(\vct q,\theta)\in\Q^d\times\Q$, the structure reports~$Z=\{s\in\Sigma:\vct{q}\cdot\vct{y}_s>\theta\}$~in time~$\softO(|\Sigma|/g_{\mathrm{rep}}+1+|Z|)$. For~$g_{\mathrm{rep}}=|\Sigma|$, the preprocessing and query costs are~$\softO(|\Sigma|^k)$~and~$\softO(1+|Z|)$, respectively.
\end{lemma}

\subsection{Envelope Cells}
\label{subsec:geom-envelope}

For $d\ge3$, we organize near-maximal pairs by sampling sources and partitioning the upper envelope of their affine functions (Definition~\ref{def:normalization}) into cells, each with a designated maximizing source. The perturbation and the bounding region containing all target positions are fixed across samples. We index sources by $s$ and targets by $i$, in either orientation of the matrix.

\begin{definition}
\label{def:envelope}
Let $U,N_\Sigma\ge1$ be integers, let $\varnothing\ne\Sigma_0\subseteq[N_\Sigma]$, and let $\vct{y}_s\in\Q^d$ ($s\in\Sigma_0$) satisfy $\|\vct{y}_s\|_\infty\le U$. The instance parameter $\lambda$ also bounds $\log_2(N_\Sigma+2)$ and $\log_2(U+2)$.

Apply once the uniform perturbation of Appendix~\ref{subsec:geom-perturb}, obtaining source points $\wt{\vct{y}}_s$ shared by all samples. For every nonempty $\Sigma'\subseteq\Sigma_0$, define
$$f_s(\vct{u})=\vct{y}_s\cdot(\vct{u},1),\qquad \wt f_s(\vct{u})=\wt{\vct{y}}_s\cdot(\vct{u},1),\qquad E_{\Sigma'}(\vct{u})=\max_{s\in\Sigma'}\wt f_s(\vct{u}),$$
and let the envelope polytope $\cP(\Sigma')$ be the epigraph of $E_{\Sigma'}$ clipped by the $d$ fixed rational bounding halfspaces constructed in Appendix~\ref{app:geom-envelope}. These halfspaces are shared by every sample, and the lower boundary over $[-U,U]^{d-1}$ is the graph of $E_{\Sigma'}$. For $s\in\Sigma'$, the facet of source $s$ is the possibly empty subset of $\cP(\Sigma')$ where its source constraint $\wt f_s(\vct{u})\le z$ is tight.
\end{definition}

\begin{lemma}[Proof in Appendix~\ref{app:geom-envelope}]
\label{lem:envfacts}
The perturbed envelope in Definition~\ref{def:envelope} satisfies:
\begin{enumerate}[label=(\alph*)]
\item $|\wt f_s(\vct{u})-f_s(\vct{u})|\le1/8$~for every~$s\in\Sigma_0$~and every~$\vct{u}\in\R^{d-1}$~with~$\|\vct{u}\|_\infty\le U$.
\item Let~$\vct u\in\Q^{d-1}$~be an admissible target with~$\|\vct u\|_\infty\le U$, and let~$\varnothing\ne\Sigma'\subseteq\Sigma_0$. Every~$s\in\Sigma'$~that maximizes~$\wt f_s(\vct{u})$~over~$\Sigma'$~also maximizes~$f_s(\vct{u})$~over~$\Sigma'$.
\item For every nonempty~$\Sigma'\subseteq\Sigma_0$, the~$|\Sigma'|+d$~constraints of~$\cP(\Sigma')$~are in general position, and~$\cP(\Sigma')$~is bounded and has nonempty interior. Its constraint coefficients and vertices have~$\Oh_d(\lambda)$~bits.
\end{enumerate}
\end{lemma}

The perturbed sources, envelope vertices, and their affine values at vertices or targets satisfying the input bounds all have $\Oh_d(\lambda)$ bits. Lemmas~\ref{lem:support} and~\ref{lem:reporttree} therefore apply with a uniform bit bound.

To decompose the envelope into cells, we apply Clarkson's recursive vertex construction~\cite[Section~3]{clarkson1988} to triangulate the source facets of the sampled envelope and project the simplices onto the target coordinates. Each projected cell has an owner whose function equals the sampled maximum on the entire cell, and its conflict set consists of the sources that exceed this owner somewhere on the cell. These sets contain every source that can improve a target's sampled maximum.

\begin{definition}
\label{def:cells}
Let~$\varnothing\ne\Sigma'\subseteq\Sigma\subseteq\Sigma_0$. Triangulate the source facets of~$\cP(\Sigma')$~by the fixed rule of Definition~\ref{def:pulling-cells} in Appendix~\ref{app:geom-cells}, and project its simplices onto their first~$d-1$~coordinates. These projected simplices are the canonical cells. A cell~$\Delta$~inherits its owner~$s_\Delta\in\Sigma'$~from its source facet. The cells and their order depend only on~$\Sigma'$, with the perturbation and bounding halfspaces fixed. The conflict set of~$\Delta$~relative to the ground set~$\Sigma$~is
$$C_\Delta=\{s'\in\Sigma:\ \wt f_{s'}(\vct{u})>\wt f_{s_\Delta}(\vct{u})\ \text{for some}\ \vct{u}\in\Delta\}.$$

\end{definition}

\begin{lemma}[Proof in Appendix~\ref{app:geom-cells}]
\label{lem:cells}
Let~$\varnothing\ne\Sigma'\subseteq\Sigma\subseteq\Sigma_0$. The canonical cells of~$\cP(\Sigma')$~satisfy:
\begin{enumerate}[label=(\alph*)]
\item Every cell~$\Delta$~is a closed~$(d-1)$-dimensional simplex with~$d$~rational vertices, and its owner~$s_\Delta$~satisfies~$\wt f_{s_\Delta}(\vct{u})=E_{\Sigma'}(\vct{u})$~for every~$\vct{u}\in\Delta$. The union of the cells contains~$[-U,U]^{d-1}$.
\item For every cell~$\Delta$~with owner~$s_\Delta$, we have~$C_\Delta=\bigcup_{\bar{\vct{v}}\in\vertset(\Delta)}\{s\in\Sigma:\wt f_s(\bar{\vct{v}})>\wt f_{s_\Delta}(\bar{\vct{v}})\}$~and~$C_\Delta\cap\Sigma'=\varnothing$.
\item $\cP(\Sigma')$~has at most~$\Cenv|\Sigma'|^{k}$~canonical cells, where~$\Cenv\ge1$~depends only on~$d$.
\item Over all nonempty~$\Sigma'\subseteq\Sigma$, at most~$(|\Sigma|+d)^{d^2+1}$~distinct owner--cell pairs~$(s_\Delta,\Delta)$~occur. For a fixed ground set~$\Sigma$, each such pair has a fixed conflict set, independent of the sample in which it occurs.
\end{enumerate}
\end{lemma}

Because the cells are closed, a target on a shared boundary may be placed in any cell containing it. To find such a cell, we use the face-by-face ray-shooting descent of Matousek and Schwarzkopf~\cite[proof of Theorem~10]{matousekSchwarzkopf1993} and Chan~\cite[proof of Theorem~4.1]{chan1996output}, using the fixed triangulation order for the vertices. The descent then identifies a cell in the enumerated list.

\begin{lemma}
\label{lem:envelope}
Let~$\varnothing\ne\Sigma'\subseteq\Sigma_0$. A Las Vegas algorithm enumerates and locates the canonical cells with expected preprocessing time~$\softO(|\Sigma'|^k)$. It returns
\begin{enumerate}[label=(\alph*)]
\item the list of all canonical cells of~$\cP(\Sigma')$~in the order of Definition~\ref{def:cells}, each with its owner and vertices;
\item a structure that, given an admissible target~$\vct u\in\Q^{d-1}$~with~$\|\vct u\|_\infty\le U$, locates a canonical cell~$\Delta$~of~$\cP(\Sigma')$~containing~$\vct{u}$, and returns its index in the list and its owner in time~$\softO(1)$.
\end{enumerate}
The list in part~(a) is determined by~$\Sigma'$.
\end{lemma}

Appendix~\ref{app:geom-cells} proves Lemma~\ref{lem:envelope}, including the exact representation and comparison of cell indices.

\subsection{Sampling Bounds}
\label{subsec:geom-sample}

We resample whenever a cell has too many conflicts. A uniform sample passes these tests with high probability by Clarkson--Shor sampling~\cite{clarkson1987,clarksonShor1989}, with a union bound over the finite set of canonical configurations fixed by the shared perturbation.

\paragraph{Sampling parameters.}
For the original instance size $N=m+n\ge1$, we fix a bound $G$ on the number of doubling prefix levels and an initial prefix size $s_0$ by setting
\begin{equation}
\label{eq:sampling}
G=\lceil\log_2(N+8)\rceil,\qquad s_0=2\Csamp G,
\end{equation}
where $\Csamp$ depends only on $d$ and is chosen in Appendix~\ref{app:geom-potential}.

\begin{lemma}[Proof in Appendix~\ref{app:geom-potential}]
\label{lem:sample}
In the setup of Definition~\ref{def:envelope}, assume~$|\Sigma_0|\le N$ and use the parameters~\eqref{eq:sampling}. Let~$\varnothing\ne\Sigma\subseteq\Sigma_0$, and let~$\Sigma'$~be a uniformly random subset of~$\Sigma$~of a given size~$|\Sigma'|\ge1$. With probability at least~$1-1/(8G)$, every canonical cell~$\Delta$~of~$\cP(\Sigma')$~satisfies~$|C_\Delta|\le \Csamp|\Sigma|G/|\Sigma'|$, where the conflict sets are taken relative to~$\Sigma$.
\end{lemma}

\section{Biased Exponential Products and the Main Theorem}
\label{sec:biased}

Let $d\ge3$ be fixed and set $k=k(d)$ as in~\eqref{eq:Fd}. Assuming a Las Vegas algorithm for $\KP_{d-1}$ with expected running time $\softO(F_{d-1}(m,n))$, we solve $\BP_{d-1}$ by covering its significant pairs with disjoint blocks, with total expected construction and evaluation time $\softO(F_d(m,n))$.

For an instance with row set~$I_0=[m]$, column set~$J_0=[n]$, and logarithmic parameter~$\lambda$~as in Section~\ref{subsec:bit-model}, we evaluate each block~$I\times J$~against the restricted input vector~$\vct{w}_J$~and add its contribution to~$\wh{\vct{o}}$, initially zero and indexed by~$I_0$. The block tolerances ensure that their accumulated errors and the omitted tail remain within $\eta$.

\begin{proposition}[Block decomposition]
\label{prop:bpdecomposition}
Under the preceding inductive hypothesis, let~$N=m+n$~and set the window~$L$, anchor window~$L_{\mathrm{anc}}$, and block tolerance~$\etablk$~as in~\eqref{eq:params} below. There is a Las Vegas algorithm that constructs and evaluates a finite collection~$\mathcal D_{\mathrm{blk}}$~of nonempty blocks~$D=I_D\times J_D$~with the following properties.
\begin{enumerate}[label=(\alph*)]
\item \emph{Coverage and anchors.} The blocks are pairwise disjoint subsets of~$I_0\times J_0$, and every pair outside them has score less than~$-L+1/8$. Each block is evaluated directly or has an anchor: either a column~$j_0\in J_D$~with~$x_{ij_0}\ge-L_{\mathrm{anc}}$~for every~$i\in I_D$, or a row~$i_0\in I_D$~with~$x_{i_0j}\ge-L_{\mathrm{anc}}$~for every~$j\in J_D$.
\item \emph{Block evaluation.} Each block product is approximated to coordinatewise error at most~$\etablk$. Evaluating an anchor block involves one~$\KP_{d-1}$~call and one computation of exact batch maxima, in expected time~$\softO(F_{d-1}(|I_D|,|J_D|))$; direct evaluation costs~$\softO(|I_D||J_D|)$~time.
\item \emph{Total cost and error.} The construction and evaluation have expected running time~$\softO(F_d(m,n))$. Extend each block output by zero to~$I_0$~and let~$\wh{\vct{o}}$~be their sum. Then
$$\|\wh{\vct{o}}-\mat{S}\vct{w}\|_\infty\le n(\etablk+W\e^{-L+1/8})\le\eta/2.$$
\end{enumerate}
\end{proposition}

\subsection{Orientations, Hits, and Parameters}
\label{subsec:bp-setup}

Because our recursive decomposition samples whichever side of a subproblem is more favorable, and reverses the roles of the two sides on light cells, we distinguish between the fixed row/column indices $I,J$ and the current source/target sets $\Sigma,\Theta$. The two orientations are as follows:
$$\begin{array}{c|cc}
&\text{sources }\Sigma&\text{targets }\Theta\\
\text{column orientation}&J&I\\
\text{row orientation}&I&J
\end{array}$$
In the geometric procedures, we index sources by $s\in\Sigma$ and targets by $i\in\Theta$ in either orientation, as in Section~\ref{subsec:geom-envelope}. Write $(\vct p_s,c_s)$ for the source data and $(\vct u_i,b_i)$ for the target data in the current orientation. With $\vct y_s=(\vct p_s,c_s)$ and $f_s(\vct u)=\vct u\cdot\vct p_s+c_s$, the quantity $b_i+f_s(\vct u_i)$ is the score of the corresponding original row--column pair. Reversing the orientation exchanges the two data families. Each block product maps the original column weights to the original row outputs, through $\mat T^{\top}$ for a column anchor and $\mat T$ for a row anchor (Lemma~\ref{lem:anchor}).

For each orientation, fix the perturbation and bounding halfspaces of Definition~\ref{def:envelope} before sampling, with~$\Sigma_0=J_0$~in the column orientation and~$\Sigma_0=I_0$~in the row orientation, with~$N_\Sigma=N$. The instance bounds satisfy that definition for both sources and target positions, so the perturbed functions and the rule for forming canonical cells are fixed across all recursive calls. Lemmas~\ref{lem:envfacts} and~\ref{lem:envelope}(b) guarantee the perturbation bounds and point-location queries at every node.

\paragraph{Shared parameters.}
The sampling parameters $G$, $\Csamp$, and $s_0$ are those in~\eqref{eq:sampling}. For the entire biased-product instance, set
\begin{equation}
\label{eq:params}
L=\Theta\!\left(\log\frac{NW}{\eta}\right),\qquad L_{\mathrm{anc}}=L+1,\qquad
\etablk=\Theta(\eta/N),
\end{equation}
with integer $L\ge4$. The constants in the window and tolerance satisfy
$$N\bigl(\etablk+W\e^{-L+1/8}\bigr)\le\eta/2.$$
Scores below the window contribute at most the tail term above, and the extra unit in $L_{\mathrm{anc}}$ allows for the perturbation. Every node restricts the original index sets with the perturbations, windows, and tolerances fixed. Thus the numerical factor in every local running time is a single polynomial in the original parameter $\lambda$ (Appendix~\ref{app:biased-arithmetic}).

\paragraph{Hits.}
We select blocks using the perturbed functions~$\wt f_s$ and evaluate them with the original scores~$b_i+f_s(\vct u_i)$. The perturbation error of at most~$1/8$ (Lemma~\ref{lem:envfacts}) relates the geometric hit threshold to the original score window.

\begin{lemma}
\label{lem:window}
A source $s\in\Sigma_0$ is a hit for target $i$ if $b_i+\wt f_s(\vct u_i)\ge-L$.
\begin{enumerate}[label=(\alph*)]
\item A hit has original score at least $-L-1/8$.
\item A source that is not a hit has original score below $-L+1/8$.
\item Let $\varnothing\ne\Sigma'\subseteq\Sigma\subseteq\Sigma_0$, and let $\Delta$ be a canonical cell of $\cP(\Sigma')$ containing $\vct u_i$. If no source in $\Sigma'$ hits $i$, every hit in $\Sigma$ belongs to $C_\Delta$.
\end{enumerate}
\end{lemma}

\begin{proof}
Parts~(a) and~(b) follow from the perturbation bound $|\wt f_s(\vct u_i)-f_s(\vct u_i)|\le1/8$ (Lemma~\ref{lem:envfacts}(a)). For part~(c), the owner $s_\Delta\in\Sigma'$ has score below the hit threshold at $\vct u_i$, so every hit $s$ satisfies $\wt f_s(\vct u_i)>\wt f_{s_\Delta}(\vct u_i)$ and belongs to $C_\Delta$ by Definition~\ref{def:cells}.
\end{proof}

\subsection{Anchor Blocks and Direct Blocks}
\label{subsec:anchor}

\begin{definition}
\label{def:anchor}
A block is a set~$I\times J$~of pairs with~$\varnothing\ne I\subseteq I_0$~and~$\varnothing\ne J\subseteq J_0$. Its matrix is~$\mat{S}_{IJ}=(S_{ij})_{i\in I,\,j\in J}$. A column anchor of the block is a column~$j_0\in J$~with~$x_{ij_0}\ge-L_{\mathrm{anc}}$~for all~$i\in I$, and a row anchor is a row~$i_0\in I$~with~$x_{i_0j}\ge-L_{\mathrm{anc}}$~for all~$j\in J$. An anchor block is a block with a specified column or row anchor.
\end{definition}

\begin{figure}[H]
\centering
\begin{tikzpicture}[font=\normalsize,x=1cm,y=1cm]
\begin{scope}[shift={(0,0)}]
\node at (1.65,3.2) {(a) \textbf{Column anchor} $j_0$};
\fill[MidnightBlue!20] (1.1,0) rectangle (1.65,2.2);
\draw[step=0.55,black!35,line width=0.3pt] (0,0) grid (3.3,2.2);
\draw[line width=0.7pt] (0,0) rectangle (3.3,2.2);
\draw[MidnightBlue,line width=1pt] (1.1,0) rectangle (1.65,2.2);
\node[above] at (1.65,2.35) {columns $J$};
\node[rotate=90] at (-0.5,1.1) {rows $I$};
\node[below,MidnightBlue] at (1.375,-0.05) {$j_0$};
\node[align=center] at (1.65,-1.0)
  {kernel rows $J$, columns $I$\\factorization uses $\mat{T}^{\top}$};
\end{scope}
\begin{scope}[shift={(7.3,0)}]
\node at (1.65,3.2) {(b) \textbf{Row anchor} $i_0$};
\fill[Orange!22] (0,0.55) rectangle (3.3,1.1);
\draw[step=0.55,black!35,line width=0.3pt] (0,0) grid (3.3,2.2);
\draw[line width=0.7pt] (0,0) rectangle (3.3,2.2);
\draw[Orange!85!black,line width=1pt] (0,0.55) rectangle (3.3,1.1);
\node[above] at (1.65,2.35) {columns $J$};
\node[rotate=90] at (-0.5,1.1) {rows $I$};
\node[right,Orange!85!black] at (3.35,0.825) {$i_0$};
\node[align=center] at (1.65,-1.0)
  {kernel rows $I$, columns $J$\\factorization uses $\mat{T}$};
\end{scope}
\end{tikzpicture}
\caption{Two possible anchor types for an original block~$\mat{S}_{IJ}$. The highlighted scores lie in~$[-L_{\mathrm{anc}},0]$. (a) Subtracting a column anchor's score cancels the row bias; each kernel row then has its maximum over~$I$, so~$\mat{T}$~has indices~$J\times I$~and appears transposed. (b) Subtracting a row anchor's score cancels the column bias; the maxima are over~$J$~and~$\mat{T}$~keeps indices~$I\times J$. Lemma~\ref{lem:anchor} proves both factorizations and estimates their diagonal factors.}
\label{fig:anchor-directions}
\end{figure}
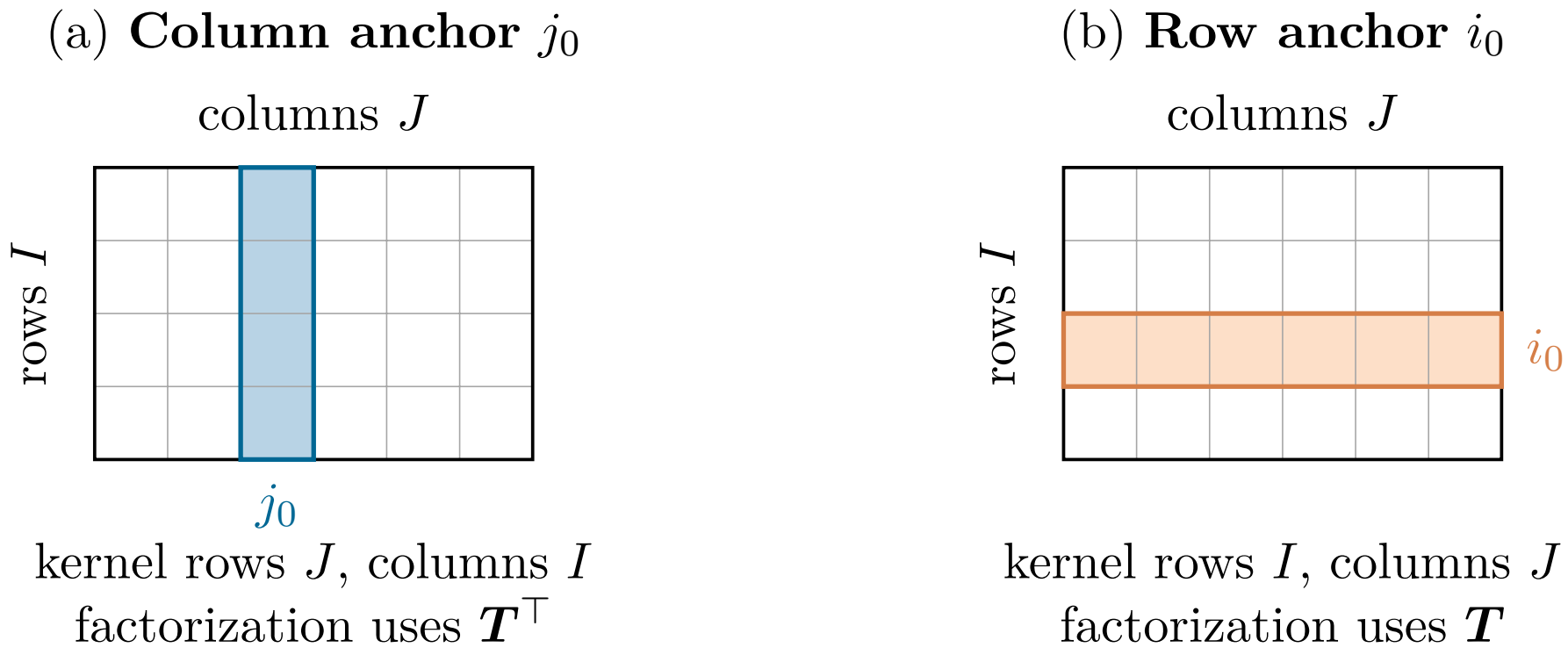

The product of a block $I\times J$ is $\mat S_{IJ}\vct w_J\in\R^I$. Subtracting a common anchor score reduces it to a shifted-kernel product in dimension $d-1$, using the transpose for a column anchor and the forward product for a row anchor (Figure~\ref{fig:anchor-directions}). Small blocks are evaluated directly.

\begin{lemma}[Anchor factorization]
\label{lem:anchor}
An anchor block~$I\times J$~can be evaluated to any error tolerance~$\eta'\in\Q\cap(0,1]$~by a Las Vegas algorithm in expected running time
$$\softO\bigl(F_{d-1}(|I|,|J|)\bigr).$$
The algorithm computes exact batch maxima once in dimension~$d-1$~and uses one~$\KP_{d-1}$~call.
\end{lemma}

\begin{proof}
\emph{Column anchor.} Suppose $j_0\in J$ anchors the block, so $-L_{\mathrm{anc}}\le x_{ij_0}\le0$ for all $i\in I$. Subtracting its score cancels the row bias $b_i$ and yields $x_{ij}-x_{ij_0}=\vct u_i\cdot\vct p'_j+c'_j$, where $\vct p'_j=\vct p_j-\vct p_{j_0}$ and $c'_j=c_j-c_{j_0}$ for $j\in J$. Define
$$M'_j=\max_{i\in I}\vct{p}'_j\cdot \vct{u}_i,\qquad T_{ji}=\e^{\vct{p}'_j\cdot \vct{u}_i-M'_j}\qquad(j\in J,\ i\in I).$$
Thus~$\mat{T}$~is the shifted kernel of Definition~\ref{def:KP} in dimension~$d-1$~with rows~$\vct{p}'_j$~($j\in J$), columns~$\vct{u}_i$~($i\in I$), and shifts~$M'_j$. Define the vectors $\vct{d}_{\mathrm{row}}=(d_i^{\mathrm{row}})_{i\in I}$ and $\vct{d}_{\mathrm{col}}=(d_j^{\mathrm{col}})_{j\in J}$ by
$$d_i^{\mathrm{row}}=\e^{L_{\mathrm{anc}}+x_{ij_0}},\qquad d_j^{\mathrm{col}}=\e^{c'_j+M'_j-L_{\mathrm{anc}}},$$
so the identity~$x_{ij}=(L_{\mathrm{anc}}+x_{ij_0})+(\vct{p}'_j\cdot \vct{u}_i-M'_j)+(c'_j+M'_j-L_{\mathrm{anc}})$~implies~$S_{ij}=d_i^{\mathrm{row}}T_{ji}d_j^{\mathrm{col}}$, that is,
$$\mat{S}_{IJ}\vct{w}_J=\vct{d}_{\mathrm{row}}\odot\mat{T}^{\top}(\vct{d}_{\mathrm{col}}\odot\vct{w}_J).$$
We have~$1\le d_i^{\mathrm{row}}\le \e^{L_{\mathrm{anc}}}$~because~$x_{ij_0}\in[-L_{\mathrm{anc}},0]$; if~$i^*\in I$~is a maximizing index for~$M'_j$, then~$c'_j+M'_j=x_{i^*j}-x_{i^*j_0}\le0+L_{\mathrm{anc}}$, hence~$0<d_j^{\mathrm{col}}\le1$.

\emph{Row anchor.} Symmetrically, for a row anchor~$i_0\in I$, set~$\vct{u}'_i=\vct{u}_i-\vct{u}_{i_0}$~and define the shifted kernel
$$M'_i=\max_{j\in J}\vct{u}'_i\cdot \vct{p}_j,\qquad T_{ij}=\e^{\vct{u}'_i\cdot \vct{p}_j-M'_i}\qquad(i\in I,\ j\in J).$$
Set
$$d_i^{\mathrm{row}}=\e^{b_i-b_{i_0}+M'_i},\qquad d_j^{\mathrm{col}}=\e^{x_{i_0j}}.$$
Subtracting the anchor score cancels the column bias, so the block product equals
$$\mat{S}_{IJ}\vct{w}_J=\vct{d}_{\mathrm{row}}\odot\mat{T}(\vct{d}_{\mathrm{col}}\odot\vct{w}_J).$$
By the same maximum comparison as above,~$0<d_i^{\mathrm{row}}\le \e^{L_{\mathrm{anc}}}$, and by the anchor condition,~$\e^{-L_{\mathrm{anc}}}\le d_j^{\mathrm{col}}\le1$.

Proposition~\ref{prop:batchsupport} computes the maxima in expected time $\softO(F_{d-1}(|I|,|J|))$. We pass them to the inductive kernel algorithm; its call and the diagonal multiplications have the same expected time bound by Appendix~\ref{app:biased-arithmetic}.
\end{proof}

\begin{lemma}
\label{lem:direct}
The product of a block~$I\times J$~can be approximated to any error tolerance~$\eta'\in\Q\cap(0,1]$~deterministically in running time~$\softO(|I||J|)$.
\end{lemma}

\begin{proof}
Since every score is nonpositive, Lemma~\ref{lem:scalarexp} approximates each exponential with $\softO(1)$ bit operations. Multiplying by the weights and summing gives the block product to error $\eta'$.
\end{proof}

\subsection{Random Prefix Chains}
\label{subsec:chains}

The top-level procedure \textsc{Decompose} processes thin nodes with full prefixes and balanced nodes with truncated prefixes. In both cases, \textsc{Chain} builds the prefix chain and \textsc{Assign} groups targets by their first hits, and in truncated mode, the final prefix's envelope also locates the remaining targets and defines the recursive subproblems.

A node is a block with a designated sampling side. Its source and target data follow the orientation convention of Section~\ref{subsec:bp-setup}. Conflict sets are relative to $\Sigma$, and the support and reporting structures are built on the fixed perturbed points $\wt{\vct y}_s$.

\begin{procedurebox}
\paragraph{Procedure~\textnormal{\textsc{Chain}}.}
The input is a node with~$|\Sigma|>s_0$, a top size~$s_0<s_{\mathrm{top}}\le|\Sigma|$, and a reporting structure~$\cR_\Sigma$~for~$(\wt{\vct{y}}_s)_{s\in\Sigma}$. The procedure runs in full mode if~$s_{\mathrm{top}}=|\Sigma|$~and in truncated mode otherwise.
\begin{enumerate}
\item Draw a uniformly random permutation of $\Sigma$ and store the source positions. Let $P_\ell$ be its prefix of size $s_\ell$, where $s_{\ell+1}=\min\{2s_\ell,s_{\mathrm{top}}\}$, and stop at the first $\ell_{\max}$ with $s_{\ell_{\max}}=s_{\mathrm{top}}$. For $\ell<\ell_{\max}$, put $B_\ell=P_{\ell+1}\setminus P_\ell$.
\item For~$\ell=0,1,\ldots,\ell_{\max}$, where the level~$\ell=\ell_{\max}$~is included only in truncated mode, perform the following steps.
\begin{enumerate}
\item Build the envelope structures of~$\cP(P_\ell)$~of Lemma~\ref{lem:envelope}.
\item For each canonical cell~$\Delta$~of~$\cP(P_\ell)$, in the order of Definition~\ref{def:cells}, use~$\cR_\Sigma$ to report the $d$ sets $\{s\in\Sigma:\wt{\vct{y}}_s\cdot(\bar{\vct{v}},1)>\wt{\vct{y}}_{s_\Delta}\cdot(\bar{\vct{v}},1)\}$ for the vertices $\bar{\vct{v}}\in\vertset(\Delta)$ (Lemma~\ref{lem:cells}(b)). Let~$C_\Delta$~be the union of these sets, with duplicates removed. Then check the conditions
$$\text{(V1)}\quad|C_\Delta|\le\frac{\Csamp|\Sigma|G}{s_\ell},\qquad\text{(V2)}\quad|C_\Delta\cap B_\ell|\le16\Csamp G\ \text{ if }\ell<\ell_{\max}.$$
\end{enumerate}
At the first cell that violates~(V1) or~(V2), discard all structures built in steps~1 and~2 and return to step~1.
\item Return the prefixes, the batches, the envelope structures of the levels processed in step~2, and all their canonical cells with their conflict sets.
\end{enumerate}
\end{procedurebox}

An attempt of Procedure~\textsc{Chain} is one pass through steps~1 and~2; it is accepted when it reaches step~3. Since the source permutation determines the canonical cells, their order, and their conflict sets (Definition~\ref{def:cells} and Lemma~\ref{lem:envelope}(a)), acceptance with exact reporting depends only on this permutation, independently of the randomness used in geometric preprocessing.

\begin{procedurebox}
\paragraph{Procedure~\textnormal{\textsc{Assign}}.}
The input is the output of an accepted attempt of Procedure~\textsc{Chain} for a node, together with the target set~$\Theta$~of the node. The mode is that of the chain.
\begin{enumerate}
\item For each set~$\Sigma^*\in\{P_0,B_0,B_1,\ldots,B_{\ell_{\max}-1}\}$, build the structure of Lemma~\ref{lem:support} for the points~$(\wt{\vct{y}}_s)_{s\in\Sigma^*}$.
\item For each target $i\in\Theta$, apply the first applicable rule.
\begin{enumerate}
\item[(A1)] Query the structure of~$P_0$~with the vector~$(\vct{u}_i,1)$. It returns~$\max_{s\in P_0}\wt f_s(\vct{u}_i)$~and a maximizing source~$s_*\in P_0$. If~$b_i+\wt f_{s_*}(\vct{u}_i)\ge-L$, assign the target the label~$(\mathrm{init},s_*)$.
\item[(A2)] Query the structures of~$B_0,B_1,\ldots$~in this order with the vector~$(\vct{u}_i,1)$, and let~$\ell<\ell_{\max}$~be the least index such that the returned maximizer~$s_*\in B_\ell$~satisfies~$b_i+\wt f_{s_*}(\vct{u}_i)\ge-L$, if such an index exists. Locate~$\vct{u}_i$~in~$\cP(P_\ell)$~by Lemma~\ref{lem:envelope}(b), which identifies a canonical cell~$\Delta$ containing $\vct{u}_i$, and assign the target the label~$(\ell,\Delta,s_*)$.
\item[(A3)] In full mode, call the target a tail target. In truncated mode, locate~$\vct{u}_i$~in~$\cP(P_{\ell_{\max}})$, which identifies a canonical cell~$\Delta$ containing $\vct{u}_i$, and assign the target the label~$(\mathrm{fin},\Delta)$.
\end{enumerate}
\item Sort the labeled targets into groups with equal labels. The source set of a group is~$\Sigma$~for a label~$(\mathrm{init},s_*)$, and~$C_\Delta$~for labels~$(\ell,\Delta,s_*)$~and~$(\mathrm{fin},\Delta)$.
\end{enumerate}
\end{procedurebox}

We represent the cell~$\Delta$ in a target label by its index in the list of Lemma~\ref{lem:envelope}(a), where \textsc{Chain} has already stored its conflict set.

\paragraph{Level sums.}
For~$\ell<\ell_{\max}$~we have~$s_\ell=s_02^\ell<s_{\mathrm{top}}\le|\Sigma|$, and~$s_{\ell_{\max}}=s_{\mathrm{top}}$. Hence~$2^{\ell_{\max}-1}<|\Sigma|/s_0$, and since~$s_0\ge4$, the number of levels satisfies~$\ell_{\max}+1<\log_2|\Sigma|\le\log_2N<G$. Since the sizes~$s_\ell$~with~$\ell<\ell_{\max}$~double from one level to the next and are less than~$s_{\mathrm{top}}$,
\begin{equation}
\label{eq:levelsums}
\sum_{\ell<\ell_{\max}}s_\ell^{k}\le2s_{\mathrm{top}}^{k},\qquad\sum_{\ell\le\ell_{\max}}s_\ell^{k}\le3s_{\mathrm{top}}^{k},
\end{equation}
and the same bounds hold with~$k-1$~in place of~$k$~if~$k\ge2$. For~$k=1$~the sums of~$s_\ell^{k-1}$~over~$\ell<\ell_{\max}$~and over~$\ell\le\ell_{\max}$~are~$\ell_{\max}$~and~$\ell_{\max}+1$, both less than~$G$.

\begin{lemma}
\label{lem:assign}
After an accepted attempt of Procedure~\textnormal{\textsc{Chain}}, Procedure~\textnormal{\textsc{Assign}} assigns at most one label to each target. In truncated mode every target receives a label, and in full mode every target without a label is a tail target. Moreover, the following hold.
\begin{enumerate}[label=(\alph*)]
\item For a group with label~$(\mathrm{init},s_*)$~or~$(\ell,\Delta,s_*)$, the source~$s_*$~lies in the source set of the group, and it anchors the group's block. For a label~$(\ell,\Delta,s_*)$, moreover,~$s_*\in C_\Delta\cap B_\ell$.
\item Consider a target $i\in\Theta$ and a source $s\in\Sigma$. If $i$ is a tail target or belongs to a labeled group whose source set excludes $s$, their original pair has score below $-L+1/8$.
\item The Cartesian products of each target group with its nonempty source set are pairwise disjoint subsets of the pairs of the node, and every other pair of the node has score less than~$-L+1/8$.
\end{enumerate}
Step~1 has expected running time~$\softO(s_{\mathrm{top}}^k)$; steps~2 and~3 require~$\softO(|\Theta|)$~time.
\end{lemma}

\begin{proof}
The first applicable rule assigns each target at most one label, and (A3) assigns every remaining target in truncated mode. A support query returns a hit precisely when its queried set contains one (Lemma~\ref{lem:support}).

\emph{Part~(a).} For a group labeled $(\mathrm{init},s_*)$ or $(\ell,\Delta,s_*)$, the assigned source hits every target of the group, so $b_i+f_{s_*}(\vct u_i)\ge-L-1/8\ge-L_{\mathrm{anc}}$ by Lemma~\ref{lem:window}(a). In an initial group, $s_*\in P_0\subseteq\Sigma$. In a later group, the preceding prefix $P_\ell$ has no hit at $\vct u_i$, so Lemma~\ref{lem:window}(c) implies $s_*\in C_\Delta\cap B_\ell$. Thus $s_*$ is an anchor in the group's source set.

\emph{Part~(b).} A tail target has no hit in~$P_0\cup B_0\cup\cdots\cup B_{\ell_{\max}-1}=P_{\ell_{\max}}=\Sigma$, so all its scores are strictly below~$-L+1/8$~by Lemma~\ref{lem:window}(b). An initial label~$(\mathrm{init},s_*)$ includes all sources in~$\Sigma$. For a label~$(\ell,\Delta,s_*)$, the proof of part~(a) shows that every hit lies in~$C_\Delta$. For a label~$(\mathrm{fin},\Delta)$, rule~(A3) was applied, so the target has no hit in~$P_0\cup B_0\cup\cdots\cup B_{\ell_{\max}-1}=P_{\ell_{\max}}$, and~$\vct{u}_i\in\Delta$; again every hit lies in~$C_\Delta$~by Lemma~\ref{lem:window}(c). In both cases, every source~$s\notin C_\Delta$~has original score below~$-L+1/8$~by Lemma~\ref{lem:window}(b).

\emph{Part~(c).} Every target has at most one label, so the groups have pairwise disjoint target sets, and these blocks are pairwise disjoint. Every pair outside these blocks either has a tail target or joins a labeled target to a source outside its group's source set, so its score is less than~$-L+1/8$~by part~(b).

\emph{Cost.} Since~$|P_0|=s_0$~and~$|B_\ell|\le s_\ell$, the sum of their~$k$-th powers is~$\Oh(s_{\mathrm{top}}^k)$~by~\eqref{eq:levelsums}, so Lemma~\ref{lem:support} implies the preprocessing bound. Processing each target requires at most~$G$~support queries and one cell-location query, for a total of~$\softO(|\Theta|)$~time including the final sorting.
\end{proof}

For the running-time analysis, let~$g_{\mathrm{rep}}$~be the group size of~$\cR_\Sigma$~in Lemma~\ref{lem:reporttree} and set~$c_{\mathrm{rep}}=\lceil|\Sigma|/g_{\mathrm{rep}}\rceil+1$. A query returning~$Z$~then costs~$\softO(c_{\mathrm{rep}}+|Z|)$~time.

\begin{lemma}
\label{lem:accept}
\label{lem:chaincost}
Each attempt of~\textnormal{\textsc{Chain}} is accepted with probability greater than~$3/4$, independently of earlier attempts. For every fixed permutation, its expected running time, over the remaining random choices, is
$$\softO\!\left(|\Sigma|+\sum_{\ell=0}^{\ell_{\max}}\bigl(c_{\mathrm{rep}}s_\ell^k+|\Sigma|s_\ell^{k-1}\bigr)\right).$$
Summing over all attempts increases the bound by at most a constant factor.
\end{lemma}

\begin{proof}
\emph{Acceptance.} At each level,~$P_\ell$~is a uniform subset of~$\Sigma$. By Lemma~\ref{lem:sample}, the probability of violating (V1) is at most~$1/(8G)$ per level and less than~$1/8$ over the fewer than~$G$~levels.

For (V2), condition on~$P_\ell$~and a cell~$\Delta$~satisfying (V1). The batch~$B_\ell$~is a uniform subset of~$\Sigma\setminus P_\ell$~of size at most~$\min\{s_\ell,|\Sigma|-s_\ell\}$. Since~$C_\Delta\cap P_\ell=\varnothing$, the random variable~$X=|C_\Delta\cap B_\ell|$~is hypergeometric, with mean at most~$2\Csamp G$. Lemma~\ref{lem:hypergeom} implies
$$\Prob{X\ge16\Csamp G}\le(\e/8)^{16\Csamp G}.$$
There are~$\Oh_d(GN^k)$~cell--level pairs, and a union bound with the sampling constants in Appendix~\ref{app:geom-potential} gives probability less than~$1/8$ for any such failure. Each attempt is therefore accepted with probability greater than~$3/4$, independently of earlier attempts because acceptance depends only on the newly drawn permutation.

\emph{Cost.} After drawing the permutation in~$\softO(|\Sigma|)$~time, we construct the envelope at each level~$\ell$~in expected time~$\softO(s_\ell^k)$~and perform~$d$~reports for each of its~$\Oh_d(s_\ell^k)$~cells. The cells satisfying (V1) contribute~$\softO(c_{\mathrm{rep}}s_\ell^k+|\Sigma|s_\ell^{k-1})$~time at that level, including forming unions and counting batch intersections. A cell violating (V1) contributes at most an additional~$\softO(|\Sigma|)$~before the attempt stops. This bound holds for every fixed permutation and conditionally on all earlier attempts, so the constant acceptance probability increases the expected total cost by at most a constant factor.
\end{proof}

\paragraph{Geometric cost of a prefix chain.}
After preprocessing~$\cR_\Sigma$, running~\textsc{Chain} and~\textsc{Assign}, including failed attempts, has expected cost
\begin{equation}
\label{eq:prefix-geometry-cost}
\softO\bigl(|\Theta|+|\Sigma| s_{\mathrm{top}}^{k-1}+c_{\mathrm{rep}}s_{\mathrm{top}}^k\bigr).
\end{equation}
This follows for both full and truncated chains by combining Lemmas~\ref{lem:chaincost} and~\ref{lem:assign} with~\eqref{eq:levelsums}, using the logarithmic number of levels when~$k=1$. For evaluating the resulting blocks, each cell has at most logarithmically many anchors and the target groups are disjoint, so concavity relates the sum of their lower-dimensional kernel costs to the node sizes.

\paragraph{Exponents in the block-cost analysis.}
Set $k'=k(d-1)$, $\alpha_d=k'/(k'+1)$, and $\gamma_d=(k-k')/(k'+1)$, where $\alpha_d$ is the exponent in the $(d-1)$-dimensional bound $F_{d-1}$ and $\gamma_d$ governs how the total block cost at level $\ell$ scales with the prefix size $s_\ell$.

\begin{lemma}
\label{lem:levelgroups}
Run Procedure~\textnormal{\textsc{Assign}} after an accepted attempt of Procedure~\textnormal{\textsc{Chain}} on a node with source set~$\Sigma$~and target set~$\Theta$. For every group labeled~$(\mathrm{init},s_*)$~or~$(\ell,\Delta,s_*)$, let~$D$~be its block, with sides~$\Theta_D$~and~$\Sigma_D$. The target sets are disjoint, so~$\sum_D|\Theta_D|\le|\Theta|$; at most~$s_0$~blocks have an initial label~$(\mathrm{init},s_*)$, and each has~$\Sigma_D=\Sigma$. For each level~$\ell<\ell_{\max}$, the collection~$\mathcal D_\ell$~of blocks labeled~$(\ell,\Delta,s_*)$~satisfies
$$|\mathcal D_\ell|=\softO(s_\ell^k),\qquad\sum_{D\in\mathcal D_\ell}|\Sigma_D|=\softO(|\Sigma|s_\ell^{k-1}),$$
$$\sum_{D\in\mathcal D_\ell}\bigl(|\Theta_D||\Sigma_D|\bigr)^{\alpha_d}=\softO\!\left(\bigl(|\Sigma||\Theta|\bigr)^{\alpha_d} s_\ell^{\gamma_d}\right).$$
All these anchor blocks together satisfy
\begin{equation}
\label{eq:prefix-block-cost}
\sum_D F_{d-1}(|\Theta_D|,|\Sigma_D|)
=\softO\!\left(|\Theta|+|\Sigma| s_{\mathrm{top}}^{k-1}+(|\Theta||\Sigma|)^{\alpha_d} s_{\mathrm{top}}^{\gamma_d}\right).
\end{equation}
This also bounds the expected time to evaluate the blocks to tolerance~$\etablk$~and accumulate their outputs. For odd $d$, $\gamma_d=0$, so the product term is independent of $s_{\mathrm{top}}$, while for even $d$, $\alpha_d=(k-1)/k$ and $\gamma_d=1/k$.
\end{lemma}

\begin{proof}
By Lemma~\ref{lem:assign}, the target groups are disjoint and at most~$s_0$~initial groups use the full source set~$\Sigma$. At level~$\ell$, each of the~$\Oh_d(s_\ell^k)$~cells has at most~$\Oh_d(G)$~anchors by (V2), and each associated source set has size~$\Oh_d(|\Sigma|G/s_\ell)$ by (V1). The first two estimates follow by summing over these cells.

For nonempty $\mathcal D_\ell$, concavity and disjointness imply
$$\begin{aligned}
\sum_{D\in\mathcal D_\ell}(|\Theta_D||\Sigma_D|)^{\alpha_d}
&\le\max_{D\in\mathcal D_\ell}|\Sigma_D|^{\alpha_d}\,|\mathcal D_\ell|^{1-\alpha_d}
\left(\sum_{D\in\mathcal D_\ell}|\Theta_D|\right)^{\alpha_d}\\
&=\softO\bigl((|\Sigma||\Theta|)^{\alpha_d} s_\ell^{k(1-\alpha_d)-\alpha_d}\bigr)
=\softO\bigl((|\Sigma||\Theta|)^{\alpha_d} s_\ell^{\gamma_d}\bigr).
\end{aligned}$$
Summing over the logarithmically many levels proves~\eqref{eq:prefix-block-cost}, with target terms totaling at most~$|\Theta|$, source terms $\softO(|\Sigma| s_{\mathrm{top}}^{k-1})$, and product terms $\softO((|\Theta||\Sigma|)^{\alpha_d} s_{\mathrm{top}}^{\gamma_d})$. The initial blocks contribute at most $s_0F_{d-1}(|\Theta|,|\Sigma|)$, which also satisfies~\eqref{eq:prefix-block-cost} because $s_0=\softO(1)$ and $k-1,\gamma_d\ge0$.

Lemma~\ref{lem:anchor} bounds each block evaluation by its $F_{d-1}$ term. Summing these conditional expectations and accumulating the outputs gives~\eqref{eq:prefix-block-cost}.
\end{proof}

\subsection{The Full Random Prefix}
\label{subsec:fullprefix}

\textsc{FullPrefix} specializes~\textsc{Chain} and~\textsc{Assign} to~$s_{\mathrm{top}}=|\Sigma|$, then evaluates the resulting anchor groups.

\begin{procedurebox}
\paragraph{Procedure~\textnormal{\textsc{FullPrefix}}.}
The input is a node with source set~$\Sigma$~and target set~$\Theta$.
\begin{enumerate}
\item If~$|\Sigma|\le s_0$, evaluate the block of the node as a direct block by Lemma~\ref{lem:direct} with~$\eta'=\etablk$, add the result to~$\wh{\vct{o}}$, and stop.
\item Build the reporting structure~$\cR_\Sigma$~with group size~$g_{\mathrm{rep}}=|\Sigma|$.
\item Run Procedure~\textsc{Chain} with the top size~$s_{\mathrm{top}}=|\Sigma|$, which is full mode, and the structure~$\cR_\Sigma$. Then run Procedure~\textsc{Assign} with the target set~$\Theta$.
\item For every group with a label~$(\mathrm{init},s_*)$~or~$(\ell,\Delta,s_*)$, evaluate the group's block with anchor~$s_*$, by Lemma~\ref{lem:anchor} with~$\eta'=\etablk$, and add the result to~$\wh{\vct{o}}$.
\end{enumerate}
\end{procedurebox}

\begin{lemma}
\label{lem:fullprefix}
Procedure~\textnormal{\textsc{FullPrefix}} has expected running time~$\softO(|\Sigma|^k+|\Theta|)$.
\end{lemma}

\begin{proof}
If~$|\Sigma|\le s_0$, direct evaluation costs~$\softO(|\Theta|)$~time. Otherwise, the reporting structure is built in expected time~$\softO(|\Sigma|^k)$~and returns a report~$Z$~in~$\softO(1+|Z|)$~time, with~$c_{\mathrm{rep}}=2$~because~$g_{\mathrm{rep}}=|\Sigma|$. Substituting~$s_{\mathrm{top}}=|\Sigma|$ in~\eqref{eq:prefix-geometry-cost} gives expected time~$\softO(|\Theta|+|\Sigma|^k)$ for the chain and assignment. Equation~\eqref{eq:prefix-block-cost} bounds anchor evaluation by this expression as well, since
\begin{equation*}
(|\Theta||\Sigma|)^{\alpha_d} |\Sigma|^{\gamma_d}
=|\Theta|^{\alpha_d}(|\Sigma|^k)^{1/(k'+1)}\le |\Theta|+|\Sigma|^k.\qedhere
\end{equation*}
\end{proof}

\subsection{Truncated Prefixes and the Heavy/Light Recursion}
\label{subsec:truncated}

At a recursive node $v$ with block $I\times J$, write $m_v=\min\{|I|,|J|\}$ and $n_v=\max\{|I|,|J|\}$. If $m_v^k\le n_v$, \textsc{FullPrefix} with the smaller side as sources takes near-linear time. Otherwise $k\ge2$, and we use the larger side as sources with a truncated prefix. Define
$$\Phi(m,n)=(mn)^{k/(k+1)},\qquad F_d(m,n)=\Phi(m,n)+m+n.$$
For balanced sizes $m,n$ with $m^k>n$ and $n^k>m$, we choose the prefix scale $\kappa_0$ to balance the reporting and light-cell cost $n\kappa_0^{k-1}$ against the heavy-child potential bound $m(n/\kappa_0)^{(k-1)/k}$. Setting both expressions equal to $\Phi(m,n)$ gives
$$\kappa_0=(m^k/n)^{1/(k^2-1)},\qquad \kappa=\lceil\kappa_0\rceil,\qquad n_\kappa=n/\kappa,\qquad s_f=s_0\kappa.$$
This choice balances $n\kappa_0^{k-1}=m(n/\kappa_0)^{(k-1)/k}=\Phi(m,n)$. Rounding to $\kappa$ gives
\begin{equation}
\label{eq:balance-estimates}
\begin{gathered}
2\le\kappa\le\min\{2\kappa_0,n\},\qquad \Phi(m,n)>\max\{m,n\},\\
n\kappa^{k-1}\le2^{k-1}\Phi(m,n),\qquad
m(n/\kappa)^{(k-1)/k}\le\Phi(m,n).
\end{gathered}
\end{equation}
If $s_f\ge n$, then $\kappa_0>1$ and $\kappa<2\kappa_0$ give $n<2s_0\kappa_0$. Raising this inequality to the power $k^2-1$ and substituting $\kappa_0^{k^2-1}=m^k/n$ yields $n^{k^2}<(2s_0)^{k^2-1}m^k$, hence
\begin{equation}
\label{eq:fullsample}
n^k\le(4\Csamp G)^km,\qquad
n\le(4\Csamp G)^2\ \text{ when }n\ge m.
\end{equation}
In the latter case, direct evaluation costs $mn\le(4\Csamp G)^2\Phi(m,n)$. These estimates also apply to the batch-maxima algorithm in Appendix~\ref{app:geom-batch}.

For a final cell $\Delta$, let $m_\Delta$ and $n_\Delta$ count its unassigned targets and conflicting sources. We process light cells with $m_\Delta\le\Gamma n_\kappa^{1/k}$ by reversing the orientation and applying \textsc{FullPrefix}, and recurse on the heavy cells; setting $\Gamma=4^{k+1}$ ensures that the heavy children's potentials sum to at most $\Gamma^{-1/(k+1)}=1/4$ of their parent's potential in the next lemma.

\begin{lemma}[Heavy cells; proof in Appendix~\ref{app:geom-potential}]
\label{lem:heavy}
For balanced sizes $m,n$ and parameters as in~\eqref{eq:balance-estimates}, let~$(m_\Delta,n_\Delta)_{\Delta\in\mathcal D}$~be a finite list of pairs of positive integers with~$\sum_{\Delta\in\mathcal D}m_\Delta\le m$~and~$n_\Delta\le n_\kappa$~for all~$\Delta$. Call~$\Delta$~heavy if~$m_\Delta>\Gamma n_\kappa^{1/k}$. Then
$$\sum_{\Delta\ \mathrm{heavy}}\Phi(m_\Delta,n_\Delta)\le\tfrac14\,\Phi(m,n),\qquad\sum_{\Delta\ \mathrm{heavy}}(m_\Delta+n_\Delta)\le\bigl(1+\Gamma^{-1}\bigr)\Phi(m,n),$$
and~$m_\Delta n_\Delta\le mn/2$~for every~$\Delta\in\mathcal D$.
\end{lemma}

\begin{procedurebox}
\paragraph{Procedure~\textnormal{\textsc{Decompose}}.}
The input is the block $I\times J$ at a recursive node $v$.
\begin{enumerate}
\item Let $m_v=\min\{|I|,|J|\}$ and $n_v=\max\{|I|,|J|\}$.
\item \emph{Thin node.} If $m_v^k\le n_v$, run \textsc{FullPrefix} with the smaller side as sources and stop.
\item \emph{Balanced node.} Use the larger side as sources $\Sigma$ and the smaller side as targets $\Theta$. Set $\kappa=\lceil(m_v^k/n_v)^{1/(k^2-1)}\rceil$, $n_\kappa=n_v/\kappa$, and $s_f=s_0\kappa$. If $s_f\ge n_v$, evaluate the block directly with tolerance $\etablk$ by Lemma~\ref{lem:direct}, add its output to $\wh{\vct o}$, and stop.
\item Build $\cR_\Sigma$ with $g_{\mathrm{rep}}=\kappa$, run \textsc{Chain} with $s_{\mathrm{top}}=s_f$ in truncated mode, and apply \textsc{Assign} to $\Theta$. Evaluate the groups with assigned anchors as in step~4 of \textsc{FullPrefix}.
\item For every nonempty group with label~$(\mathrm{fin},\Delta)$, let~$\Theta_\Delta$~be its target set, and set~$m_\Delta=|\Theta_\Delta|$~and~$n_\Delta=|C_\Delta|$.
\begin{itemize}
\item If~$C_\Delta=\varnothing$, do nothing.
\item If~$C_\Delta\ne\varnothing$~and~$m_\Delta^k\le\Gamma^k n_\kappa$, call the cell light and reverse the parent orientation. Run \textsc{FullPrefix} with source set $\Sigma_{\mathrm{new}}=\Theta_\Delta$ and target set $\Theta_{\mathrm{new}}=C_\Delta$.
\item Otherwise, call the cell heavy and recurse on its block with \textsc{Decompose}.
\end{itemize}
\end{enumerate}
\end{procedurebox}

\begin{lemma}
\label{lem:decompose}
Procedure~\textnormal{\textsc{Decompose}} is a Las Vegas algorithm. Applied to~$I_0\times J_0$, it evaluates pairwise disjoint direct or anchor blocks, each to coordinatewise error at most~$\etablk$. Every omitted pair has score less than~$-L+1/8$, and the procedure has expected running time~$\softO(F_d(m,n))$.
\end{lemma}

\begin{proof}
\emph{Coverage.} By Lemma~\ref{lem:assign}, each assignment partitions its node's pairs into anchor blocks, blocks associated with final cells, and discarded pairs with scores below~$-L+1/8$. A full prefix evaluates its anchor blocks, and a truncated prefix also passes each nonempty block from a final cell to~\textsc{FullPrefix} or recursion. Direct evaluation includes every pair in a block, and changing orientation leaves its original row--column pairs unchanged. Induction therefore proves that the evaluated blocks are disjoint and cover every pair with score at least $-L+1/8$. Their individual errors are at most $\etablk$ by Lemmas~\ref{lem:anchor} and~\ref{lem:direct}.

\emph{Local cost.} A thin leaf is processed in expected time~$\softO(m_v+n_v)$~by Lemma~\ref{lem:fullprefix}. If step~3 evaluates a balanced node directly, the running time is~$\softO(\Phi(m_v,n_v))$~by~\eqref{eq:fullsample}.

At a node reaching step~4, the final prefix has size~$s_f=s_0\kappa$. By (V1) and~$s_0=2\Csamp G$, its final groups satisfy
$$n_\Delta\le\Csamp n_vG/s_f=n_\kappa/2,\qquad\sum_\Delta m_\Delta\le m_v,$$
where the second inequality follows from disjointness of the target groups. By~\eqref{eq:balance-estimates}, we have~$\kappa\le n_v$,~$n_v\kappa^{k-1}=\Oh_d(\Phi(m_v,n_v))$, and~$m_v+n_v=\Oh(\Phi(m_v,n_v))$. Building the reporting structure with group size~$\kappa$ thus costs~$\softO(\Phi(m_v,n_v))$ in expectation. Its query parameter is~$c_{\mathrm{rep}}=\lceil n_\kappa\rceil+1=\Oh(n_\kappa)$, so~\eqref{eq:prefix-geometry-cost} with~$s_{\mathrm{top}}=s_0\kappa$ implies the same expected cost for the chain and assignment.

For the anchor blocks, the cost in~\eqref{eq:prefix-block-cost} is
$$\softO\bigl(m_v+n_v\kappa^{k-1}+(m_vn_v)^{\alpha_d}\kappa^{\gamma_d}\bigr).$$
For odd~$d$, the last term is $\Oh_d(\Phi(m_v,n_v))$ because~$k'=k$ and~$\gamma_d=0$. For even~$d$, use~$k'=k-1$,~$\gamma_d=1/k$~and~$\kappa\le2\kappa_0$~to obtain
$$(m_vn_v)^{(k-1)/k}\kappa^{1/k}
\le2^{1/k}\Phi(m_v,n_v)\left(\frac{m_v}{n_v^k}\right)^{1/(k(k^2-1))}
\le2\Phi(m_v,n_v).$$
There are~$\softO(\kappa^k)$~final cells. A nonempty light group satisfies~$m_\Delta^k\le\Gamma^kn_\kappa$~and~$n_\Delta\le n_\kappa/2$, so its call to~\textsc{FullPrefix}, with the parent's targets as sources, requires expected time~$\softO(n_\kappa)$, for a total of~$\softO(n_v\kappa^{k-1})$~over all light cells. These local bounds include failed sampling attempts and hold conditionally on the input to each node.

\emph{Summing the recursion.} By Lemma~\ref{lem:heavy}, the heavy children have total potential at most one quarter of their parent's and total linear size $\Oh_d(\Phi(m_v,n_v))$; each child also satisfies $m_\Delta n_\Delta\le m_vn_v/2$. Summing over the recursion as in Appendix~\ref{app:geom-potential} yields total expected time $\softO(F_d(m,n))$.
\end{proof}

\subsection{Error Analysis and the Proof of Proposition~\ref{prop:BP}}
\label{subsec:biasederr}

\begin{proof}[Proof of Propositions~\ref{prop:bpdecomposition} and~\ref{prop:BP}]
Consider a $\BP_{d-1}$ instance with the notation of this section. The algorithm fixes the perturbations and bounding halfspaces for both orientations and the parameters~\eqref{eq:sampling} and~\eqref{eq:params} as in Section~\ref{subsec:bp-setup}, sets~$\wh{\vct{o}}=\vct{0}$, runs Procedure~\textsc{Decompose} on $I_0\times J_0$, and returns the accumulated vector $\wh{\vct o}$.

\emph{Correctness.} By Lemma~\ref{lem:decompose}, the evaluated blocks are disjoint and each has error at most $\etablk$. In any row, at most $n$ block errors contribute, and each omitted pair has weight below $\e^{-L+1/8}$. Hence
$$\|\wh{\vct o}-\mat S\vct w\|_\infty
\le n\bigl(\etablk+W\e^{-L+1/8}\bigr)\le\eta/2$$
by~\eqref{eq:params}. The total error, including final rounding, is at most $\eta$.

\emph{Cost.} The shared perturbations, accumulation of block outputs, and final rounding require~$\softO(N)$~time plus the linear terms in the block bounds (Appendix~\ref{app:biased-arithmetic}). Adding the cost of \textsc{Decompose} from Lemma~\ref{lem:decompose} gives total expected time $\softO(F_d(m,n))$.
\end{proof}

\subsection{Proof of Theorem~\ref{thm:main}}
\label{subsec:mainproof}

\begin{proof}[Proof of Theorem~\ref{thm:main}]
For $d\ge2$, apply Theorem~\ref{thm:KP} to the products defined in Section~\ref{sec:prelim}, using $U=W=\lceil B\rceil$ and tolerance $\eta=\eps/(8(B+1))$. The returned ratios are $\eps$-accurate by Lemma~\ref{lem:shift}. The kernel algorithms are Las Vegas and have total expected cost
$$r\,\softO(F_d(m,n)),$$
which also includes preparing the weight vectors and computing the final ratios, at cost $r\,\softO(m+n)$. For fixed $r$, this is $\softO(F_d(m,n))$.

Adding a zero coordinate embeds a one-dimensional instance in the plane with identical scores. The planar algorithm therefore gives the required bound for $d=1$, since $F_2(m,n)=\Oh(m+n)=\Oh(F_1(m,n))$.
\end{proof}

\section[A Static-Query Comparison with Semigroup Halfplane Aggregation]{A Static-Query Comparison with\\ Semigroup Halfplane Aggregation}
\label{sec:semigroup}

Corollary~\ref{cor:online-d2} preprocesses fixed keys and values in~$\softO(n)$~time and space and answers each query in~$\softO(1)$~time. Because truncating the exponentially decaying tail reduces each query to aggregating keys whose scores are close to the row maximum, it is natural to compare this bound with exact halfplane aggregation.

\paragraph{The comparison model.}
In the semigroup arithmetic model, preprocessing stores linear forms with nonnegative integer coefficients in the point weights, which belong to a commutative semigroup. A query combines stored forms by the semigroup operation, and the scheme must work for every weight assignment and every query in the prescribed set of ranges. Space counts the stored forms, and query cost counts the distinct forms used in an answer, allowing nonnegative integer coefficients (combining $\ell$ distinct forms requires at least $\ell-1$ binary semigroup operations).

Auxiliary geometric computation is free, with an adjoined identity element serving as the answer to an empty range. The lower bound applies to faithful semigroups, in which identically equal linear forms involve the same set of variables. Exact rational summation using only additions of weights fits this model.

\begin{lemma}
\label{lem:semigroup-finite-precision}
Let~$d\ge2$~be a fixed integer.  For all large enough~$n$, there is a labeled
set~$P_n\subseteq[0,1]^d$~of~$n$~points with~$\Oh_d(\log n)$-bit rational
coordinates, and a collection of halfspaces with~$\Oh_d(\log n)$-bit integer
coefficients, such that every faithful-semigroup storage scheme using
$n_{\mathrm{st}}\ge n$~stored forms has worst-case query cost
$$
 \wt\Omega_d\!\left(\frac{n^{d/(d+1)}}{n_{\mathrm{st}}^{1/(d+1)}}\right).
$$
These finite-precision halfspaces realize every halfspace range on
$P_n$.
\end{lemma}

\begin{proof}
Arya, Mount, and Xia~\cite[Theorem~4.1(i)]{aryaMountXia2012} prove this lower bound for scattered point sets, and their Lemma~4.6 shows that independent uniform points in $[0,1]^d$ are scattered with probability $1-\mathrm o(1)$. We use such a set whose order type survives rounding to a dyadic grid with $\Oh_d(\log n)$-bit coordinates.

Set $\delta_{\mathrm{sep}}=n^{-(d+3)}$. A uniform point lies within distance $\delta_{\mathrm{sep}}$ of a fixed proper affine subspace with probability $\Oh_d(\delta_{\mathrm{sep}})$. A union bound over the successive affine heights of all ordered $(d+1)$-tuples shows that, with probability $1-\Oh_d(n^{-2})$, every orientation determinant has magnitude at least $\delta_{\mathrm{sep}}^d$. Fix a realization $P$ satisfying both this event and the scattered-set condition.

Round each coordinate to the nearest multiple of $2^{-p}$, with
$$2^{-p}\le\frac{\delta_{\mathrm{sep}}^d}{4d\,d!},\qquad p=\Oh_d(\log n).$$
By multilinearity, each orientation determinant changes by at most $2d\,d!\,2^{-p}\le\delta_{\mathrm{sep}}^d/2$. Thus the rounded set $P'$ has the same order type as $P$, hence the same labeled halfspace ranges. Since the semigroup storage/query tradeoff depends only on these ranges, the lower bound transfers to $P'$.

Finally, $P'$ lies on a grid with denominator $2^p=n^{\Oh_d(1)}$. Lemma~\ref{lem:halfspace-encoding} in Appendix~\ref{app:halfspace-encoding} represents each range by an $\Oh_d(\log n)$-bit integer halfspace.
\end{proof}

\begin{corollary}
\label{cor:semigroup-query-separation}
For polynomial coordinate and value bounds, inverse-polynomial additive error tolerance, polylogarithmic coordinate encoding lengths, and a fixed number of value channels, planar attention admits a deterministic static structure with~$\softO(n)$~preprocessing time and space and~$\softO(1)$~query time. In contrast, a faithful-semigroup scheme
for exact planar halfspace aggregation with~$\softO(n)$~stored forms has
worst-case query cost
$$
 \wt\Omega(n^{1/3}),
$$
even for points and halfspaces whose coordinates and coefficients have
$\Oh(\log n)$-bit encodings.
\end{corollary}

\begin{proof}
Apply Corollary~\ref{cor:online-d2} for attention and Lemma~\ref{lem:semigroup-finite-precision} with~$d=2$~and~$n_{\mathrm{st}}=\softO(n)$~for halfspaces. Both structures are preprocessed before the query arrives.
\end{proof}

\section{Conclusion and Further Directions}
\label{sec:conclusion}

We have shown that softmax attention in fixed dimension $d$ can be computed at arbitrary positive temperature within the best known batch maximum-inner-product bound in $m$ and $n$,
$$\softO\bigl((mn)^{k/(k+1)}+m+n\bigr),\qquad k=\lfloor d/2\rfloor,$$
and deterministically in near-linear time in the plane. Together with the dimension-preserving reduction in Section~\ref{sec:reduction}, this identifies the current best exponents for the two problems.

The planar data structure also separates attention from exact range aggregation. After $\softO(n)$ preprocessing and storage, each query requires $\softO(1)$ time, whereas faithful-semigroup halfplane aggregation with near-linear storage has a polynomial worst-case query lower bound (Section~\ref{sec:semigroup}).

The central open question is whether the agreement between the current bounds reflects equality of the optimal exponents. A dimension-preserving reduction from attention to batch maximum inner product with a subpolynomial additional factor in the running time would establish that equality.

\paragraph{Optimal exponents.}
For fixed $d\ge2$, consider instances with $m=n$. Let $\alpha_{\ATT}(d)$ be the infimum of exponents $\alpha$ for which attention with $r=d$ value channels can be computed by a Las Vegas algorithm in expected time $n^{\alpha+\mathrm o(1)}$ whenever coordinate and value magnitudes are polynomially bounded, $\eps=n^{-\Oh(1)}$, and encoding lengths are polylogarithmic in $n$. The algorithm may depend on the fixed exponents in these bounds. Define $\alpha_{\BMIP}(d)$ analogously for $\BMIP_d$ with polynomially bounded integer coordinates. Our reduction and upper bound imply
$$\alpha_{\BMIP}(d)\le\alpha_{\ATT}(d)\le2-\frac{2}{\lfloor d/2\rfloor+1},$$
with equality to $1$ for $d=2,3$.

\begin{conjecture}[Dimensional complexity of attention]
\label{conj:exponents}
For every fixed integer~$d\ge4$,
$$\alpha_{\ATT}(d)\le\alpha_{\BMIP}(d),$$
and hence~$\alpha_{\ATT}(d)=\alpha_{\BMIP}(d)$.
\end{conjecture}

Such a black-box reduction would transfer future improvements for batch maximum inner product directly to attention, but constructing it for $d\ge4$ remains open. For lower bounds, the known unconditional results of the form $n^{1+\Omega(1)}$ apply to restricted models, including Erickson's polyhedral partitioning algorithms~\cite{erickson1996halfspace} (Section~\ref{subsec:related}).

\paragraph{Precision dependence and determinism.}
The factors suppressed by soft-O may be reduced through faster integer arithmetic, improved polynomial bases or interpolation nodes, and techniques that remove logarithmic losses in related geometric problems~\cite{chan2012partition,chanZheng2022}. To derandomize the algorithms, we need deterministic replacements for the random prefixes, the source order in the batch-maxima procedure~\textsc{ThinMax} (Appendix~\ref{app:geom-batch}), and the geometric preprocessing. Deterministic constructions of cuttings and convex hulls~\cite{matousek1992reporting,chazelle1993hull} offer starting points for these replacements.

\paragraph{Offline aggregation.}
For offline halfspace aggregation, all queries are given in advance. A faithful-semigroup lower bound with exponent above~$2-2/(\lfloor d/2\rfloor+1)$~would establish a batch analogue of the comparison in Section~\ref{sec:semigroup}. For simplex ranges, Chazelle~\cite{chazelle1997offline} proved an offline semigroup lower bound of~$\wt\Omega(n^{2-2/(d+1)})$~for~$n$~points and~$n$~simplices.

\paragraph{Implications for AI.}
Our near-linear algorithms in dimensions two and three motivate architectures with many micro-heads, each with small query and key dimensions and few value channels. By distributing representation across these heads, such architectures could use the algorithms to process longer contexts, with exact factorizations of small fixed rank offering a related route to low-dimensional scores~\cite{gupta2026}. For a fixed key--value cache, the planar structure answers queries in $\softO(1)$ time from stored aggregate moments (Corollary~\ref{cor:online-d2}), suggesting that an extension to growing caches could reduce the repeated reads of keys and values that limit autoregressive decoding bandwidth~\cite{shazeer2019}. Developing these applications calls for studying the modeling capacity of many small heads, reducing the arithmetic cost, and updating the cached geometric information efficiently, and could inspire new designs for both AI models and hardware architectures.

\appendix
\section{Numerical Stability and Precision}
\label{app:arithmetic}

This appendix verifies that all intermediate quantities appearing in our geometric primitives, polynomial evaluations, and dimension reductions have bit lengths polynomial in the instance parameter $\lambda$ of Section~\ref{subsec:bit-model}. The table locates the algorithms and their arithmetic analyses.
\begin{center}
\begin{tabular}{@{}p{.28\linewidth}p{.30\linewidth}p{.34\linewidth}@{}}
\toprule
Component & Algorithm and correctness & Bit complexity\\
\midrule
Planar kernel products and online queries
& Section~\ref{sec:d2}
& Appendix~\ref{app:planar-arithmetic}\\[3pt]
Anchor blocks
& Section~\ref{subsec:anchor}
& Appendix~\ref{app:biased-arithmetic}\\[3pt]
Temperature interpolation
& Section~\ref{subsec:kernel-to-biased}
& Appendix~\ref{app:induction-arithmetic}\\[3pt]
Ray shooting
& Appendix~\ref{subsec:geom-cuttings}
& Appendix~\ref{app:ray-arithmetic}\\[3pt]
Support and reporting
& Appendices~\ref{app:geom-support} and~\ref{app:reporting}
& Appendix~\ref{subsec:geom-perturb}\\[3pt]
Envelope cells and location
& Appendix~\ref{app:geom-cells}
& Appendices~\ref{app:geom-envelope} and~\ref{app:ray-shifts}\\[3pt]
Batch Max-IP
& Appendix~\ref{app:geom-batch}
& Appendices~\ref{subsec:geom-perturb}--\ref{app:ray-shifts}\\
\bottomrule
\end{tabular}
\end{center}

\subsection{Input Size, Temperature, and Reduction Cost}
\label{app:basic-arithmetic}

Reading and rounding a rational coordinate to a prescribed dyadic precision requires time polynomial in its encoding length and the number of precision bits. For an irrational inverse temperature, use a positive dyadic $\wt\beta$ with $|\wt\beta-\beta|\le\min\{\beta/2,\eps/(4dB^3)\}$. The scores change by at most $\eps/(4B)$, so the output changes by at most $\eps/2$ by Lemma~\ref{lem:softmaxpert}, leaving tolerance $\eps/2$ for the attention algorithm. This approximation needs $\Oh_d(\log(1/\eps)+\log(B+2)+|\log\beta|+1)$ bits. For $\beta=d^{-1/2}$, it is computable within the same polynomial bit cost. Thus temperature scaling and input preparation require $\softO(1)$ time per entry.

The reductions in Section~\ref{sec:reduction} consist of scaling input coordinates, computing inner products of length $d$, and rounding. The resulting numbers have magnitudes polynomial in $m+n,d,B$ and denominators of logarithmic encoding length. Consequently all these numbers have polynomially bounded encoding lengths in $\log(m+n+d+B)$, and the additional work is $\softO((m+n)d)$. In particular, this remains $n^{1+\mathrm o(1)}$ when $m=n$, $d=n^{\mathrm o(1)}$, and $B=n^{\Oh(1)}$.

\subsection{Arithmetic for Planar Kernel Products}
\label{app:planar-arithmetic}

For each call to Lemma~\ref{lem:kernel}, use the instance parameter $\lambda$ of Section~\ref{subsec:bit-model}.

\paragraph{Geometric rounding.}
For the rowwise tolerance $\eta_{\mathrm{row}}$ of Section~\ref{subsec:d2grid}, choose dyadic parameters
$$\delta=\Theta\!\left(\frac{\eta_{\mathrm{row}}}{N(U+1)}\right),\qquad \eta_{\mathrm{ent}}=\Theta(\eta_{\mathrm{row}}/N)$$
with small absolute constants satisfying the error estimates below. Both parameters have $\Oh(\lambda)$ bits.

\begin{lemma}
\label{lem:gridding}
Let~$\wt{\vct{q}}_i,\wt{\vct{k}}_j$~be obtained by rounding each coordinate of the original~$\vct{q}_i,\vct{k}_j$~to~$\delta\Z$~with error at most~$\delta$, and set
$$\wt h_i=\max_{j\in[n]}\wt{\vct{q}}_i\cdot\wt{\vct{k}}_j.$$
For every~$i\in[m]$, the original kernel~$\mat{T}$~satisfies
$$\sum_{j=1}^n\left|T_{ij}-\e^{\wt{\vct{q}}_i\cdot\wt{\vct{k}}_j-\wt h_i}\right|\le\frac{\eta_{\mathrm{row}}}{16}.$$
\end{lemma}

\begin{proof}
Rounding changes each inner product by~$\Oh((U+1)\delta)$. Lemma~\ref{lem:shift} bounds the change in each shifted exponential by~$\Oh((U+1)\delta)$ as well, giving row error~$\Oh(N(U+1)\delta)\le\eta_{\mathrm{row}}/16$~for a small enough constant in~$\delta$.
\end{proof}

\paragraph{Collinear keys.}
\label{subsec:d2degenerate}
For collinear rounded keys, translate one key to the origin and write $\vct k_j=t_j\vct u$ with $\|\vct u\|_\infty=1$, using $t_j$ from a coordinate of maximum spread or setting every $t_j=0$ when all keys coincide. For query $i$, let $\alpha_i=\vct q_i\cdot\vct u$ and $h_i=\max_j\alpha_i t_j$. In the order of increasing $t_j$, the selected range $\{j:\alpha_i t_j\ge h_i-L\}$ is a prefix, suffix, or the whole array. On this range, $s_{ij}=\alpha_i t_j-h_i+3L\in[2L,3L]$, so expanding the polynomial below in powers of $t_j$ reduces forward products and online queries to prefix sums of $t_j^\ell w_j$ for $0\le\ell\le g$, and transpose products to interval additions of weighted query coefficients. Every omitted shifted score is below $-L$, as required for the error estimates of Section~\ref{subsec:d2cost}.

\paragraph{Taylor approximation.}
Write $\exp_a(s)=\sum_{\ell=0}^a s^\ell/\ell!$.

\begin{lemma}
\label{lem:taylor}
For~$H\ge0$,~$s\in[0,H]$~and an integer~$a\ge0$,
$$\exp_a(s)\ge1,\qquad
0\le \e^s-\exp_a(s)\le \e^H\left(\frac{\e H}{a+1}\right)^{a+1}.$$
In particular, with the parameters of Section~\ref{subsec:d2grid},
$$\e^{-3L}\bigl(\e^s-\exp_g(s)\bigr)\le\eta_{\mathrm{ent}}\qquad(0\le s\le3L).$$
\end{lemma}

\begin{proof}
The Taylor terms are nonnegative. The standard remainder formula~\cite[Eq.~(1.4.36)]{dlmf}, together with~$(a+1)!\ge((a+1)/\e)^{a+1}$, implies the first bound. For a large enough constant in~$g=\Theta(L)$, the rescaled remainder is at most~$2^{-g}\le\eta_{\mathrm{ent}}$.
\end{proof}

\paragraph{Scalar exponentials.}
\label{app:scalar-exponentials}
\begin{lemma}
\label{lem:scalarexp}
For~$\lambda\ge1$, let~$x\in\Q$~and~$\delta_x\in\Q\cap(0,1]$~have encoding lengths~$\Oh(\lambda)$, and suppose~$\max\{x,0\}=\Oh(\lambda)$. In~$\poly(\lambda)$~time, one can compute a nonnegative dyadic~$\omega$~such that~$0\le\e^x-\omega\le\delta_x$~and its numerator and denominator have~$\Oh(\lambda)$~bits.
\end{lemma}

\begin{proof}
If~$x\le-\lceil\log_2(8/\delta_x)\rceil$, then~$\e^x\le\delta_x/8$ and we return~$\omega=0$. Otherwise~$|x|=\Oh(\lambda)$, so evaluating~$\Oh(\lambda)$~Taylor terms by Lemma~\ref{lem:taylor} produces an interval~$[A,A+\rho]$ containing~$\e^{|x|}$, with~$A\ge1$ and~$\rho\le\delta_x/8$. For~$x<0$, the reciprocal endpoints enclose~$\e^x$ in an interval of width at most~$\rho$, since~$A\ge1$. Rounding the lower endpoint down to a dyadic grid of mesh~$\Theta(\delta_x)\le\delta_x/2$ yields a nonnegative~$\omega$ satisfying~$0\le\e^x-\omega\le\rho+\delta_x/2\le5\delta_x/8<\delta_x$. The Taylor sums and endpoints have polynomial bit length, and the computation requires~$\poly(\lambda)$~time. The dyadic output has~$\Oh(\lambda)$-bit numerator and denominator because~$\log(1/\delta_x)=\Oh(\lambda)$~and~$\max\{x,0\}=\Oh(\lambda)$.
\end{proof}

For the planar kernel, $\e^{-3L}\ge2^{-5L}$, so an absolute tolerance of $\eta_{\mathrm{ent}}2^{-5L}$ ensures the relative error for $\omega$ in Section~\ref{subsec:d2moment}, with $\omega>0$ for a small enough constant in $\eta_{\mathrm{ent}}$.

\paragraph{Rowwise kernel error.}
Every omitted rounded exponential weight is less than~$\e^{-L}\le\eta_{\mathrm{ent}}$. For an entry with $j\in\mathcal K_i$, the polynomial error is at most~$\eta_{\mathrm{ent}}$ by Lemma~\ref{lem:taylor}, and the approximation to the common factor contributes at most
$$(\e^{-3L}-\omega)\exp_g(s_{ij})\le\eta_{\mathrm{ent}} \e^{-3L}\e^{3L}=\eta_{\mathrm{ent}}.$$
Thus the row error relative to the rounded exponential weights is at most~$3n\eta_{\mathrm{ent}}\le3\eta_{\mathrm{row}}/16$. Centering leaves the rounded shifted scores unchanged, so the triangle inequality and Lemma~\ref{lem:gridding} imply the error estimate relative to the original~$\mat T$.

\paragraph{Product error.}
For either product, set
$$\eta_{\mathrm{row}}=\frac{\eta}{8N(W+1)}.$$
We round the input weights to a working vector~$\vct w$ or~$\vct z$ on a common dyadic grid with coordinatewise error at most~$\eta/(8N)$, so the rounded entries have magnitude at most~$W+1$. The corresponding sweep then evaluates~$\mat P\vct w$ or~$\mat P^{\top}\vct z$ exactly, and we round each output coordinate to within~$\eta/8$.

By Lemma~\ref{lem:kernelerr}, the rowwise kernel error is at most~$\eta_{\mathrm{row}}$, whose contribution for the original input vector satisfies
$$|((\mat{T}-\mat{P})\vct{w}^{\mathrm{in}})_i|\le W\eta_{\mathrm{row}}\le\frac{\eta}{8},\qquad
|((\mat{T}-\mat{P})^{\top}\vct{z}^{\mathrm{in}})_j|\le mW\eta_{\mathrm{row}}\le\frac{\eta}{8}.$$
The extra factor~$m$~in the transpose bound comes from summing one entrywise error from each row. Since~$0\le P_{ij}\le1$, the subsequent rounding of the input vector contributes at most~$n\eta/(8N)$~for the forward product and~$m\eta/(8N)$~for the transpose product. Each is at most~$\eta/8$, so the total error including output rounding is at most~$3\eta/8<\eta$.

\paragraph{Attention and online queries.}
For planar attention, set $W=\lceil B\rceil$ and $\eta=\eps/(8(B+1))$ in the kernel calls. By Lemma~\ref{lem:shift}, the ratio error is at most $(B+1)\eta/(1-\eta)<\eps/4$, so the total remains below $\eps$ after output rounding to error $\eps/16$.

For the static structure, use $N=n+1$ and $\eta_{\mathrm{row}}=\eps/(16(W+1))$. Round the keys and each arriving query with the geometric mesh $\delta$, and round the values with error at most $\eps/16$. Let $\vct t$ be the original shifted-kernel row and $\vct p$ its row in $\mat P$. Then $\|\vct t-\vct p\|_1\le\eta_{\mathrm{row}}$ and $\sum_jp_j\ge1-\eta_{\mathrm{row}}>0$. The ratio error for an original value channel of magnitude at most $W$ is at most
$$\frac{2W\eta_{\mathrm{row}}}{1-\eta_{\mathrm{row}}}+\frac{\eps}{16}.$$
The second term is the value-rounding error under the nonnegative normalized weights. Rounding the output to error $\eps/16$ keeps the total below $\eps$.

\paragraph{Exact moment sums.}
The geometric and weight grids have common dyadic denominators of~$\Oh(\lambda)$~bits, and the polynomial degree is~$g=\Oh(\log(NW/\eta))=\Oh(\lambda)$. Before centering, scale the rounded keys of Lemma~\ref{lem:gridding}~to integer coordinates. Their polygon's centroid is a ratio of sums of fixed-degree products of these coordinates, so the shoelace formula expresses all centered keys with polynomial bit length and a common denominator. The collinear case has the same property, since its parameters are differences of grid coordinates and one shared direction vector.

Because the weighted key monomials and row coefficients involve powers through degree~$g$~and Taylor factorials, their numerators and common denominators have polynomial bit length. Summing at most~$N$~signed terms over these denominators adds only~$\Oh(\log N)$~numerator bits, so the sums and the polynomially many terms in each weighted sum have bit lengths polynomial in $\lambda$. The remaining geometric computations have the same bound because they consist of fixed-degree arithmetic and fixed-dimensional linear systems.

Each sweep requires~$\softO(N)$~bit operations and storage, as required by Lemma~\ref{lem:kernel}, including product rounding and attention divisions by denominators bounded away from zero. For the static structure, persistence stores~$\softO(n)$~nodes, and a query visits~$\softO(1)$~nodes to combine the moments with polynomially many coefficients. The common-denominator bounds hold uniformly over queries satisfying the input bounds, proving Corollary~\ref{cor:online-d2}, and the query workspace can be reused throughout the stream.

\subsection{Arithmetic for Biased Block Products}
\label{app:biased-arithmetic}

We now establish the bit-complexity bounds suppressed by $\softO$ in Lemmas~\ref{lem:anchor} and~\ref{lem:direct}. All blocks belong to one original biased-product instance and share its fixed window~$L$~and tolerance~$\etablk$. All precision bounds refer to the parameter $\lambda$ of this original instance (Section~\ref{subsec:bit-model}).

\paragraph{Anchor products.}
The transformed coordinates in Lemma~\ref{lem:anchor} are input coordinates or their differences. Hence the exact inner-product maxima and the rational exponents defining the diagonal factors have bit length polynomial in~$\lambda$. The factors are bounded by~$\e^{\Oh(L)}$, so we can approximate them by Lemma~\ref{lem:scalarexp} with a number of bits polynomial in~$\lambda$~and the target precision. For a block with~$N_{IJ}=|I|+|J|$, approximate the factors and kernel product to a common dyadic error~$\delta$. Including final rounding, the total error is~$\Oh(\e^{2L_{\mathrm{anc}}}N_{IJ}(W+1)\delta)$. It suffices to use
$$\Oh_d\bigl(L+\log N_{IJ}+\log W+\log(1/\eta')\bigr)$$
additional bits to ensure block error at most $\eta'$. The scaled weights have magnitude at most~$\e^{\Oh(L)}W$~and polynomial bit length, so the lower-dimensional kernel call has only a polynomial increase in its logarithmic parameters. The remaining scalar computations require~$\softO(|I|+|J|)$~time.

\paragraph{Direct products and accumulation.}
For a direct block, approximate each exponential and round each weighted term to a single dyadic grid before summing. The common denominator keeps intermediate bit lengths polynomial, and a suitable grid mesh ensures the error bound of Lemma~\ref{lem:direct} in~$\softO(|I||J|)$~time. In both kinds of blocks, round the block outputs to one grid shared by the whole decomposition, with mesh proportional to~$\etablk$ as in Section~\ref{subsec:bit-model}. Each output row receives at most~$n$~block contributions, so the common denominator and all intermediate numerators have polynomial bit length in the original~$\lambda$.

\paragraph{Group indices and accumulation.}
The labels distinguishing initial, first-hit, and final groups contain only a constant number of source, cell, and level indices. Label comparisons and the computation of sampling parameters involve numbers of polynomial bit length in~$\lambda$. Their running time, together with accumulation and final rounding, is bounded by the linear terms in the block and geometric estimates.

\subsection{Precision under Composition and Recursion}
\label{app:induction-arithmetic}

Here $\lambda$ is the parameter of Section~\ref{subsec:bit-model} for the current kernel-product or biased-product instance.

\paragraph{Interpolation coefficients and sums.}
In Proposition~\ref{prop:KPstep}, round each exact coefficient $\mathcal L_{i\ell}$ to a single dyadic mesh $\delta_c=\Theta(\eta/(NW\nint))$, with a small absolute constant, and denote the result by $\wh{\mathcal L}_{i\ell}$. Then
$$|\wh{\mathcal L}_{i\ell}-\mathcal L_{i\ell}|\le\delta_c,\qquad
\sum_\ell|\wh{\mathcal L}_{i\ell}|\le9^{\nint}.$$
Before interpolation, round the original weights toward zero to a common dyadic grid of mesh $\Oh(\eta/N)$. The rounded weights have magnitude at most $W$ and change the exact kernel product by $\Oh(\eta)$, since kernel entries are at most one. We apply the interpolation scheme to these rounded weights, using the rounded coefficients in both the forward output combination and the transpose input weights. Because each entry of the biased matrix lies in $(0,1]$, coefficient rounding adds at most $NW\nint\delta_c$ to either product, and the bound on the sum of absolute coefficients ensures the estimates for errors from the biased products in the main proof. With sufficiently small grid constants and final rounding error, these additional errors sum to at most $\eta/2$.

\paragraph{One reduction step.}
The normalized directions and exact maxima have encoding lengths $\Oh_d(\lambda)$, since normalization uses a fixed number of rational operations per coordinate and each maximum equals an input inner product. Each interpolation coefficient is a product of polynomially many rationals of polynomial encoding length, and the number of nodes and groups is polynomial in $\lambda$. The coefficients therefore have polynomial encoding length, and their absolute sum requires only polynomially many additional precision bits in the biased calls. The transformed coordinates, weights, and error tolerances have the same polynomial bound. With weights and outputs on common dyadic grids, summing contributions adds logarithmically many bits in the number of terms, establishing the numerical cost in Proposition~\ref{prop:KPstep}. These estimates also apply to the diagonal factors of anchor blocks, whose exponents are bounded above by the score window and whose entries below the target tolerance can be rounded to zero (Appendix~\ref{app:biased-arithmetic}).

\paragraph{Recursion at a fixed dimension.}
The shared parameters of Section~\ref{subsec:bp-setup} and the exact geometry in Appendix~\ref{app:geometric-arithmetic} ensure polynomial operand lengths at every node of one biased-product instance.

\paragraph{Recursion on the dimension.}
Each lower-dimensional call decreases the dimension and may replace $\lambda$ by a fixed polynomial in $\lambda$. There are at most $d-2$ reductions before the planar base case, so their composition is a polynomial whose degree depends only on the fixed dimension. The numerical factors accumulated along this constant-length chain have the same property. Hence the hidden factors remain polynomial in the original encoding lengths and logarithmic parameters.

\paragraph{Computing attention.}
For Theorem~\ref{thm:main}, the tolerance $\eta=\eps/(8(B+1))$ adds $\Oh(\log(B+1))$ precision bits. The kernel outputs have polynomial encoding length, so each division and output rounding requires time polynomial in the original numerical parameters.

\subsection{Exact Geometric Computation}
\label{subsec:geom-model}
\label{app:geometric-arithmetic}
\label{app:ray-arithmetic}

For fixed $d\ge2$, we use the notation of Section~\ref{sec:geom} and the instance parameter $\lambda$ of Section~\ref{subsec:bit-model}.

\begin{lemma}
\label{lem:geom-bitmodel}
Suppose a geometric algorithm performs $T$ arithmetic or combinatorial operations and stores $S$ objects. If each geometric quantity is obtained from $\Oh_d(1)$ input points or constraints by a bounded-depth, bounded-degree algebraic computation, and each combinatorial operation acts on encodings of bit length polynomial in $\lambda$, then it has an exact implementation with running time $\softO(T)$ and space $\softO(S)$. The implementation has the same expected bounds and is deterministic or Las Vegas according to the original algorithm.
\end{lemma}

\begin{proof}
The fixed-dimensional linear-algebra computations involve rationals of $\Oh_d(\lambda)$ bits. More generally, bounded-degree algebraic numbers obtained at bounded depth have defining coefficients of $\Oh_d(\lambda)$ bits, and exact sign tests, root isolation, and comparisons require time polynomial in $\lambda$~\cite[Chapters~8 and~10]{basuPollackRoy2006}. Applying these bounds to each operation proves the claim, also after taking expectations.
\end{proof}

Hull vertices and vertices in the low-dimensional hierarchies are intersections of original constraints~\cite{preparataShamos1985,dobkinKirkpatrick1990,clarksonShor1989}, as are the vertices of the first triangulation in Appendix~\ref{subsec:geom-cuttings}. Its local arrangement vertices are intersections of original hyperplanes and facets of a first-stage simplex, using two fixed levels of linear algebra. The fixed-dimensional linear-programming predicates~\cite[Sections~2 and~3.2]{megiddo1984} and auxiliary algebraic decompositions also have a fixed number of elimination levels. Since recursive nodes reuse subsets of the original constraints and the same bounding simplex, the same depth bounds hold at every recursive node and the bit lengths remain $\Oh_d(\lambda)$.

Distinct rational values obtained by a bounded number of these operations differ by at least $2^{-\Oh_d(\lambda)}$. If each value changes by less than one quarter of this gap, every strict comparison has the same outcome and every perturbed minimizer or maximizer remains valid for the original values. This is a standard use of exact geometric computation~\cite{yap1997}.

\subsection{Uniform Perturbations and Caps}
\label{subsec:geom-perturb}

\begin{lemma}
\label{lem:perturb}
Let $\lambda\ge1$ be an integer and let $(\vct{y}_h)_{h\in\mathcal I}$ be points in $\Q^d$ of bit length at most $\lambda$, indexed by distinct positive integers at most $2^\lambda$. For a large enough positive integer $\cpt$ depending only on $d$, set $\dpert=2^{-\cpt\lambda}$ and
$$\wt{\vct{y}}_h=\vct{y}_h+\bigl(\dpert h,(\dpert h)^2,\ldots,(\dpert h)^d\bigr).$$
\begin{enumerate}[label=(\alph*)]
\item The perturbed points have $\Oh_d(\lambda)$ bits and satisfy $\|\wt{\vct{y}}_h-\vct{y}_h\|_\infty\le2^{-(\cpt-1)\lambda}$.
\item In every signed coordinate permutation $\pi$, the hyperplanes $z=f^\pi_{\wt{\vct{y}}_h}(\vct{u})$ are in general position.
\item Every intersection of $d$ of these hyperplanes has $\Oh_d(\lambda)$-bit coordinates.
\end{enumerate}
The constant $\cpt$ may be increased to satisfy any fixed exponential separation bound of the form $2^{-\Oh_d(\lambda)}$.
\end{lemma}

\begin{proof}
For at most $d$ normals, test a square minor containing the constant coordinate and distinct remaining coordinates; for $d+1$ hyperplanes, test their augmented determinant. As a polynomial in $\dpert$, the determinant has a highest-degree coefficient equal to a generalized Vandermonde determinant $(h^\ell)$ with distinct positive indices and distinct nonnegative exponents. Descartes' rule of signs implies that this coefficient is nonzero~\cite[Section~1.11(ii)]{dlmf}. Clearing denominators gives a nonzero integer polynomial of bounded degree with $\Oh_d(\lambda)$-bit coefficients. For sufficiently small $\dpert=2^{-\cpt\lambda}$, its lowest-degree nonzero term dominates the rest. The coefficient bounds are uniform, so one value of $\dpert$ works for every index subset and signed coordinate permutation. The perturbation estimate follows from the definition, and the intersection coordinates have $\Oh_d(\lambda)$ bits by Lemma~\ref{lem:geom-bitmodel}.
\end{proof}

We close the region above an envelope with explicit cap halfspaces.
\begin{definition}
\label{def:container}
Let $U_Y,U_u\ge1$ be integers and set
$$\acap=d^2(U_Y+2),\qquad
\Ztop=d(U_Y+2)(U_u+1),\qquad
\Zcap=\Ztop+1+(d-1)\acap U_u.$$
Let $\vct{e}_h$ be the $h$-th unit vector of $\R^{d-1}$ for $h<d$, and let $\vct{e}_d=-(1,\ldots,1)$. For a signed coordinate permutation $\pi$, set
$$\vct{c}^\pi_h=\pi^{-1}(-\acap\vct{e}_h,\Zcap)\qquad(h\in[d]).$$
For a nonempty finite set $\mathcal Y$ of points in $\R^d$ with $\|\vct{y}\|_\infty\le U_Y+1$ and cap perturbations satisfying $\|\wt{\vct{c}}_h-\vct{c}^\pi_h\|_\infty\le1/(8d(U_u+1))$, write $g_h(\vct{u})=(\pi\wt{\vct{c}}_h)\cdot(\vct{u},1)$ and define
$$P^\pi(\mathcal Y;\wt{\vct{c}})
=\{(\vct{u},z)\in\R^{d-1}\times\R:f^\pi_{\vct{y}}(\vct{u})\le z\ (\vct{y}\in\mathcal Y),\
z\le g_h(\vct{u})\ (h\in[d])\}.$$
The lower constraints are the source constraints, the upper ones are the cap constraints, and $\vct{x}_{\mathrm{top}}=(\vct{0},\Ztop)$.
\end{definition}

\begin{lemma}
\label{lem:container}
The capped polytope $P=P^\pi(\mathcal Y;\wt{\vct{c}})$ satisfies:
\begin{enumerate}[label=(\alph*)]
\item $(\vct{u},\Ztop)\in\intr P$ whenever $\|\vct{u}\|_\infty\le U_u$;
\item $P$ is bounded;
\item a vertical downward ray from an interior point $(\vct{u},z_0)$ exits at $z=\max_{\vct{y}\in\mathcal Y}f^\pi_{\vct{y}}(\vct{u})$, and every exit constraint with positive directional inner product is the constraint of a maximizing source.
\end{enumerate}
\end{lemma}

\begin{proof}
On the target box, the source functions lie below $\Ztop$, while every perturbed cap lies above it with a fixed positive margin. For boundedness, note that
$$\max_h\vct{e}_h\cdot\vct{u}\ge\|\vct{u}\|_\infty/(d-1).$$
For the chosen $\acap$, the minimum cap lies below every source function whenever $\|\vct u\|_\infty$ is sufficiently large, uniformly over the allowed cap perturbations. Their intersection is therefore bounded in both $\vct u$ and $z$. A downward ray exits through a source constraint at the largest source value.
\end{proof}

\subsection{Support Queries and Envelope Construction}
\label{app:geom-support}
\label{app:geom-envelope}

\begin{proof}[Proof of Lemma~\ref{lem:support}]
For each normalizing permutation, we form the cap halfspaces of Definition~\ref{def:container} with $U_Y=2^\lambda$ and $U_u=1$, indexing the caps by $1,\ldots,d$ and the sources by distinct subsequent integers. Applying Lemma~\ref{lem:perturb} with the same source indexing and a uniform $\Oh_d(\lambda)$ bit bound puts the combined constraints in general position. The caps together with any one source define the bounding simplex for Proposition~\ref{prop:rayshoot}, whose structure we build using the interior point $\vct{x}_{\mathrm{top}}$ from Lemma~\ref{lem:container}(a).

For a nonzero query $\vct q$, use its permutation $\pi_{\vct q}$ from Definition~\ref{def:normalization} and shoot vertically downward from $(\vct u_{\vct q},\Ztop)$. Lemma~\ref{lem:container} gives a maximizer of the perturbed scores. Distinct original scores have gaps at least $2^{-\Oh_d(\lambda)}$, and the perturbation changes each score by less than one quarter of this gap uniformly over queries satisfying the input bounds. The returned index therefore maximizes the original scores as well, and we evaluate its original score exactly. Since there are a fixed number of coordinate groups, Proposition~\ref{prop:rayshoot} gives preprocessing time $\softO(|\Sigma|^k)$ and query time $\softO(1)$, deterministically for $d\le3$.
\end{proof}

\begin{proof}[Envelope construction and proof of Lemma~\ref{lem:envfacts}]
Use the identity permutation and $U_Y=U_u=U$. Index the caps by $h\in[d]$ and the sources by $d+s$, and apply Lemma~\ref{lem:perturb} once to all these points with a uniform $\Oh_d(\lambda)$ bit bound. Define $g_h(\vct{u})=\wt{\vct{c}}_h\cdot(\vct{u},1)$ and set
$$\cP(\Sigma')=\bigl\{(\vct{u},z):\wt f_s(\vct{u})\le z\ (s\in\Sigma'),\quad z\le g_h(\vct{u})\ (h\in[d])\bigr\}.$$
The bounding halfspaces $z\le g_h(\vct{u})$ and the height $\Ztop$ are shared by every sample. By Lemma~\ref{lem:container}, each $(\vct{u},\Ztop)$ lies in the interior when $\|\vct{u}\|_\infty\le U$, so the lower boundary over this box is exactly the graph of $E_{\Sigma'}$.

The cap coefficients, indices, and source coordinates have $\Oh_d(\lambda)$ bits. Set the perturbation so that its effect on every source value in $[-U,U]^{d-1}$ is at most $1/8$ and is less than one quarter of the gaps between distinct original values at every rational target satisfying the input bounds. These gaps are at least $2^{-\Oh_d(\lambda)}$, so one perturbation constant depending only on the dimension suffices and also satisfies the cap perturbation bound, proving parts~(a) and~(b). For part~(c), every subset remains in general position, its envelope polytope is bounded by Lemma~\ref{lem:container}, and its vertex bit lengths follow from Lemma~\ref{lem:geom-bitmodel}. The interior point $(\vct u,\Ztop)$ needed for ray shooting depends only on the target and the fixed height, so it has $\Oh_d(\lambda)$ bits and is computable in $\softO(1)$ time.
\end{proof}

\subsection{Interior Shifts for Cell Location}
\label{app:ray-shifts}

For a query satisfying the bounds of Lemma~\ref{lem:envelope}, the descent in Appendix~\ref{app:geom-cells} follows a ray from the least vertex $\vct{v}_F$ of a face $F$ through a point $\vct{x}\ne\vct{v}_F$ in that face. Write $\vct{w}_{\mathrm{ray}}=\vct{x}-\vct{v}_F$. For a shift $\delta_{\mathrm{sh}}\in(0,1)$ specified below, we move the ray origin into the interior by setting $\vct{v}_{\mathrm{sh}}=(1-\delta_{\mathrm{sh}})\vct{v}_F+\delta_{\mathrm{sh}}\vct{x}_{\mathrm{top}}$. For every constraint with $\vct{a}_H\cdot\vct{w}_{\mathrm{ray}}>0$, its exit parameter changes from $\tau_H=(b_H-\vct{a}_H\cdot\vct{v}_F)/(\vct{a}_H\cdot\vct{w}_{\mathrm{ray}})$ to
$$\tau_H^{\mathrm{sh}}
=\tau_H-\delta_{\mathrm{sh}}
\frac{\vct{a}_H\cdot(\vct{x}_{\mathrm{top}}-\vct{v}_F)}
{\vct{a}_H\cdot\vct{w}_{\mathrm{ray}}}.$$
Every descent has at most $d-1$ steps, each using one original vertex and constraint, so its coordinates and exit parameters have $\Oh_d(\lambda)$ bits. Their nonzero gaps and denominators are at least $2^{-\Oh_d(\lambda)}$. A dyadic $\delta_{\mathrm{sh}}\in(0,1)$ with $\Oh_d(\lambda)$ bits then preserves every strict ordering among the exit parameters, so each shifted minimizer also minimizes the original parameters. Since the bounds are uniform over samples and targets, a single shift suffices for all location queries.

\subsection{Integer Encodings of Halfspace Ranges}
\label{app:halfspace-encoding}

\begin{lemma}
\label{lem:halfspace-encoding}
Fix $d\ge2$ and an integer $D\ge1$. Let $P'=\{\vct{z}_i/D:i\in[n]\}\subseteq[0,1]^d$ with $\vct{z}_i\in\{0,\ldots,D\}^d$. Every halfspace range on $P'$ is realized by a halfspace with a nonzero integer normal and coefficients of $\Oh_d(\log(D+2))$ bits.
\end{lemma}

\begin{proof}
Without loss of generality, the selected range is nonempty and proper. On the integer points $\vct z_i$, maximize the nonnegative separation margin subject to $\|(\vct a,c)\|_\infty\le1$; its positive maximum occurs at a vertex determined by $d+2$ tight constraints with integer coefficients of magnitude $\Oh_d(D+1)$. Cramer's rule followed by clearing denominators yields integer $\vct a,c$ of magnitude $(D+1)^{\Oh_d(1)}$ (Lemma~\ref{lem:geom-bitmodel}), and rescaling the normal by $D$ realizes the original range with the required encoding length.
\end{proof}

\section{Geometric Constructions and Sampling Proofs}
\label{app:geom-proofs}

This appendix proves the geometric data structure bounds and sampling estimates stated in Section~\ref{sec:geom}, together with the exact batch-maxima bound of Proposition~\ref{prop:batchsupport}; Appendix~\ref{app:arithmetic} details their exact rational implementations. Throughout, we follow the notation of Section~\ref{sec:geom} with fixed $d\ge2$ and $k=\lfloor d/2\rfloor$.

\subsection{Classical Polytope Facts}\label{subsec:geom-classical}

We use the active sets of faces to label the canonical cells. Relative to defining halfspaces $\cH$, the active set of a face $F$ is $\mathcal S_F=\{H\in\cH:F\subseteq\partial H\}$, with the standard properties below~\cite[Propositions~2.2, 2.3, and~2.16, and Theorem~2.7]{ziegler1995}.

\begin{lemma}
\label{lem:simple}
Let~$\cH$~be a finite set of closed halfspaces in~$\R^d$~in general position whose intersection~$P$~is bounded and has nonempty interior. For~$\mathcal T\subseteq\cH$~define~$\mathcal A_{\mathcal T}=\bigcap_{H\in\mathcal T}\partial H$, with~$\mathcal A_\varnothing=\R^d$, and~$F_{\mathcal T}=P\cap\mathcal A_{\mathcal T}$.
\begin{enumerate}[label=(\alph*)]
\item Every vertex of~$P$~lies on exactly~$d$~of the hyperplanes~$\partial H$,~$H\in\cH$.
\item If~$F_{\mathcal T}\ne\varnothing$, then~$F_{\mathcal T}$~contains a vertex of~$P$,~$|\mathcal T|\le d$,~$F_{\mathcal T}$~is a face of~$P$~of dimension~$d-|\mathcal T|$,~$\aff F_{\mathcal T}=\mathcal A_{\mathcal T}$, and~$\mathcal T=\{H\in\cH:F_{\mathcal T}\subseteq\partial H\}$. Every nonempty face~$F$~of~$P$~equals~$F_{\mathcal T}$~for exactly one set~$\mathcal T$, its active set~$\mathcal S_F$. In particular~$P\cap\aff F=F$.
\item The number of chains~$F^{(0)}\supset F^{(1)}\supset\cdots\supset F^{(d-1)}$~of faces of~$P$~with~$\dim F^{(i)}=d-1-i$~is at most~$d!\,|\vertset(P)|$. Moreover,~$|\vertset(P)|=\Oh_d(|\cH|^k)$~\cite{mcmullen1970,seidel2017hull}.
\end{enumerate}
\end{lemma}

The face identities follow from simplicity and independence of the active normals. Ordering the $d$ facets at each vertex gives the flag count in part~(c), and McMullen's Upper Bound Theorem gives the vertex bound by polarity~\cite{mcmullen1970,seidel2017hull}.

\begin{proposition}
\label{prop:facelattice}
Let~$d\ge3$. Let~$\cH$~be a set of~$|\cH|\ge d+1$~closed halfspaces in~$\R^d$~in general position whose intersection~$P$~is bounded and has nonempty interior, and let~$d+1$~members of~$\cH$~be given whose intersection is bounded and has nonempty interior. A Las Vegas algorithm computes all vertices of~$P$, each with the~$d$~members of~$\cH$~whose bounding hyperplanes contain it, in expected~$\Oh(|\cH|^{k}+|\cH|\log(|\cH|+2))$~arithmetic operations~\cite[Theorems~4.7 and~4.8]{clarksonShor1989}.
\end{proposition}

Starting from the given bounding simplex, insert the remaining halfspaces in uniformly random order. Each feature is defined by $\Oh_d(1)$ variable constraints, and an intersection with $q$ such constraints has complexity $\Oh_d((q+1)^k)$. Thus the analysis of~\cite[Theorem~3.9]{clarksonShor1989} applies with the fixed constraints present in every sample. For rational inputs, Lemma~\ref{lem:geom-bitmodel} implies expected running time $\softO(|\cH|^k)$.

\subsection{Constructing Cuttings with a Fixed Bounding Simplex}
\label{subsec:geom-cuttings}

\begin{proposition}[Ray shooting]
\label{prop:rayshoot}
Let $\cH$ be a finite collection of closed halfspaces in $\R^d$ with rational coefficients, in general position, and let $P=\bigcap_{H\in\cH}H$. Suppose that an interior point $\vct{x}_{\mathrm{in}}\in\intr P$ and a subset $\mathcal B\subseteq\cH$ of $d+1$ members with bounded intersection are given. Let $\lambda\ge\log_2(|\cH|+2)$ bound the bit lengths of the constraint coefficients and of $\vct{x}_{\mathrm{in}}$. A Las Vegas algorithm builds a data structure in expected time $\softO(|\cH|^k)$; for $d\le3$, these bounds hold deterministically. Given a ray $\{\vct{x}+\tau_{\mathrm{ray}}\vct{w}_{\mathrm{ray}}:\tau_{\mathrm{ray}}\ge0\}$ with $\vct{x}\in\intr P$, $\vct{w}_{\mathrm{ray}}\ne\vct{0}$, and coordinates of bit length at most $\lambda$, the structure returns in time $\softO(1)$ a constraint $H\in\cH$ such that $\vct{a}_H\cdot\vct{w}_{\mathrm{ray}}>0$ and $\partial H$ contains the ray's exit point from $P$.
\end{proposition}

We give the details of the classical constructions of~\cite{clarksonShor1989,matousekSchwarzkopf1993} to establish their bounds in our bit-complexity model (Section~\ref{subsec:bit-model}).

The ray-shooting algorithm uses the same bounding simplex for every sample and recursive node. Let~$\cH$~be a finite collection of closed halfspaces in $\R^d$ in general position with a common interior point~$\vct{x}_{\mathrm{in}}$, and let~$\mathcal B\subseteq\cH$~consist of~$d+1$~halfspaces with bounded intersection~$B$. Write~$\mathcal A=\cH\setminus\mathcal B$. For~$R\subseteq\mathcal A$, define
$$Q_R=B\cap\bigcap_{H\in R}H.$$
Because the fixed constraints are present in every sample,~$Q_R$~is always a full-dimensional polytope containing~$\vct{x}_{\mathrm{in}}$~in its interior. We triangulate by coning the lexicographically least vertex of each positive-dimensional face to the triangulations of the facets that omit it. A resulting full-dimensional simplex has vertices~$\vct{v}_0,\ldots,\vct{v}_d$~along a chain~$F_0=Q_R\supset F_1\supset\cdots\supset F_d$, where~$\dim F_i=d-i$,~$\vct{v}_i$~is the least vertex of~$F_i$, and~$\vct{v}_i\notin F_{i+1}$ for $0\le i<d$. Denote the triangulation by~$\mathcal T_{\mathrm{cut}}(R)$, describing each simplex by the ordered active sets of its vertices.

\begin{lemma}
\label{lem:bounded-configurations}
Let $R_0\subseteq\mathcal A$, let $\sigma\in\mathcal T_{\mathrm{cut}}(R_0)$, and let $\vct{v}_0,\ldots,\vct{v}_d$ be its vertices in pulling-chain order. Define
$$D_\sigma=\mathcal A\cap\bigcup_{i=0}^d\{H\in\cH:\vct{v}_i\in\partial H\},\qquad K_\sigma=\{H\in\mathcal A:\sigma\not\subseteq H\}.$$
Then~$|D_\sigma|\le d(d+1)$, and for every~$R\subseteq\mathcal A$,
$$\sigma\in\mathcal T_{\mathrm{cut}}(R)\quad\Longleftrightarrow\quad D_\sigma\subseteq R\ \text{ and }\ R\cap K_\sigma=\varnothing.$$
Moreover,~$|\mathcal T_{\mathrm{cut}}(R)|=\Oh_d((|R|+1)^k)$, where~$k=\lfloor d/2\rfloor$.
\end{lemma}

\begin{proof}
By Lemma~\ref{lem:simple}, each~$\vct{v}_i$~lies on exactly~$d$~hyperplanes of~$\mathcal B\cup R_0$. General position of~$\cH$~implies that these are also all the hyperplanes of~$\cH$~through~$\vct{v}_i$. Hence~$|D_\sigma|\le d(d+1)$, and every sample in which~$\sigma$~occurs contains~$D_\sigma$~and avoids~$K_\sigma$.

For the converse, suppose that~$D_\sigma\subseteq R$~and~$R\cap K_\sigma=\varnothing$. Then~$\sigma\subseteq Q_R$, and each~$\vct{v}_i$~has the same complete active set. Let~$\mathcal S_i$~be the active set of the original face~$F_i$, and set
$$F'_i=Q_R\cap\bigcap_{H\in\mathcal S_i}\partial H.$$
The constraints of~$\mathcal S_i$~belong to~$\mathcal B\cup D_\sigma$. The affinely independent points~$\vct{v}_i,\ldots,\vct{v}_d$~lie in~$F'_i$~and span~$\aff F_i$, so~$\dim F'_i=d-i$. Consequently the~$F'_i$~form the same sequence of facet incidences, with~$\vct{v}_i\notin F'_{i+1}$ for $0\le i<d$.

We show that~$\vct{v}_i$~is the least point of~$F'_i$. A bounded polytope is the convex hull of its vertices, so~$\vct{v}_i$~is the least point of the original face~$F_i$. If~$\vct{y}\in F'_i$~were lexicographically smaller than~$\vct{v}_i$, the point~$\vct{v}_i+t(\vct{y}-\vct{v}_i)$~would satisfy all constraints active at~$\vct{v}_i$~for~$t\in[0,1]$. Every other constraint of~$\mathcal B\cup R_0$~is strict at~$\vct{v}_i$~and holds for all small enough~$t>0$. The same point also satisfies the equalities of~$\mathcal S_i$. This would be a point of~$F_i$~smaller than~$\vct{v}_i$, a contradiction. Thus~$\vct{v}_i$~is the bottom vertex of~$F'_i$, and the original pulling chain occurs in~$Q_R$.

Finally, the full pulling chains are a subset of the complete face chains counted in Lemma~\ref{lem:simple}(c), with~$Q_R$~prepended. By that lemma, their number is at most~$d!\,|\vertset(Q_R)|=\Oh_d((|R|+d+1)^k)=\Oh_d((|R|+1)^k)$.
\end{proof}

The set~$K_\sigma$~contains all halfspaces that prevent~$\sigma$~from occurring in a sample. To build a cutting, we also count the hyperplanes crossing its interior,
$$c_\sigma=|\{H\in\mathcal A:\partial H\cap\intr\sigma\ne\varnothing\}|.$$
A hyperplane crossing the interior of~$\sigma$~has points of~$\sigma$~on both sides, so~$c_\sigma\le|K_\sigma|$. The time for refining simplices with large~$c_\sigma$ depends on a moment estimate obtained by comparing a sample with a smaller uniform sample.

\begin{lemma}
\label{lem:bounded-moments}
Let~$n=|\mathcal A|\ge1$, and let~$R$~be a uniformly random~$q$-element subset of~$\mathcal A$, where~$q$ is an integer with $1\le q\le n$. For every fixed real~$\nu\ge1$,
$$\E\sum_{\sigma\in\mathcal T_{\mathrm{cut}}(R)}\max\{1,q|K_\sigma|/n\}^{\nu}=\Oh_{d,\nu}(q^k).$$
The estimate also applies to interior crossings, with~$c_\sigma$~in place of~$|K_\sigma|$.
\end{lemma}

\begin{proof}
Set~$b_{\mathrm{def}}=d(d+1)$. For~$q<4b_{\mathrm{def}}$, the number of simplices and every summand are bounded by constants depending only on~$d,\nu$, so the claim follows.

For~$q\ge4b_{\mathrm{def}}$, let~$q_-=\lfloor q/2\rfloor$. For a fixed configuration~$\sigma$, let~$a=|D_\sigma|$~and~$b=|K_\sigma|$, and write~$p_j(\sigma)$~for its probability of occurring in a uniformly random~$j$-element sample. Configurations with~$p_q(\sigma)=0$~contribute zero to the desired expectation.

For the remaining configurations, drawing a uniform~$q$-element sample and then a uniform~$q_-$-element subset is equivalent to drawing the smaller sample first and extending it uniformly. Lemma~\ref{lem:bounded-configurations} implies
$$p_q(\sigma)\frac{(q_-)_a}{(q)_a}=p_{q_-}(\sigma)\frac{\binom{n-q_- -b}{q-q_-}}{\binom{n-q_-}{q-q_-}},$$
where~$(j)_a=j(j-1)\cdots(j-a+1)$~and an impossible selection has probability zero. Since~$a\le b_{\mathrm{def}}$~and~$q_-\ge2b_{\mathrm{def}}$, we have~$(q)_a/(q_-)_a\le4^{b_{\mathrm{def}}}$. The probability that the extension avoids~$K_\sigma$~is at most~$\e^{-(q-q_-)b/(n-q_-)}\le \e^{-qb/(2n)}$. It follows that
$$p_q(\sigma)\le4^{b_{\mathrm{def}}}\e^{-qb/(2n)}p_{q_-}(\sigma).$$
For fixed~$\nu$, the function~$\max\{1,t\}^\nu \e^{-t/2}$~is bounded on~$[0,\infty)$. Multiplying this inequality by~$\max\{1,qb/n\}^\nu$~and summing over the finitely many configurations yields
$$\E\sum_{\sigma\in\mathcal T_{\mathrm{cut}}(R)}\max\{1,q|K_\sigma|/n\}^{\nu}\le \Oh_{d,\nu}(1)\sum_\sigma p_{q_-}(\sigma)=\Oh_{d,\nu}(q^k),$$
by Lemma~\ref{lem:bounded-configurations}. The last assertion follows from~$c_\sigma\le|K_\sigma|$. This is the double-sampling argument underlying the configuration bounds of~\cite{clarksonShor1989,chazelleFriedman1990}, with~$\mathcal B$~fixed in every sample.
\end{proof}

\begin{lemma}
\label{lem:verified-cuttings}
Under the preceding assumptions, let~$n=|\mathcal A|\ge1$ and let $q$ be an integer with $1\le q\le n$. A Las Vegas algorithm constructs~$\Oh_d(q^k)$~closed full-dimensional simplices whose union contains~$Q_{\mathcal A}$, each meeting~$Q_{\mathcal A}$, such that the interior of each simplex is crossed by at most~$n/q$~hyperplanes of~$\mathcal A$. It also computes their closed conflict lists
$$\mathcal A_\tau=\{H\in\mathcal A:\partial H\cap\tau\ne\varnothing\},\qquad |\mathcal A_\tau|\le n/q+d(d+1).$$
The algorithm uses~$n q^{\Oh_d(1)}$~arithmetic and combinatorial operations in expectation. Every vertex is obtained from~$\Oh_d(1)$~original constraints by linear algebra of depth at most two.
\end{lemma}

\begin{proof}
We draw a uniform~$q$-element sample~$R\subseteq\mathcal A$, enumerate~$Q_R$ and its pulling triangulation, and count the interior crossings~$c_\sigma$ for each~$\sigma\in\mathcal T_{\mathrm{cut}}(R)$ in~$n q^{\Oh_d(1)}$~time. Set~$\kappa_\sigma=qc_\sigma/n$ and refine each simplex with~$\kappa_\sigma>1$ as follows.

\emph{Local refinement.} Let~$\mathcal C_\sigma$ be the set of~$c_\sigma$~hyperplanes crossing~$\intr\sigma$, which contains every hyperplane that can cross the interior of a subsimplex of~$\sigma$, and set~$u_\sigma=\lceil\kappa_\sigma\rceil$. We draw a uniform sample of size
$$h_\sigma=\min\{c_\sigma,C_d u_\sigma\lceil\log_2(u_\sigma+1)\rceil\}$$
from~$\mathcal C_\sigma$, with a positive integer~$C_d$ depending only on~$d$ and satisfying the sampling estimate below. Integer division and bit-length comparisons compute this size, which is~$\Oh_d(\kappa_\sigma\log(\kappa_\sigma+1))$. We form the sample's arrangement inside~$\sigma$, triangulate the closures of its full-dimensional cells by pulling, and verify that every resulting simplex has at most~$n/q$~interior crossings. If the test fails, we repeat the local sampling.

The ranges on~$\mathcal C_\sigma$~consist of the hyperplanes crossing the interior of a full-dimensional simplex. For a simplex with vertices~$\vct{w}_0,\ldots,\vct{w}_d$, the crossing condition is
$$\min_i(\vct{a}_H\cdot \vct{w}_i-b_H)<0<\max_i(\vct{a}_H\cdot \vct{w}_i-b_H).$$
On any~$h$~hyperplanes, these ranges have~$h^{\Oh_d(1)}$~distinct traces, determined by the signs of~$(d+1)h$~linear forms in the~$d(d+1)$~vertex coordinates. Thus the range space has VC dimension depending only on~$d$.

If~$h_\sigma=c_\sigma$, the full sample satisfies the crossing bound. Otherwise the random~$\eps$-net theorem~\cite{hausslerWelzl1987}, with~$\eps=1/\kappa_\sigma$, shows that~$h_\sigma$~independent draws intersect every range of size greater than~$c_\sigma/\kappa_\sigma=n/q$~with probability at least~$1/2$.

Deleting repetitions from the independent draws and filling uniformly from the unused elements gives a uniform~$h_\sigma$-element subset by symmetry. It intersects every range met by the independent draws and is therefore an~$\eps$-net with probability at least~$1/2$. Since the interiors of the triangulated arrangement cells have no sample crossings, every such net satisfies the crossing bound, so local refinement requires at most two attempts in expectation.

If the local sample has~$h$~hyperplanes, the arrangement inside~$\sigma$~has~$\Oh_d((h+1)^d)$~vertices. At each vertex, at most~$d$~original hyperplanes and at most~$d+1$~facet hyperplanes of~$\sigma$~are incident. There are thus~$\Oh_d(1)$~complete face chains through each vertex, including incidences between cells, and~$\Oh_d((h+1)^d)$~simplices in the pulling triangulations. This count remains valid in the presence of dependencies among the original hyperplanes and the facet hyperplanes of~$\sigma$. Since~$\log^d(\kappa+1)=\Oh_d(\kappa)$~for~$\kappa\ge1$, every accepted refinement has~$\Oh_d(\kappa_\sigma^{d+1})$~simplices.

\emph{Acceptance and cost.} After all refinements, discard every simplex~$\tau$~with~$\tau\cap Q_{\mathcal A}=\varnothing$. This is a fixed-dimensional linear-programming feasibility test on~$n+\Oh_d(1)$~constraints using~$\Oh_d(n)$~operations~\cite{megiddo1984}. Appendix~\ref{app:ray-arithmetic} specifies the predicates of bounded algebraic depth for this feasibility test. The remaining simplices cover~$Q_{\mathcal A}$, because the refinements cover~$Q_R\supseteq Q_{\mathcal A}$. Accept if at most~$C'_d q^k$~simplices remain, and otherwise draw a new outer sample~$R$, where the constant $C'_d$ depends only on $d$ and satisfies the estimate below.

Conditional on~$R$, the number of simplices before discarding is at most
$$\Oh_d(1)\sum_{\sigma\in\mathcal T_{\mathrm{cut}}(R)}\max\{1,\kappa_\sigma\}^{d+1}.$$
Its expectation is~$\Oh_d(q^k)$ by Lemma~\ref{lem:bounded-moments}, and Markov's inequality gives outer acceptance probability at least~$1/2$ for a large enough~$C'_d$. Each attempt, including the local retries, uses~$n q^{\Oh_d(1)}$~operations in expectation, since~$\kappa_\sigma\le q$~and all candidate lists are checked by enumeration. Repetition therefore increases the expected cost by at most a factor of two.

Each remaining simplex has at most~$n/q$~interior crossings. A hyperplane that meets a closed simplex while avoiding its interior supports a face containing a vertex. By general position, at most~$d$~original hyperplanes pass through any vertex, so at most~$d(d+1)$~additional hyperplanes belong to the closed conflict list.

\emph{Coordinates.} Appendix~\ref{app:ray-arithmetic} constructs every cutting vertex from $\Oh_d(1)$ original constraints through two levels of linear algebra, with polynomial exact arithmetic cost.
\end{proof}

\begin{proof}[Proof of Proposition~\ref{prop:rayshoot}]
For~$d\le3$, the deterministic algorithms described in Appendix~\ref{subsec:geom-model} achieve the required real-RAM complexity~\cite[Lemma~2.1 and Corollary~2.2]{chan1996output}. For~$d\ge4$, we use the static recursion of Matousek and Schwarzkopf~\cite[Lemma~7]{matousekSchwarzkopf1993}, with the cuttings constructed above. The implementation uses the same bounding simplex at every node and closed conflict lists, with bounded coordinate complexity.

\emph{Recursive structure.} Every node keeps~$\mathcal B$~and stores a variable subset~$\mathcal A'\subseteq\cH\setminus\mathcal B$, representing~$Q_{\mathcal A'}=B\cap\bigcap_{H\in\mathcal A'}H$. For~$n'=|\mathcal A'|$~above a constant threshold, set~$q=\lfloor(n')^\alpha\rfloor$, where the positive rational~$\alpha$ depends only on~$d$ and satisfies the preprocessing estimate below, and construct the cutting of Lemma~\ref{lem:verified-cuttings}. The child for~$\tau$~stores its closed conflict list~$\mathcal A'_\tau$, together with~$\mathcal B$. As in~\cite[Lemmas~7 and~11]{matousekSchwarzkopf1993}, a query descends through a simplex containing the last point of the ray in the union of the cutting simplices. This union is compact and contains~$Q_{\mathcal A'}$. A leaf scans its constraints and returns one minimizing the exit parameter
$$\frac{b_H-\vct{a}_H\cdot \vct{x}}{\vct{a}_H\cdot \vct{w}_{\mathrm{ray}}}\qquad(\vct{a}_H\cdot \vct{w}_{\mathrm{ray}}>0).$$

\emph{Exact auxiliary queries.} To select the last simplex met by the ray, we use the sign-vector method of~\cite[Lemmas~11--12]{matousekSchwarzkopf1993}. For each simplex facet~$H$, define
$$U_H=b_H-\vct{a}_H\cdot \vct{x},\qquad V_H=\vct{a}_H\cdot \vct{w}_{\mathrm{ray}}.$$
The signs of~$U_H,V_H$~and~$U_HV_{H'}-U_{H'}V_H$~determine the ray's intersection interval with each simplex and the order of all finite endpoints, including the cases~$V_H=0$~and equal exit parameters. They come from~$q^{\Oh_d(1)}$~polynomials of degree at most two in the~$2d$~ray parameters. Construct a sign-invariant cylindrical algebraic decomposition, including cells of every dimension, and associate each cell representing a ray that meets the union with a simplex of largest exit parameter; break ties by simplex index~\cite{basuPollackRoy2006}.

This decomposition requires~$q^{\Oh_d(1)}$~operations because the dimension is fixed. At query time, binary search among the ordered roots at each of the~$2d$~levels locates the ray in~$\Oh_d(\log(q+2))$~exact geometric operations. Thus we obtain an exact implementation of the point-location structure of~\cite{cegs1991} used in the cited recursion.

\emph{Closed conflicts and ray exits.} Let~$\vct{x}^*$~be the exit point from the current node polytope. The selected simplex~$\tau$~contains a point~$\vct{x}'_{\mathrm{ray}}$~of the ray at or beyond~$\vct{x}^*$, and Lemma~\ref{lem:verified-cuttings} ensures the existence of a point~$\vct{x}'_P\in\tau\cap Q_{\mathcal A'}$.

The query origin remains interior to every node polytope, since their constraints are drawn from~$\cH$. For any~$H\in\mathcal A'$~tight at~$\vct{x}^*$, we have~$\vct{a}_H\cdot \vct{w}_{\mathrm{ray}}>0$~and
$$\vct{a}_H\cdot \vct{x}'_{\mathrm{ray}}\ge b_H\ge \vct{a}_H\cdot \vct{x}'_P.$$
The segment joining these two points lies in~$\tau$~and intersects~$\partial H$, so~$H\in\mathcal A'_\tau$. Together with the fixed constraints~$\mathcal B$, the child contains every constraint tight at~$\vct{x}^*$. Its polytope contains the parent's and has the same ray exit, so induction along the query path proves that the leaf returns a valid original constraint.

\emph{Complexity.} An accepted node has at most~$C'_d q^k$~children, each with at most~$n'/q+c_d$~variable constraints, where~$c_d=d(d+1)$. For small enough~$\alpha$, its expected preprocessing cost is~$\Oh_d((n')^k)$, since~$k\ge2$. Let $S(t)$ be the worst expected preprocessing time of any node with at most $t$ variable constraints. Independent random choices for the children give
$$S(n')\le C'_d q^k S(n'/q+c_d)+\Oh_d((n')^k).$$
Set the leaf threshold so that~$n'>2c_d$~and~$q\ge2$~at every internal node. Then
$$(n'/q+c_d)-2c_d\le(n'-2c_d)/q.$$
For these shifted node sizes, the storage analysis of~\cite[Lemma~7]{matousekSchwarzkopf1993} implies~$S(n')=\softO((n'+1)^k)$, and the single-path query analysis implies~$\softO(1)$~operations per query. The resampling procedures terminate with probability one, and the node sizes decrease above the leaf threshold.

Lemma~\ref{lem:geom-bitmodel}, applied to the representations in Appendix~\ref{app:ray-arithmetic}, converts these operation counts to bit complexity. Preprocessing is deterministic for $d\le3$ and Las Vegas in higher dimensions.
\end{proof}

\subsection{Strict Halfspace Reporting}\label{app:reporting}

\begin{proof}[Proof of Lemma~\ref{lem:reporttree}]
Partition $\Sigma$, in its fixed enumeration order, into $\lceil|\Sigma|/g_{\mathrm{rep}}\rceil$ groups of size at most $g_{\mathrm{rep}}$. Within each group, build the binary tree of~\cite[Observation~2.3]{chan1996lpq} and use Lemma~\ref{lem:support} to test its nodes. At a node containing $\Sigma'$, the query continues precisely when $\max_{s\in\Sigma'}\vct{q}\cdot\vct{y}_s>\theta$. A leaf stores the original source index, so repeated points are reported separately.

For a group of size $g$, the tree requires $\softO(g^k)$ expected time to preprocess, for a total of $\softO(|\Sigma|g_{\mathrm{rep}}^{k-1})$ over all groups. A query visits the group roots and only $\Oh(\log(g_{\mathrm{rep}}+1))$ additional nodes per reported index, each requiring one support query. Its total time is $\softO(|\Sigma|/g_{\mathrm{rep}}+1+|Z|)$. The algorithm is deterministic for $d\le3$.\qedhere
\end{proof}

\subsection{Canonical Cells and Point Location}
\label{app:geom-cells}

Work in the setup of Definition~\ref{def:envelope}, with~$d\ge3$.

\begin{definition}
\label{def:pulling-cells}
Let~$\varnothing\ne\Sigma'\subseteq\Sigma_0$~and~$\cP=\cP(\Sigma')$. Order the vertices of~$\cP$~lexicographically by their coordinates. For a nonempty face~$F$~of~$\cP$, let~$\vct{v}_F$~be its least vertex; it exists by Lemma~\ref{lem:simple}(b). A pulling chain is a sequence of faces~$F^{(0)}\supset F^{(1)}\supset\cdots\supset F^{(d-1)}$~such that~$F^{(0)}$~is the nonempty facet of a source~$s\in\Sigma'$~and, for~$1\le i\le d-1$,~$F^{(i)}$~is a facet of~$F^{(i-1)}$~with~$\vct{v}_{F^{(i-1)}}\notin F^{(i)}$. Its projected simplex is
$$\Delta=\conv\{\bar{\vct{v}}_{F^{(0)}},\bar{\vct{v}}_{F^{(1)}},\ldots,\bar{\vct{v}}_{F^{(d-1)}}\},$$
with owner~$s_\Delta=s$~and label
$$\bigl(s_\Delta,\mathcal S_{\vct{v}_{F^{(0)}}},\ldots,\mathcal S_{\vct{v}_{F^{(d-1)}}}\bigr),$$
where each active set is the sorted tuple of its constraint indices. The canonical cells are listed in the lexicographic order of these labels.
\end{definition}

\begin{proof}[Proof of Lemma~\ref{lem:cells}]
Write~$\cP=\cP(\Sigma')$. The recursive vertex triangulation of~\cite[Section~3]{clarkson1988}, using the least vertex on each face, triangulates every source facet into the simplices of Definition~\ref{def:pulling-cells}. For a cell $\Delta$, projection from the graph $z=\wt f_{s_\Delta}(\vct{u})$ to $\vct{u}$ is an affine bijection. Hence each simplex projects to a closed $(d-1)$-simplex with rational vertices, and its owner realizes the sampled envelope on the entire cell.

For every $\vct{u}\in[-U,U]^{d-1}$, the point $(\vct{u},E_{\Sigma'}(\vct{u}))$ lies on the lower boundary of $\cP$ by Definition~\ref{def:envelope}. Hence it belongs to a simplex on a source facet, whose projection is a canonical cell containing the target, proving the coverage in part~(a).

The suffix vertices of a pulling chain have the same affine hull as the corresponding face; the identity $\cP\cap\aff F=F$ in Lemma~\ref{lem:simple} then determines the chain from its vertex active sets. Thus distinct chains have distinct labels, and the labeled list depends only on the sample.

For part~(b), the affine function $\wt f_s-\wt f_{s_\Delta}$ is positive somewhere on a simplex exactly when it is positive at a vertex. The owner is maximal among the sampled sources on its entire facet, so $C_\Delta\cap\Sigma'=\varnothing$.

For part~(c), every pulling chain is a complete face chain. By Lemma~\ref{lem:simple}(c), their number is at most
$$d!\,|\vertset(\cP(\Sigma'))|
=\Oh_d((|\Sigma'|+d)^k)
=\Oh_d(|\Sigma'|^k).$$
Let $\Cenv$ be the implied constant in this bound, increased to at least one.

For part~(d), each vertex is the unique intersection of the $d$ hyperplanes in its active set. Every such set is a subset of the $|\Sigma|+d$ fixed source and cap constraints. Thus the owner and $d$ vertex active sets have at most
$$|\Sigma|(|\Sigma|+d)^{d^2}\le(|\Sigma|+d)^{d^2+1}$$
possible encodings. Each encoding determines the vertices and owner, and hence the cell; because the source functions are fixed, its conflict set is fixed as well.
\end{proof}

\begin{proof}[Proof of Lemma~\ref{lem:envelope}]
Write~$\cP=\cP(\Sigma')$~and let~$\cH$~be its set of~$|\Sigma'|+d$~constraints. By Lemma~\ref{lem:envfacts}(c), these constraints are in general position and their intersection is bounded. The point~$\vct{x}_{\mathrm{top}}=(\vct{0},\Ztop)$~is interior by the cap inequalities in Appendix~\ref{app:geom-envelope}. To descend from a face $F$ through a point $\vct x\in F\setminus\{\vct v_F\}$, put $\vct w_{\mathrm{ray}}=\vct x-\vct v_F$ and extend the ray to its exit point $\vct x'=\vct v_F+\tau_*\vct w_{\mathrm{ray}}$, where
$$\tau_H=\frac{b_H-\vct a_H\cdot\vct v_F}{\vct a_H\cdot\vct w_{\mathrm{ray}}},\qquad
\tau_*=\min_{\vct a_H\cdot\vct w_{\mathrm{ray}}>0}\tau_H\ge1.$$
By Lemma~\ref{lem:simple}(b), every minimizing constraint defines a facet $F'=F\cap\partial H$ omitting $\vct v_F$, and $\vct x=(1-1/\tau_*)\vct v_F+(1/\tau_*)\vct x'$ is the convex-combination relation used in the descent.

\emph{Faces.} We begin by enumerating the vertices and their active constraints with Proposition~\ref{prop:facelattice}. Use the~$d$~cap constraints and the source constraint of some~$s\in\Sigma'$~as the initial constraints. Their intersection is~$\cP(\{s\})$, which is bounded and has nonempty interior by Lemma~\ref{lem:envfacts}(c). The output consists of all vertices~$\vct{v}$~of~$\cP$~with their active sets~$\mathcal S_{\vct{v}}$.

These active sets determine every face. By Lemma~\ref{lem:simple}(b), a set~$\mathcal T\subseteq\cH$~is the active set of a nonempty face if and only if~$\mathcal T\subseteq\mathcal S_{\vct{v}}$~for some vertex~$\vct{v}$; the vertices of~$F_{\mathcal T}$~are exactly those with~$\mathcal T\subseteq\mathcal S_{\vct{v}}$. List the~$2^d|\vertset(\cP)|$~pairs~$(\mathcal T,\vct{v})$~with~$\mathcal T\subseteq\mathcal S_{\vct{v}}$~and sort by~$\mathcal T$, represented as a sorted tuple of indices. This groups the vertices of each nonempty face~$F$~and identifies its least vertex~$\vct{v}_F$.

The facets of~$F$~are exactly~$F_{\mathcal S_F\cup\{H\}}$~with~$H\in\mathcal S_{\vct{v}}\setminus\mathcal S_F$~for some vertex~$\vct{v}\in F$. Each such face has dimension~$\dim F-1$~by Lemma~\ref{lem:simple}(b), and every facet of~$F$~contains a vertex whose active set contains its own. Thus, for every pair~$(\mathcal T,\vct{v})$~and every~$H\in\mathcal S_{\vct{v}}\setminus\mathcal T$, form~$(\mathcal T,\mathcal T\cup\{H\})$~and remove repetitions by sorting. There are at most~$d\,2^d|\vertset(\cP)|$~such pairs. We store the faces in a dictionary indexed by their active sets, together with these facet lists. For each~$F$~with~$\dim F\ge1$, let~$\operatorname{next}(F)$~be the first facet in this order that omits~$\vct{v}_F$.

\emph{Cells.} A depth-first search now enumerates the pulling chains, starting at each nonempty source facet and, at a face~$F$, continuing into the facets~$F'$~with~$\vct{v}_F\notin F'$. Every branch extends to a complete chain, and at most~$\dim F<d$~facets are skipped at~$F$; their active sets have the form~$\mathcal S_F\cup\{H\}$~with~$H\in\mathcal S_{\vct{v}_F}\setminus\mathcal S_F$. Each other facet starts a branch, so the search uses~$\Oh(d^2)$~steps per complete chain.

Assign each chain its label from Definition~\ref{def:pulling-cells}. The labels are distinct by the proof of Lemma~\ref{lem:cells}; sorting them produces the list in part~(a) and a dictionary from labels to positions. Each label and its vertices have $\Oh_d(\lambda)$-bit descriptions, so comparisons require time $\softO(1)$. There are at most $d!\,|\vertset(\cP)|$ chains by Lemma~\ref{lem:simple}(c).

\emph{Ray shooting.} Build the structure of Proposition~\ref{prop:rayshoot} for~$\cH$, using~$\vct{x}_{\mathrm{top}}$~as the interior point and the~$d$~cap constraints together with the source constraint used above as~$\mathcal B$. Their intersection is the bounded simplex~$\cP(\{s\})$.

\emph{Location.} Given~$\vct{u}$, first shoot a vertical ray from~$(\vct{u},\Ztop)$~in direction~$(\vct{0},-1)$. Its origin is interior by Lemma~\ref{lem:container}(a), and part~(c) identifies the answer as the source constraint of some~$\sown\in\Sigma'$~with~$\wt f_{\sown}(\vct{u})=E_{\Sigma'}(\vct{u})$. Thus the exit point~$\vct{x}^{(0)}=(\vct{u},E_{\Sigma'}(\vct{u}))$~lies in the nonempty facet~$F^{(0)}$~of~$\sown$. We then descend through faces: for~$i\ge0$, suppose we have a face~$F^{(i)}$~of dimension~$d-1-i$~and a point~$\vct{x}^{(i)}\in F^{(i)}$.
\begin{itemize}
\item If~$i=d-1$, then~$F^{(i)}$~is a vertex and~$\vct{x}^{(i)}=\vct{v}_{F^{(i)}}$; stop.
\item If~$i<d-1$~and~$\vct{x}^{(i)}=\vct{v}_{F^{(i)}}$, complete the chain by~$F^{(i+1)}=\operatorname{next}(F^{(i)})$,~$F^{(i+2)}=\operatorname{next}(F^{(i+1)})$, and so on, and stop.
\item Otherwise, set~$F=F^{(i)}$~and~$\vct{x}=\vct{x}^{(i)}$~in the formulas for the exit point above, so that~$\vct{w}_{\mathrm{ray}}=\vct{x}^{(i)}-\vct{v}_{F^{(i)}}$, and set~$\vct{v}_{\mathrm{sh}}=(1-\delta_{\mathrm{sh}})\vct{v}_{F^{(i)}}+\delta_{\mathrm{sh}}\vct{x}_{\mathrm{top}}$, with~$\delta_{\mathrm{sh}}\in(0,1)$~fixed in Appendix~\ref{app:ray-shifts}. The point~$\vct{v}_{\mathrm{sh}}$~is interior, since~$\vct{a}_H\cdot \vct{v}_{\mathrm{sh}}<b_H$~for every~$H\in\cH$. Query the structure with the ray~$\{\vct{v}_{\mathrm{sh}}+\tau_{\mathrm{ray}}\vct{w}_{\mathrm{ray}}:\tau_{\mathrm{ray}}\ge0\}$. Let~$H$~be the answer, compute~$\tau_*=\tau_H$~and~$\vct{x}^{(i+1)}=\vct{v}_{F^{(i)}}+\tau_*\vct{w}_{\mathrm{ray}}$, and let~$F^{(i+1)}=F_{\mathcal S_{F^{(i)}}\cup\{H\}}$, found in the dictionary.
\end{itemize}
Finally, find the resulting chain in the dictionary of tuples and return its position in the canonical cell list and its owner~$\sown$.

\emph{Correctness of a step.} The shift~$\delta_{\mathrm{sh}}$~from Appendix~\ref{app:ray-shifts} ensures that the constraint returned by the interior ray minimizes the original exit parameter~$\tau_H$. The exit point thus lies in a facet omitting the least vertex, with $\vct{x}^{(i)}\in\conv\{\vct{v}_{F^{(i)}},\vct{x}^{(i+1)}\}$.

\emph{Correctness of the location.} Each step selects a facet omitting the least vertex, and $\operatorname{next}$ completes the chain with the same property if the descent stops early. The resulting faces form a pulling chain, and iterating the convex-combination relation shows that
$$\vct x^{(0)}\in\conv\{\vct v_{F^{(0)}},\ldots,\vct v_{F^{(d-1)}}\}.$$
Thus $\vct u$, the projection onto the first $d-1$ coordinates, belongs to the enumerated cell of this chain, whose owner is $s_\Delta=\sown$.

\emph{Cost.} The sample determines $\Oh_d(|\Sigma'|^k)$ vertices, face incidences, and pulling chains, so enumeration, sorting, and preprocessing for ray shooting require expected time $\softO(|\Sigma'|^k)$, including the exact arithmetic of Appendix~\ref{app:geometric-arithmetic}. Locating a cell requires at most $d$ ray queries and a constant number of dictionary lookups in $\softO(1)$ total time, with coordinates obtained at bounded depth as in Appendix~\ref{app:ray-shifts}.\qedhere
\end{proof}

\subsection{Sampling and Potential Estimates}\label{app:geom-potential}

To fix the sampling parameters~\eqref{eq:sampling}, let $\cconf\ge1$ be an integer depending only on $d$ such that for every integer $x\ge1$,
$$(x+d)^{d^2+1}\lceil\log_2(x+8)\rceil\le(x+8)^{\cconf},\qquad
\Cenv x^k\lceil\log_2(x+8)\rceil\le(x+8)^{\cconf}.$$
Such a constant exists because both counts grow polynomially in $x$ times a logarithmic factor. Set
$$G=\lceil\log_2(N+8)\rceil,\qquad
\Csamp=64(\cconf+1),\qquad s_0=2\Csamp G,\qquad \Gamma=4^{k+1}.$$
Then $G\ge4$ exceeds the number of prefix levels used below, and $(N+8)^{\cconf-\Csamp}<1/8$.

\begin{proof}[Proof of Lemma~\ref{lem:sample}]
By Lemma~\ref{lem:cells}(d), at most~$(|\Sigma|+d)^{d^2+1}$~owner--cell pairs~$(s_\Delta,\Delta)$~occur over all nonempty subsets of~$\Sigma$, and each pair determines its conflict set~$C_\Delta$. These pairs and their conflict sets are determined before sampling by the shared perturbation (Definition~\ref{def:envelope}). Consider such a pair with~$|C_\Delta|>\Csamp|\Sigma|G/|\Sigma'|$. If this pair occurs in~$\cP(\Sigma')$, then~$\Sigma'\cap C_\Delta=\varnothing$~by Lemma~\ref{lem:cells}(b). The probability of this event is at most~$\e^{-|\Sigma'||C_\Delta|/|\Sigma|}$~by Lemma~\ref{lem:hypergeom}, and
$$\e^{-|\Sigma'||C_\Delta|/|\Sigma|}<\e^{-\Csamp G}\le2^{-\Csamp G}\le(N+8)^{-\Csamp}.$$
By the definition of~$\cconf$~with~$x=N$, and since~$|\Sigma|\le N$, the number of pairs is at most~$(N+d)^{d^2+1}\le(N+8)^{\cconf}/G$. A union bound over the pairs with large conflict sets implies failure probability at most $(N+8)^{\cconf-\Csamp}/G<1/(8G)$.
\end{proof}

\begin{proof}[Proof of Lemma~\ref{lem:heavy}]
Let~$\mathcal D_{\mathrm h}\subseteq\mathcal D$~be the set of heavy~$\Delta$. Since~$\sum_\Delta m_\Delta\le m$~and every heavy~$\Delta$~has~$m_\Delta>\Gamma n_\kappa^{1/k}$, we have~$|\mathcal D_{\mathrm h}|\le m/(\Gamma n_\kappa^{1/k})$. By~$n_\Delta\le n_\kappa$~and by the concavity of the power function with exponent~$k/(k+1)$, under which~$\sum_{\Delta\in\mathcal D_{\mathrm h}}m_\Delta^{k/(k+1)}\le|\mathcal D_{\mathrm h}|^{1/(k+1)}\bigl(\sum_{\Delta\in\mathcal D_{\mathrm h}}m_\Delta\bigr)^{k/(k+1)}$,
$$\begin{aligned}\sum_{\Delta\ \mathrm{heavy}}(m_\Delta n_\Delta)^{k/(k+1)}&\le n_\kappa^{k/(k+1)}\,|\mathcal D_{\mathrm h}|^{1/(k+1)}\,m^{k/(k+1)}\\&\le n_\kappa^{k/(k+1)}\Bigl(\frac{m}{\Gamma n_\kappa^{1/k}}\Bigr)^{1/(k+1)}m^{k/(k+1)}=\Gamma^{-1/(k+1)}\,m\,n_\kappa^{(k-1)/k},\end{aligned}$$
where we used~$k/(k+1)-1/(k(k+1))=(k-1)/k$. Here~$\Gamma^{1/(k+1)}=4$, and~$m\,n_\kappa^{(k-1)/k}=m(n/\kappa)^{(k-1)/k}\le\Phi(m,n)$~by \eqref{eq:balance-estimates}, proving the first bound. For the second,
$$\sum_{\Delta\ \mathrm{heavy}}(m_\Delta+n_\Delta)\le m+|\mathcal D_{\mathrm h}|\,n_\kappa\le m+\frac{m\,n_\kappa^{(k-1)/k}}{\Gamma}\le\Phi(m,n)+\frac{\Phi(m,n)}{\Gamma},$$
by \eqref{eq:balance-estimates}. Finally,~$m_\Delta n_\Delta\le m\,n_\kappa=mn/\kappa\le mn/2$, because~$\kappa\ge2$.
\end{proof}

\paragraph{Summing recursive costs.}
For a recursion satisfying the child-size bounds of Lemma~\ref{lem:heavy}, the sum of the potentials at depth $\ell$ is at most $4^{-\ell}\Phi(m,n)$, where $m,n$ are the root sizes. Consequently,
$$\sum_v\Phi(m_v,n_v)\le\frac43\Phi(m,n),\qquad
\sum_v(m_v+n_v)\le m+n+\Oh_d\!\left(\sum_{v\ \mathrm{internal}}\Phi(m_v,n_v)\right).$$
The recursion has finite depth because $m_\Delta n_\Delta\le m_vn_v/2$. If each node has conditional expected cost $\softO(\Phi(m_v,n_v)+m_v+n_v)$ with uniform numerical factors, summing these bounds and taking expectations establishes total expected time $\softO(F_d(m,n))$.

\subsection{Implementing Batch Maximum Inner Product}
\label{app:geom-batch}

Finally, we prove Proposition~\ref{prop:batchsupport} for $d\ge4$  ($k\ge2$) by adapting the envelope-sampling and heavy/light recursion framework of Section~\ref{subsec:truncated} to exact maximization. Working in the setup of Definition~\ref{def:envelope} with sources $\vct{y}_s=(\vct{p}_s,c_s)$, $s\in\Sigma_0$ and a nonempty finite list of rational targets $(\vct{u}_i)_{i\in\Theta_0}$ in $[-1,1]^{d-1}$, we use the sampling parameters~\eqref{eq:sampling} with $N\geq |\Sigma_0|+|\Theta_0|$ and bit bound $\lambda\geq \log_2(N+8)$. Both procedures below take nonempty subsets $\Theta\subseteq\Theta_0$ and $\Sigma\subseteq\Sigma_0$ and return the exact maximum of $f_s(\vct{u}_i)=\vct{u}_i\cdot\vct{p}_s+c_s$ over $s\in\Sigma$ for every $i\in\Theta$.

As in Section~\ref{subsec:bp-setup}, calls restrict the original point sets and share one perturbation and precision. Appendix~\ref{app:geometric-arithmetic} provides $\Oh_d(\lambda)$-bit representations for their geometric quantities.

\paragraph{Procedure~\textnormal{\textsc{ThinMax}}$(\Theta,\Sigma)$.}
We permute $\Sigma$ uniformly at random and process doubling prefixes, from one source to all of $\Sigma$, maintaining for each target $i\in\Theta$ its current maximum $h_i$ and a maximizing index. By duality, the reporting structure is built on lifted targets and queried by sources. In each round, we apply Lemma~\ref{lem:reporttree} with group size $g_{\mathrm{rep}}=|\Theta|$ to the points $(\vct u_i,h_i)$, then query with direction $(\vct p_s,-1)$ and threshold $-c_s$ for each source $s$ in the next batch. The values $h_i$ stay fixed during these reports; afterward, we update the maxima and maximizing indices using all reported values.

\begin{lemma}
\label{lem:thinmax}
Procedure~\textnormal{\textsc{ThinMax}}$(\Theta,\Sigma)$ returns $\max_{s\in\Sigma}f_s(\vct{u}_i)$ and a maximizing index for every $i\in\Theta$, in expected running time $\softO(|\Theta|^k+|\Sigma|)$.
\end{lemma}

\begin{proof}
The reporting inequality is exactly $f_s(\vct{u}_i)>h_i$, so each update computes the maximum over the enlarged prefix. For the cost, let $q$ and $q_+\le2q$ be the sizes of two consecutive prefixes. Conditional on the larger prefix, its preceding prefix is a uniform $q$-element subset. For a fixed target, order these $q_+$ sources by decreasing value, breaking ties by index. Every reported source precedes the first sampled source, so the expected number reported is at most
$$\frac{q_+-q}{q+1}\le1.$$
Ties can only reduce this count. There are $\Oh(\log(N+2))$ rounds, each preprocessing $|\Theta|$ points in expected time $\softO(|\Theta|^k)$. Each source is queried at most once, and the expected total number of reports is $\Oh(|\Theta|\log(N+2))$, proving the bound.
\end{proof}

\paragraph{Procedure~\textnormal{\textsc{Maxima}}$(\Theta,\Sigma)$.}
Let $m=|\Theta|$ and $n=|\Sigma|$, both positive. If $m^k\le n$, run \textsc{ThinMax} and return. If $n^k\le(4\Csamp G)^km$, use the support-query structure of Lemma~\ref{lem:support} to answer every target directly and return.

Otherwise, compute $\kappa,n_\kappa,s_f$ from Section~\ref{subsec:truncated} and build the reporting structure on the perturbed sources with group size $\kappa$. Here $s_f<n$ by \eqref{eq:fullsample}. Repeatedly draw a uniform sample of size $s_f$, build its envelope cells, and compute each conflict set by the $d$ vertex reports of Lemma~\ref{lem:cells}(b). Reject the sample at the first cell with $|C_\Delta|>n_\kappa/2$, and otherwise accept it. Locate the targets in the accepted cells, obtaining groups $\Theta_\Delta$. For each nonempty group, return the larger of the owner's original value and the maximum over $C_\Delta$. An empty conflict set needs no further call; otherwise use \textsc{ThinMax} when $|\Theta_\Delta|^k\le\Gamma^kn_\kappa$ and recurse with \textsc{Maxima} when this inequality fails.

\begin{lemma}
\label{lem:maxima}
Procedure~\textnormal{\textsc{Maxima}}$(\Theta,\Sigma)$ is a Las Vegas algorithm returning $\max_{s\in\Sigma}f_s(\vct{u}_i)$ and a maximizing index for every $i\in\Theta$. Its expected running time is $\softO(\Phi(m,n)+m+n)$, where $m=|\Theta|$ and $n=|\Sigma|$.
\end{lemma}

\begin{proof}
For a target in cell $\Delta$, the owner or some source in $C_\Delta$ maximizes the perturbed score over $\Sigma$ and, by Lemma~\ref{lem:envfacts}(b), also maximizes the original score. Thus
$$\max_{s\in\Sigma}f_s(\vct{u}_i)
=\max\Bigl\{f_{s_\Delta}(\vct{u}_i),\ \max_{s\in C_\Delta}f_s(\vct{u}_i)\Bigr\},$$
where $\max\varnothing=-\infty$. This proves correctness by induction on $mn$, since Lemma~\ref{lem:heavy} halves this product in every recursive child.

Both thin cases run in expected time $\softO(m+n)$. At a balanced node, the reporting structure requires expected preprocessing time $\softO(n\kappa^{k-1})$. Each sampling attempt has $\softO(\kappa^k)$ cells, with reporting cost $\softO(n_\kappa)$ per cell passing the conflict check and at most $\softO(n)$ for the first failing cell. The expected cost per attempt is thus $\softO(n\kappa^{k-1})$. Since each attempt is accepted with probability at least $7/8$ (Lemma~\ref{lem:sample}), the expected total over all attempts has the same order. Each light cell costs $\softO(n_\kappa)$ in expectation by Lemma~\ref{lem:thinmax}; locating targets and returning answers adds $\softO(m)$. By~\eqref{eq:balance-estimates}, the total nonrecursive work is $\softO(\Phi(m,n)+m+n)$.

The target groups are disjoint and every conflict set has size at most $n_\kappa/2$. Summing over the recursion using Lemma~\ref{lem:heavy} and Appendix~\ref{app:geom-potential} yields total expected time $\softO(F_d(m,n))$.
\end{proof}

\begin{proof}[Proof of Proposition~\ref{prop:batchsupport}]
For $d\le3$, the structure of Lemma~\ref{lem:support} answers all queries deterministically in $\softO(m+n)$ time. For $d\ge4$, partition the nonzero queries into the at most $2d$ groups of Definition~\ref{def:normalization}. In each group $\pi$, apply \textsc{Maxima} to sources $\pi\vct k_s$ ($s\in[n]$) and targets $\vct u_{\vct q_i}$, with $U=2^\lambda$ for an input bit bound $\lambda$. Each returned index maximizes the original inner product because $\vct q_i\cdot\vct k_s=\beta_{\vct q_i}f^\pi_{\vct k_s}(\vct u_{\vct q_i})$ and $\beta_{\vct q_i}>0$. Summing the running-time bounds in Lemma~\ref{lem:maxima} over the groups and evaluating the selected inner products exactly requires $\softO(F_d(m,n))$ time.
\end{proof}

\section*{AI Methodology}
\addcontentsline{toc}{section}{AI Methodology}
We used DeepSeek Harness to assist with the proof of Proposition~\ref{prop:BP}. The tool materially affected Section~\ref{sec:biased}. More details can be found in the footnote in Section~\ref{subsec:overview-prefixes}. The authors verified the correctness and originality of all content, including references.

\section*{Acknowledgment}
\addcontentsline{toc}{section}{Acknowledgment}
This work was supported by the National Key R\&D Program of China, Project Number 2025YFA1016700. This work was also supported by the National Natural Science Foundation of China (NSFC) under Grant No. 62576011.
\printbibliography[heading=bibintoc]
\end{document}